\documentclass[twocolumn]{aastex631}
\newcommand{\msun}{$\mathcal{M}_{\odot}$}    
\newcommand{\ngc}{$N_\textrm{GC}^{\textrm TOT}$}  

\newcommand{\grad}{$^{\circ}$}

\newcommand{\fwhm}{{\sc fwhm}}
\newcommand{\class}{{\sc class\_star}}
\newcommand{\fluxradius}{{\sc flux\_radius}}
\newcommand{\spread}{{\sc spread\_model}}

\newcommand{\magpsf}{{\sc mag\_psf}}
\newcommand{\magerrpsf}{{\sc magerr\_psf}}
\newcommand{\magauto}{{\sc mag\_auto}}
\newcommand{\aimage}{{\sc a\_image}}
\newcommand{\bimage}{{\sc b\_image}}

\newcommand{\rv}{$R_{\textrm{vir}}$}
\newcommand{\airmas}{{\sc airmass}}
\newcommand{\satur}{{\sc satur\_level}}
\newcommand{\median}{{\sc median}}
\newcommand{\rms}{{\sc rms}}
\newcommand{\ra}{{\sc ra}}
\newcommand{\dec}{{\sc dec}}
\graphicspath{{figures/}}
\usepackage{ragged2e}

\begin{document}

\title{The S-PLUS Fornax Project (S+FP): Mapping globular clusters systems within 5 virial radii around NGC\,1399}

\author[0000-0003-2127-2841]{Luis Lomel\'i-N\'u\~nez}
\affiliation{Universidade Federal do Rio de Janeiro, Observat\'orio do Valongo, Ladeira do Pedro Ant\^onio, 43, Sa\'ude CEP 20080-090 Rio de Janeiro, RJ, Brazil}

\author[ 	
0000-0002-0620-136X]{A. Cortesi}
\affiliation{Universidade Federal do Rio de Janeiro, Observat\'orio do Valongo, Ladeira do Pedro Ant\^onio, 43, Sa\'ude CEP 20080-090 Rio de Janeiro, RJ, Brazil}
\affiliation{Instituto de Física, Universidade Federal do Rio de Janeiro, 21941-972, Rio de Janeiro, RJ,
Brazil}

\author[0009-0007-2396-0003]{A.V. Smith Castelli}
\affiliation{Instituto de Astrof\'isica de La Plata, CONICET-UNLP, Paseo del Bosque s/n, La Plata, B1900FWA, Argentina}
\affiliation{Facultad de Ciencias Astron ómicas y Geof\'isicas, Universidad Nacional de La Plata, Paseo del Bosque s/n, La Plata, B1900FWA, Argentina}

\author[0000-0003-3153-8543]{M. L. Buzzo}
\affiliation{Centre for Astrophysics and Supercomputing, Swinburne University, John Street, Hawthorn VIC 3122, Australia}
\affiliation{European Southern Observatory, Karl-Schwarzschild-Strasse 2, 85748 Garching bei M\"unchen, Germany }
\affiliation{ARC Centre of Excellence for All Sky Astrophysics in 3 Dimensions (ASTRO 3D), Australia}

\author[0000-0002-4677-0516]{Y.D. Mayya}
\affiliation{Instituto Nacional de Astrof\'isica \'Optica y Electr\'onica, Luis Enrique Erro 1, Tonantzintla 72840, Puebla, Mexico}

\author[0000-0002-8634-4204]{Vasiliki Fragkou}
\affiliation{Universidade Federal do Rio de Janeiro, Observat\'orio do Valongo, Ladeira do Pedro Ant\^onio, 43, Sa\'ude CEP 20080-090 Rio de Janeiro, RJ, Brazil}

\author[ 	
0000-0002-8351-8854]{J. A., Alzate-Trujillo}
\affiliation{Centro de Estudios de Física del Cosmos de Arag\'on}

\author[0009-0005-6830-1832]{R. F. Haack}
\affiliation{Instituto de Astrof\'isica de La Plata, CONICET-UNLP, Paseo del Bosque s/n, La Plata, B1900FWA, Argentina}
\affiliation{Facultad de Ciencias Astron ómicas y Geof\'isicas, Universidad Nacional de La Plata, Paseo del Bosque s/n, La Plata, B1900FWA, Argentina}

\author[0000-0003-2931-2932]{J.P. Calder\'on}
\affiliation{Instituto de Astrof\'isica de La Plata, CONICET-UNLP, Paseo del Bosque s/n, La Plata, B1900FWA, Argentina}
\affiliation{Facultad de Ciencias Astron ómicas y Geof\'isicas, Universidad Nacional de La Plata, Paseo del Bosque s/n, La Plata, B1900FWA, Argentina}

\author[0000-0002-6164-5051]{A. R. Lopes}
\affiliation{Instituto de Astrof\'isica de La Plata, CONICET-UNLP, Paseo del Bosque s/n, La Plata, B1900FWA, Argentina}

\author[0000-0002-2363-5522
]{Michael Hilker}
\affiliation{European Southern Observatory, Karl-Schwarzschild-Strasse 2, 85748 Garching bei München, Germany}

\author[ 	
0000-0003-4675-3246]{M. Grossi}
\affiliation{Universidade Federal do Rio de Janeiro, Observat\'orio do Valongo, Ladeira do Pedro Ant\^onio, 43, Sa\'ude CEP 20080-090 Rio de Janeiro, RJ, Brazil}

\author{Karín Men\'endez-Delmestre}
\affiliation{Universidade Federal do Rio de Janeiro, Observat\'orio do Valongo, Ladeira do Pedro Ant\^onio, 43, Sa\'ude CEP 20080-090 Rio de Janeiro, RJ, Brazil}

\author{Thiago S. Gonçalves}
\affiliation{Universidade Federal do Rio de Janeiro, Observat\'orio do Valongo, Ladeira do Pedro Ant\^onio, 43, Sa\'ude CEP 20080-090 Rio de Janeiro, RJ, Brazil}

\author[0000-0003-3220-0165]{Ana L. Chies-Santos}
\affiliation{Instituto de F\'isica, Universidade Federal do Rio Grande do Sul, Av. Bento Gonçalves 9500, Porto Alegre, RS, Brazil}

\author[0000-0002-9891-8017]{L.~A.~Guti\'{e}rrez-Soto}
\affiliation{Instituto de Astrof\'isica de La Plata, CONICET-UNLP, Paseo del Bosque s/n, La Plata, B1900FWA, Argentina}

\author[ 	
0009-0006-0373-8168]{Ciria Lima-Dias}
\affiliation{Instituto Multidisciplinario de Investigaci\'on y Postgrado, Universidad de La Serena, Ra\'ul Bitr\'an 1305, La Serena, Chile}
\affiliation{Departamento de Astronom\'ia, Universidad de La Serena, Av. Cisternas 1200, La Serena, Chile}

\author[0000-0002-7186-7889]{S. V. Werner}
\affiliation{Centre for Extragalactic Astronomy, Durham University, South Rd., Durham, DH1 3LE, UK}
\affiliation{Institute for Computational Cosmology, Durham University, South Rd., Durham, DH1 3LE, UK}

\author[0000-0003-3537-4849]{Pedro K. Humire}
\affiliation{Instituto de Astronomia, Geofísica e Ciências Atmosféricas, Rua do Matão, 1226, Cidade Universitária, São Paulo 05508-090, Brazil}

\author[0000-0003-2435-8528]{R. C. Thom de Souza}
\affiliation{Campus Avan\c{c}ado em Jandaia do Sul, Universidade Federal do Paran\'a, Jandaia do Sul, PR, 86900-000, Brazil}
\affiliation{Programa de P\'os-gradua\c{c}\~ao em Ciência da Computa\c{c}\~ao, Universidade Estadual de Maring\'a, Maring\'a, PR}

\author[ 	
0000-0002-5045-9675]{A. Alvarez-Candal}
\affiliation{Instituto de Astrof\'isica de Andaluc\'ia, CSIC,  Apt 3004, E18080 Granada, Spain}

\author[0000-0002-5854-7426]{Swayamtrupta Panda}
\affiliation{Laborat\'orio Nacional de Astrof\'isica, MCTI, Rua dos Estados Unidos, 154, Bairro das Na\c c\~oes, Itajub\'a, MG 37501-591, Brazil}
\affiliation{International Gemini Observatory/NSF NOIRLab, Casilla 603, La Serena, Chile}

\author[ 	
0000-0002-4175-4728]{Avinash Chaturvedi}
\affiliation{Leibniz-Institut für Astrophysik Potsdam (AIP), An der Sternwarte 16, 14482 Potsdam, Germany}

\author[0000-0002-8280-4445]{E. Telles}
\affiliation{Observatório Nacional, Rua General José Cristino, 77, Bairro São Cristóvão, Rio de Janeiro 20921-400, Brazil}

\author[0000-0002-5267-9065]{C. Mendes de Oliveira}
\affiliation{Instituto de Astronomia, Geofísica e Ciências Atmosféricas, Rua do Matão, 1226, Cidade Universitária, São Paulo 05508-090, Brazil}

\author{A. Kanaan} 
\affiliation{Departamento de Física - CFM - Universidade Federal de Santa Catarina, PO BOx 476, 88040-900, Florianópolis, SC, Brazil}

\author[0000-0002-0138-1365]{T. Ribeiro}
\affiliation{Rubin Observatory Project Office, 950 N. Cherry Ave., Tucson, AZ 85719, USA}

\author[ 	
0000-0002-4064-7234]{W. Schoenell}
\affiliation{The Observatories of the Carnegie Institution for Science, 813 Santa Barbara St, Pasadena, CA 91101, USA}

\begin{abstract}

We present the largest sample ($\sim$13,000 candidates, $\sim$3000 of wich are bona-fide candidates) of globular cluster (GCs) candidates reported in the Fornax Cluster so far. 
The survey is centered on the NGC~1399 galaxy, extending out to 5 virial radii (\rv) of the cluster.
We carried out a photometric study using images observed in the 12-bands system 
of the Southern Photometric Local Universe Survey (S-PLUS), corresponding to 106 pointings, 
covering a sky area of $\sim$208~square~degrees. 
Studying the properties of spectroscopically confirmed GCs, we have designed a method to select 
GC candidates using structural and photometric parameters. 
We found evidence of color bimodality in 2 broad bands colors, namely $(g-i)_{0}$ and $(g-z)_{0}$,
while, in the narrow bands, we did not find strong statistical evidence to confirm bimodality in any color. 
We analyzed the GCs luminosity functions (GCLF) in the 12-bands of S-PLUS, and we can highlight two points: 
a) due to the relatively shallow depth of S-PLUS, it is only possible
to observe the bright end of the GCLF and, 
b) at that level, in all the bands it
can be appreciated the log-normal distribution typical for GC systems.
With the spatial coverage reached in this study, we
are able for the first time explore the large scale distribution of GCs within and around a galaxy cluster.
In particular, we noted that the GCs might be clustered along substructures, which traces the current cluster build up.
\end{abstract}


\section{Introduction} \label{sec:intro}

Globular Clusters (GCs) are among the oldest objects in the Universe, which makes them a key component for understanding the formation and assembly history of galaxies \citep[]{Ashman:1998, Brodie:2006,Forbes:2018_gcs}. 
Their relatively high luminosities (M$_V=-5$ to $-10$ mag) and compact sizes (half-light radius of a few parsecs) allow them to be readily detectable in nearby galaxies \citep{Harris:1996}. 
It has been shown that GC systems of massive galaxies, especially of the metal-rich variety,
form through in-situ processes and continue assembling during processes of merging or accretion 
\citep[]{Kruijssen:2019, Reina:2022}. 
In dense environments such as galaxy clusters, GCs can be associated with individual cluster
galaxies or with the intracluster light \citep[][]{Dawe1976, Hanes1986, White1987, Harris1987, West1995, Bassino:2003, Williams2007, Schuberth:2008, Karla:2017, Lee2022, Harris:2024}, 
and can be used to estimate the dark matter content of the
galaxy cluster \citep[][]{Diego:2023, Reina:2023}. 
A variety of GC system properties that are potentially relevant to cosmological theories of galaxy formation have been identified. These include color distribution \citep{Larsen:2001, West:2004}, luminosity function \citep{Reed:1994, Whitmore:1995}, radial density distribution \citep{Bassino:2006, Kartha:2014}, specific frequency as a function of galaxy type \citep{Harris:1981, Peng:2008, Georgiev:2010}, and the nature of their size distribution \citep{Kundu:1998, Larsen:2001, Webb:2012}.
Besides, in the past two decades, different scaling relations have been found between the GC systems and their host galaxies \citep[e.g.,][]{Caso:2024}. These relations associate for instance the total number of GCs (\ngc) with the masses of super massive central black holes of their host galaxies \citep[e.g.][]{Burkert:2010, Harris:2011, Harris:2014, Lomeli:2017, Lomeli:2022cfht} and their host galaxy’s halo virial mass \citep[e.g.][]{Spitler:2009, Hudson:2014, Forbes:2018, Burkert:2020}, offering evidence of host galaxy-GCs-halo connection. The majority of these properties have been exhaustively reviewed in \citet{Brodie:2006}, \citet{Forbes:2018_gcs} and \citet{Beasley:2020}.

Being the second nearest rich galaxy cluster, Fornax (m-M=31.51, $\sim$19~Mpc; \citealt{Blakeslee:2009}), represents a remarkable environment where the processes involved in the formation and evolution of galaxies can be analysed in detail. 
In the literature there are a variety of studies focused on the Fornax cluster, which tackle different aspects of the cluster: the central galaxy NGC~1399 \citep[e.g.][]{Iodice:2016}, 
X-ray emission \citep[e.g.][]{Jones:1997}, dwarf galaxy population \citep[e.g.][]{Munoz:2015, Yasna:2018, Venhola:2019}, ultra-diffuse galaxies \citep[e.g.][]{Zaritsky:2023} 
and atomic neutral hydrogen gas \citep[e.g.][]{Serra:2023}. In particular, the GC system of Fornax has been studied in the past by different authors using both photometric \citep[e.g.][]{Kissler:1997, Ostrov:1998, Bassino:2006, Jordan:2015, Blakeslee:2012} and spectroscopic data \citep[e.g.][]{Bergond:2007, Schuberth:2010, Fahrion:2020, Avinash:2022}. With photometric GC studies it is possible to analyse a large number of GC candidates, which enables performing statistical analysis. Yet, in such studies there is a non negligible fraction of contaminants (such as foreground stars and background galaxies).   
On the other hand, spectroscopic GC studies are much more precise, but they are expensive in terms of telescope observing time, and the total number of recovered GCs is small compared to 
photometric studies.

\begin{figure*} 
    \includegraphics[height=11cm,width=17cm]{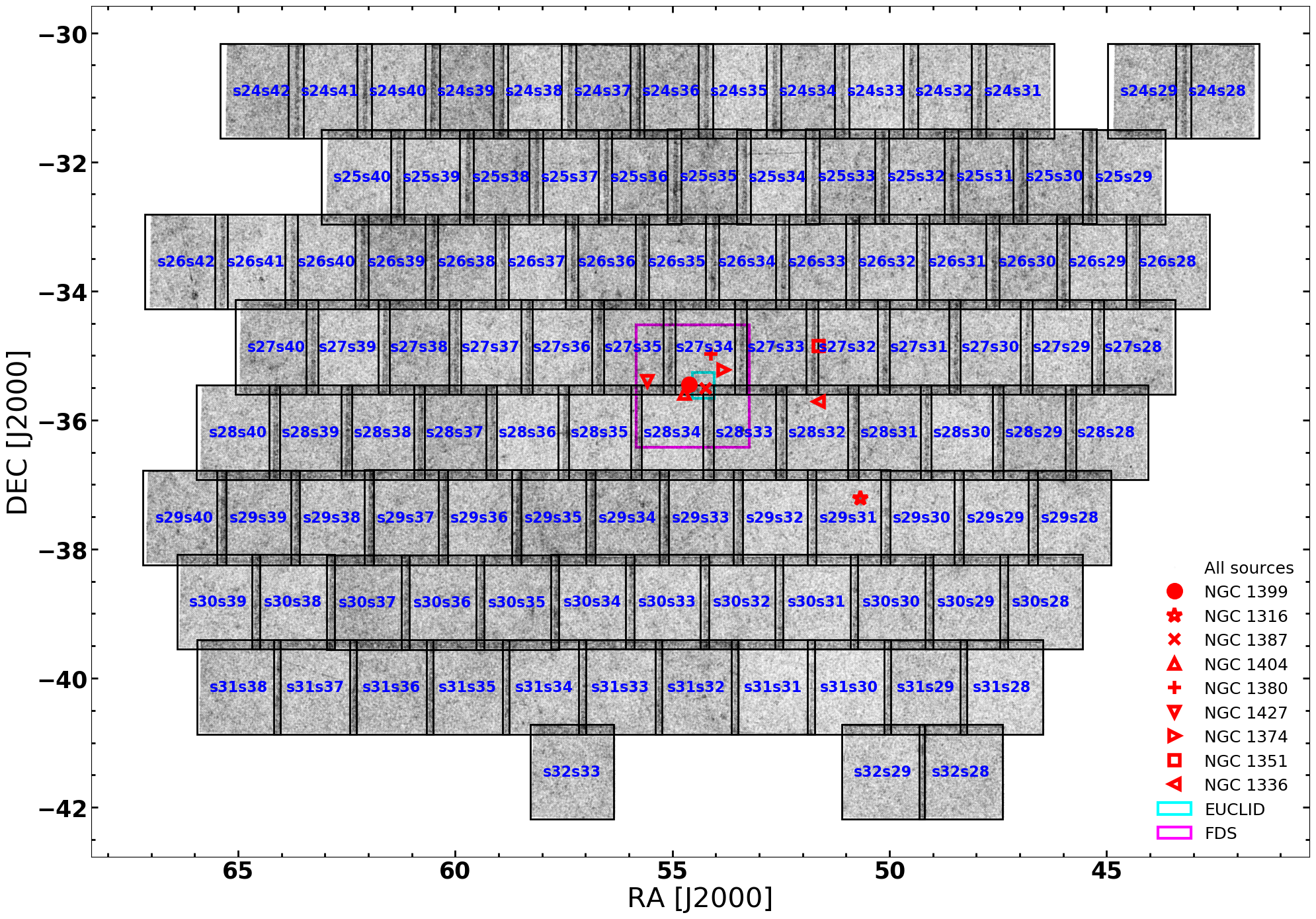}
    \caption{Data coverage of the 106 S-PLUS pointings in Fornax.
    {\it Cloud of black dots:} all detections in each FoV. A total of 3,085,787 sources were detected. A darker color implies a greater number of detected sources.
    {\it Black solid lines:} limits of each pointing.
    {\it Blue legends:} in the center of each pointing we show their key names from S-PLUS.
    {\it Red symbols:} indicate the central coordinates of some of the brightest galaxies in Fornax.
    {\it Magenta and cyan squares:} represent the FoVs of FDS  \citep{Iodice:2016} and EUCLID \citep{Saifollahi:2024} studies.    
    The image is aligned such that north is up and east is to the left.} 
    \label{figura:cover_data} 
\end{figure*}

Different works were dedicated to the analysis of the GC photometric data from the Fornax Deep Survey (FDS) taken with the VLT Survey Telescope (VST). For example, \citet[][]{Iodice:2016} found that the core of the Fornax cluster is characterised by a very extended and diffuse envelope surrounding 
the luminous galaxy NGG~1399; \citet[][]{Dabrusco:2016} report a density structure in the spatial distribution of  GC candidates in a region $\sim$0.5~deg$^{2}$ within the core of Fornax;
\citet{Cantiello:2018} obtained surface density maps, color distributions, and radial density profiles of GC candidates around NGC~1399. All of these works studied the GCs around NGC~1399 over a projected area of $\sim$10~deg$^{2}$ \citep[e.g.][]{Iodice:2016, Cantiello:2018}. In a recent work, \citet[][]{Saifollahi:2024}, using Euclid \citep[][]{Euclid:2022} observations of a 0.5~deg$^{2}$ field in the central region of Fornax, identified more than 5000 new GC candidates down to I$_{E}$ = 25.0~mag, about 1.5 mag fainter than the typical turn-over magnitude ($-$7.4 mag in V-band, \citealt[][]{Harris:1996, Jordan:2007, Villegas:2010})  of the  GC luminosity function, and investigate their spatial distribution within the intracluster field. 

In addition, there are other photometric surveys focused on the Fornax cluster with different objectives such as the Advanced Camera for Surveys Fornax Cluster Survey \citep[ACSFCS;][]{Jordan:2007fornax}, the Next Generation Fornax Survey \citep[NGFS;][]{Munoz:2015} and the Dark Energy Spectroscopic Instrument \citep[DESI;][]{Desi:2023}. In particular, the ACSFCS has targeted galaxies with the \textit{Hubble Space Telescope (HST)} and identified the GCs \citep[eg.][]{Jordan:2015} around them.
The NGFS is a deep multiwavelength survey that covers Fornax out to its virial radius (0.7 Mpc; \citealt{Drinkwater:2001}) and that have studied the dwarf galaxies in the central region of the cluster \citep[e.g.][]{Eigenthaler:2018, Yasna:2018, Evelyn:2020}.

The GCs in Fornax have been objects of different spectroscopic studies, focused on confirming them as members of the cluster by estimating their radial velocities \citep[e.g.][]{Minniti:1998, Schuberth:2010, Avinash:2022}, on the determination of ages and metallicities \citep[e.g.][]{Kissler:1998,Fahrion:2020} and on the identification of GCs belonging to the intra-cluster (IC) medium \citep[e.g.,][]{Bergond:2007, Schuberth:2008}. In particular, \citet{Avinash:2022} used spectroscopic data from the Visible Multi Object Spectrograph at the Very Large Telescope (VLT/VIMOS), covering one square degree around the central massive galaxy NGC 1399, to confirm a total of 777 GCs. Combined with previous literature radial velocity measurements of GCs in Fornax, they compiled the most extensive spectroscopic GC sample of 2341 objects in this environment. They found that red GCs are mostly concentrated around major galaxies, while blue GCs are kinematically irregular and are widely spread throughout the cluster. 

\begin{table}{}
    \setlength\tabcolsep{6.0pt}
	\begin{center}	    
	\caption{The S-PLUS Filter System.}
	\label{tabla:splus}
     \begin{scriptsize}
	\begin{tabular}{l c r r r} 
   
        \hline
Filter  &  $\lambda_{central}$ &   FWHM    &  $A_{\lambda}/A_{V}$ & Comments \\ 
        &  [\AA]           &   [\AA]       &                      &   \\  
(1)     &  (2)             &    (3)        &      (4)             & (5)  \\  
\hline  
 u      &    3577          &    325        &  1.584  &    Javalambre u     \\    
 J0378  &    3771          &    151        &  1.528  &    [OII]     \\    
 J0395  &    3941          &    103        &  1.483  &    Ca H+K     \\    
 J0410  &    4094          &    200        &  1.434  &    H$\delta$     \\    
 J0430  &    4292          &    200        &  1.364  &    G-band     \\    
 g      &    4774          &    1505       &  1.197  &    SDSS-like g     \\    
 J0515  &    5133          &    207        &  1.085  &    Mgb Triplet     \\    
 r      &    6275          &    1437       &  0.866  &    SDSS-like r     \\    
 J0660  &    6614          &    147        &  0.810  &    H$\alpha$     \\    
 i      &    7702          &    1507       &  0.646  &    SDSS-like i     \\    
 J0861  &    8611          &    410        &  0.518  &    Ca Triplet     \\    
 z      &    8882          &    1270       &  0.484  &    SDSS-like z     \\                           
	\hline
	\end{tabular}
    \end{scriptsize}
    \end{center}
           \begin{scriptsize}
            \tablecomments{
                (1) Filter name. 
                (2) Filter reference wavelength (\AA). 
                (3) Filter band-width (\AA). 
                (4) Milky Way extinction from \citet[][]{Cardelli:1989} 
                (5) Comparison with other filter systems. }
                \end{scriptsize}
\end{table}

Despite the extensive exploration of the GC system in the Fornax cluster using photometric data, all studies have focused on the central regions near NGC~1399 and have been performed in three or four photometric broad bands. In contrast, the Fornax images obtained by the Southern Photometric Local Universe Survey (S-PLUS; \citealt{Mendes:2019}), analysed in the context of the S-PLUS Fornax Project (S+FP; \citealt{Analia:2024}), provide coverage of approximately 208 square degrees. This allows for the largest study of GCs in a galaxy cluster to date, extending the study of GCs up to 5 virial radii (\rv) along the East-West direction.
In a pilot study, \citet{Buzzo:2022} assessed the effectiveness of identifying GCs in the Fornax cluster using S-PLUS images.
From the obtained photometry, they applied template fitting techniques to a sample of 115 GCs around NGC~1399 to recover photometric redshifts, as well as ages and metallicities for the GCs.
However, it should be stressed that the S-PLUS images are not deep enough (r$\sim$21.30 mag) to reach the faint end of the GC luminosity function (GCLF).

Apart from Fornax, the GC systems in the Virgo galaxy cluster has also been widely studied \citep[e.g.,][]{Hanes:1977, Cohen:1988, Harris:1991}. The ACS Virgo Cluster Survey \citep[ACSVCS,][]{Cote:2004} is an HST-ACS imaging program of 100 early-type galaxies in the Virgo Cluster. With this data, it was found that the galaxies of the Virgo cluster on average appear to have bimodal or asymmetric GCs color distributions \citep[][]{Peng:2006}; the GCLF turnover is roughly constant in bright galaxies, but it decreases slightly in dwarf galaxies  \citep[][]{Jordan:2007}. 
The Next Generation Virgo Survey \citep[NGVS/NGVS-IR,][]{Ferrarese:2012, Munoz:2014} covers 104~deg$^{2}$ in 6-bands (u*grizK$_{s}$) and extra deep observations in the g band, g=25.90~mag. These data allowed the study of the GC system until the virial radius of the Virgo cluster. In particular, in the spatial distribution of GCs in 100~deg$^{2}$ it was found a difference in concentration between red (more concentrated) and blue (more extended) GCs over the full extent of the cluster \citep[][]{Durrell:2014}; and it was found the possible existence of substructures in the GC population around the Virgo cD galaxy M87 \citep[][]{Powalka:2018}.   

\begin{figure*} 
    \centering
    \includegraphics[width=0.5\columnwidth]{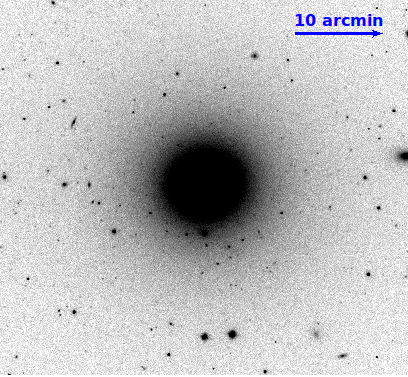}
    \includegraphics[width=0.5\columnwidth]{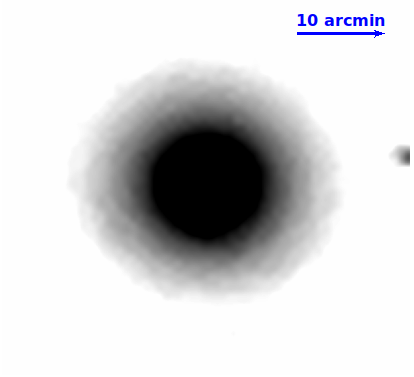}    
    \includegraphics[width=0.5\columnwidth]{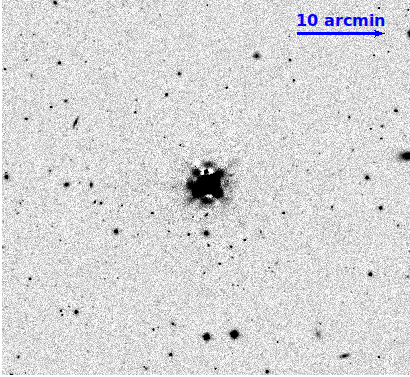}
    
    \caption{Example of the subtraction of the light profile of the galaxies. 
    {\it Left:} S-PLUS g-band image of NGC~1399. 
    {\it Center:} galaxy model.
    {\it Right:} residual image.
    All images are aligned such that north is up and east is to the left.
    } 
    \label{figura:fondo_galaxia} 
\end{figure*}

This project aims at studying the Fornax cluster, using the whole S+FP homogeneous data taken through the 12 optical S-PLUS bands. As in \cite{Analia:2024} (hereafter, Paper\,I) and \cite{Rodrigo:2024} (hereafter, Paper\,II), we consider for Fornax a redshift of z = 0.0048, taking into account that the systemic velocity of NGC~1399 is $v_{r}$= 1442 km s$^{-1}$  \citep{Maddox:2019}. In addition, we assume a distance modulus of (m$-$M) = 31.51 mag for Fornax \citep{Blakeslee:2009} and, at the corresponding distance, 1 arcsec subtends $\sim$0.1 kpc. 
{The mean half-mass radius, r$_{e}$, of GCs is $\sim$4~pc  \citep[][]{vandenBergh:1991, Jordan:2005, Webb:2012}, therefore, at the Fornax distance, GCs are unresolved sources. 
With the distance modulus used here, the peak (turnover) of the GCLF in V-band is $\sim$24~mag \citep[e.g.,][]{Kissler:1997}.
We use a cosmology with H$_{0}$=70.5, $\Omega_{0}$=0.30 and $\Omega_{l}$=0.70 throughout.
In this first paper we focus on extracting the GC catalogue, verifying its effectiveness, and presenting global results. In a following work we will focus on the detailed study of the GC stellar population properties and their relation with the cluster environment.

The paper is organised as follows. In Section\,\ref{seccion:data} we describe the S-PLUS data used in this study. In Section\,\ref{seccion:detection2}, we present the source detection and photometry. In Section \ref{seccion:sample}, we present the selection method. Results of the completeness tests are also presented. The analysis and discussion of the properties of the GC system are given in Section\,\ref{seccion:analysis}. In Section\,\ref{seccion:conclusions} we give our concluding remarks.

\section{S-PLUS Data}
\label{seccion:data}

The S-PLUS\footnote{\url{https://splus.cloud}} survey will cover $\sim$9300~square~degrees of the sky. It uses a robotic $\sim$0.8~m telescope that is located at the Cerro Tololo Inter-American Observatory (CTIO) in Chile \citep[][]{Mendes:2019}. S-PLUS has a pixel scale of 0.55~arcsec/pixel and a camera FoV of $\sim$1.5~square~degrees. It uses the Javalambre 12-filter photometric system designed for the Javalambre-Photometric Local Universe Survey\citep[J-PLUS,][]{Cenarro:2019}. The filter system is composed of the following 7 narrow-band filters: J0378, J0395, J0410, J0430, J0515, J0660, J0861. These map the [OII], Ca H+K, H$\delta$, G-band, Mgb triplet, H$\alpha$ and Ca triplet lines respectively. The system also includes the u, g, r, i, and z broad-band filters (similar to  SDSS). In Table~\ref{tabla:splus}, we list the S-PLUS filter system.

The S-PLUS Fornax cluster data are part of the Main Survey (MS) of S-PLUS
and cover a FoV of $\sim$20$\times$12~square~degrees in 106 pointings with the 12 bands. In Figure~\ref{figura:cover_data} we show the spatial distribution of all the sources detected (see Section~\ref{seccion:detection}) as a {\it cloud of black dots} in the 106 pointings delimited by the {\it black squares} ($\sim$1.4$\times$1.4~square~degrees). In the centre of each FoV the key name of the S-PLUS pointing is shown and the {\it red cross} represents the coordinates ($\alpha$=54.620941, $\delta$=-35.450657, J2000 coordinates) of NGC~1399 (E0) galaxy, the most massive galaxy (10$^{13}$~\msun, \citealt[][]{Schuberth:2010}) in Fornax defined as the cluster center. In Appendix~\ref{apendice_A}, some important features for the detection of the sources as well as key values of the header for each FoV in the i-band are listed. In the 106 pointings it was possible to detect $\sim$3~million sources. In the next section we describe the method for the detection and photometry.

The S-PLUS Fornax Project (S+FP) is a collaboration focused on exploiting the 12 bands of S-PLUS covering the Fornax Cluster. Two papers have already been published within this framework, \citet{Analia:2024} and \citet{Rodrigo:2024}, and several studies are currently being carried out. 

\section{Source detection and photometry} 
\label{seccion:detection2}

\citet{Buzzo:2022} studied NGC~1399, the Fornax cluster central galaxy, 
and assessed the effectiveness of identifying GCs using the first S-PLUS images available. 
They performed aperture photometry in an area of $\sim$14$\times$14~arcmin and used four GCs selection criteria (magnitude, concentration index, Gaia proper motion and template fitting) to select GC candidates. In this work we increase considerably the area for GC detection in Fornax using 106 FoVs of S-PLUS data, covering an area of $\sim$208~square~degrees. 
Moreover, in this work, we perform PSF photometry, specially designed for the detection and measurement of point sources with the characteristics of GCs in Fornax.
We also improve the detection and measurement of point sources from the method of Paper II,
which is focused on recovering extended sources, like galaxies. In this section we describe the detection   and photometry. 

\begin{figure*}
\includegraphics[width=0.52\columnwidth]{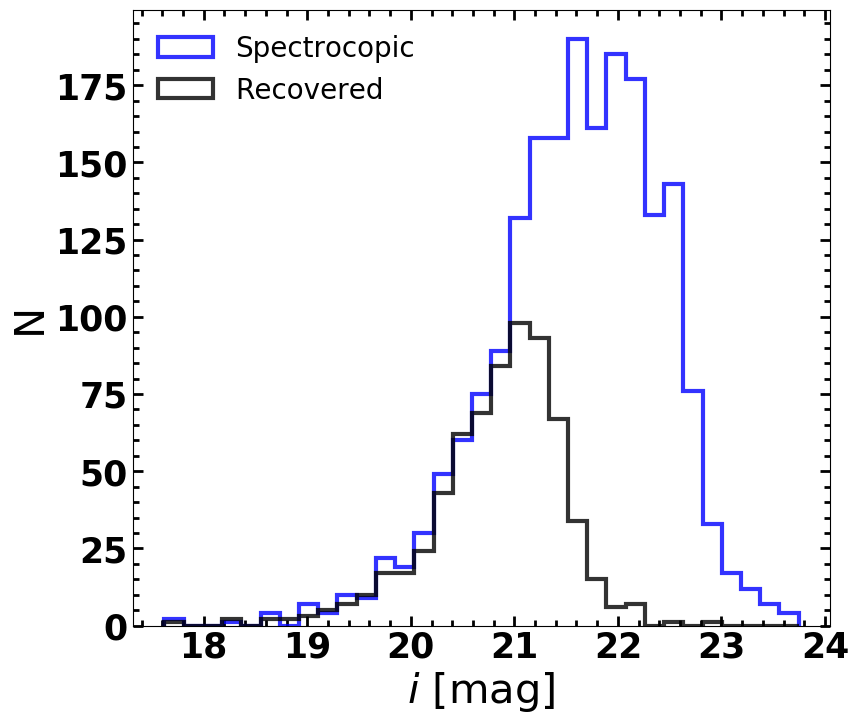}
\includegraphics[width=0.5\columnwidth]{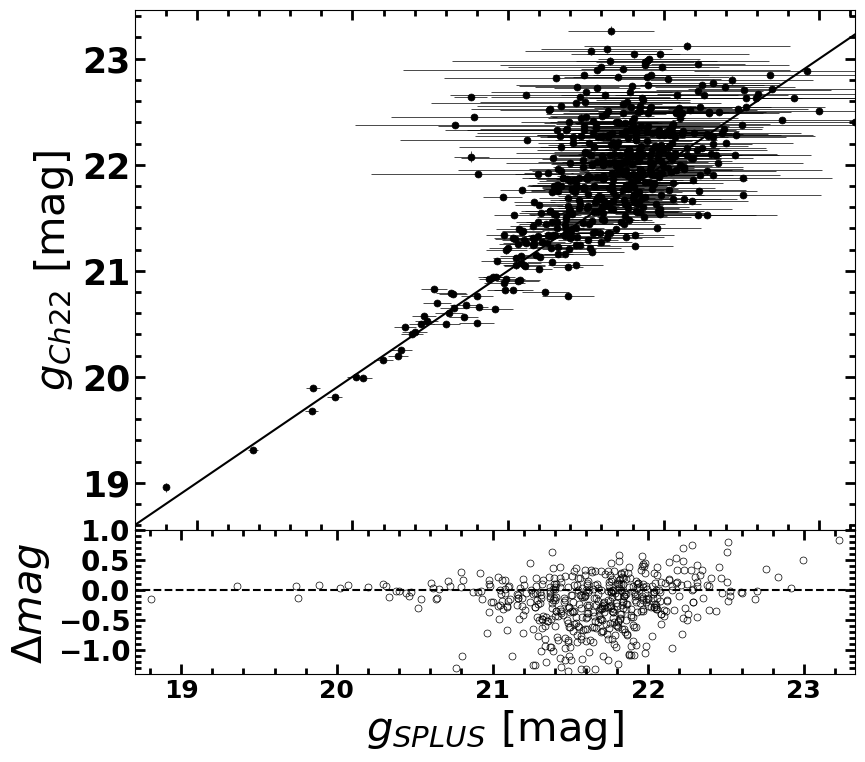}
\includegraphics[width=0.5\columnwidth]{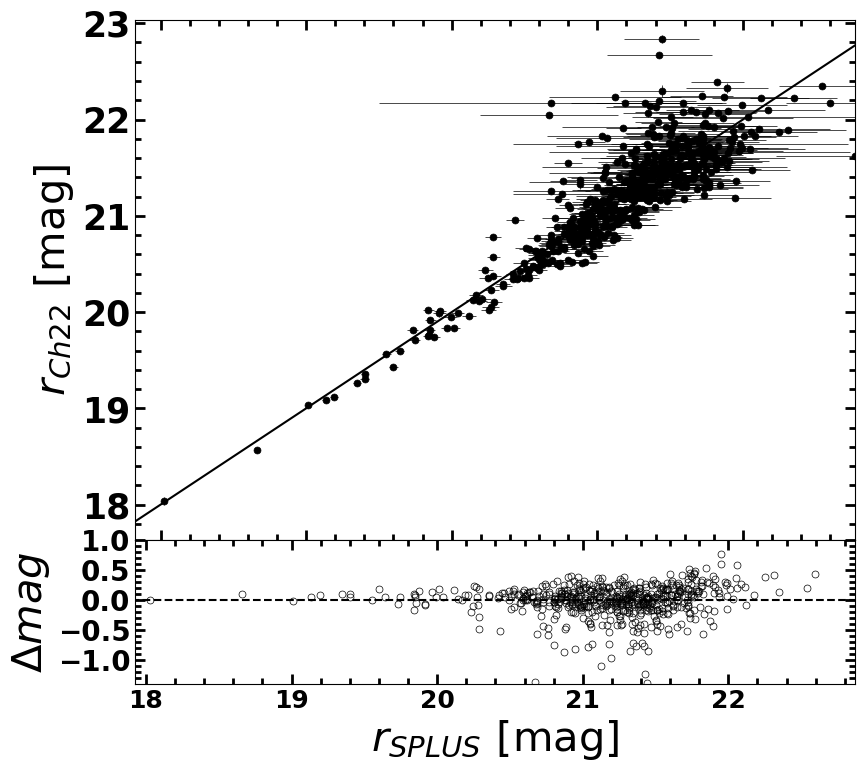}
\includegraphics[width=0.5\columnwidth]{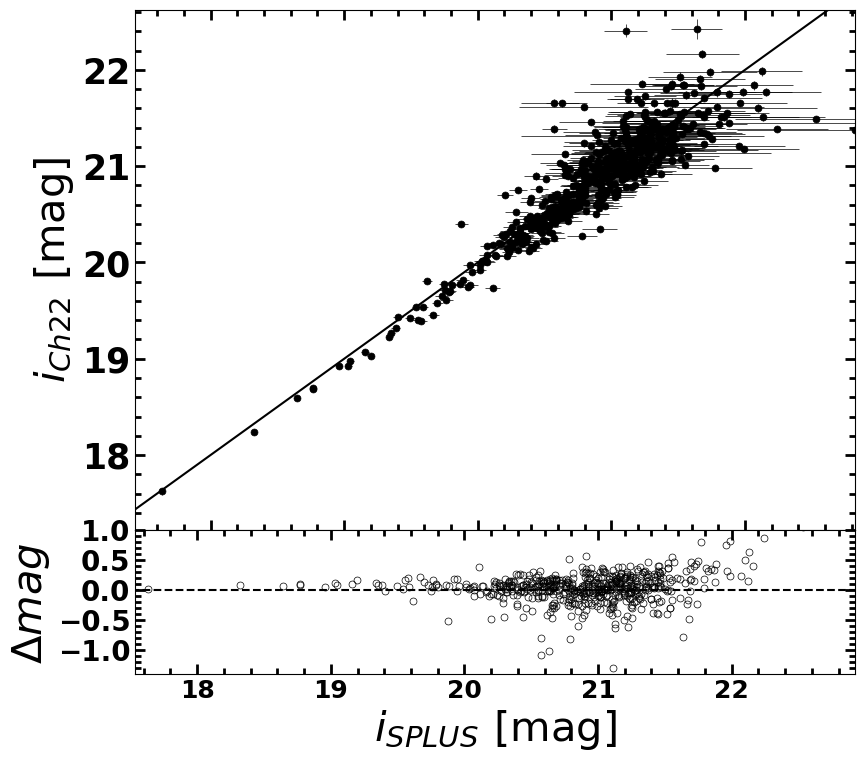}  
   \caption{{\it First panel to the left:} i-band magnitude distribution of the GC spectroscopic sample from the literature (blue histogram) and GCs recovered with S-PLUS (black histogram). From left to right, we show Fornax GCs PSF magnitude comparison in the g-band ({\it second panel}), r-band ({\it third panel}) and i-band ({\it fourth panel}). In the last three panels, we show the S-PLUS GC magnitudes versus the spectroscopic GC magnitudes (\citealt[][Ch22]{Avinash:2022}). The black solid line in the top panels is the identity line while the black dashed lines in the bottom panels mark where the differences between magnitudes ($\Delta mag= SPLUS - Ch22$) are equal to zero.
    } 
    \label{figura:mag_comparison} 
\end{figure*}

\subsection{Source detection}
\label{seccion:detection}

For source detection and posterior photometric measurements, we used a combination of  SExtractor\footnote{\url{https://www.astromatic.net/software/sextractor}} \citep{Sextractor:1996} and PSFEx\footnote{\url{https://www.astromatic.net/software/psfex}} \citep{Psfex:2011}. With the objective of recovering the largest amount of GCs that have been confirmed spectroscopically in the literature \citep[][]{Avinash:2022}, most of which reside within 0.5 \rv, we performed a series of tests with different SExtractor parameters in different runs on test images. These test images were processed to remove the light distribution of the galaxies following different methods.
The final detection was performed on images from which their median-ﬁltered version was subtracted as faint sources are detected more easily in a median-subtracted image \citep{Lomeli:2017} rather than in the original ones. 
We perform different tests in the median-ﬁltered images changing the size of the filter (e.g. 3$\times$3, 11$\times$11, etc.). The best results for recovering sources was obtained using a filter of 41$\times$41 pix$^{2}$ (Figure~\ref{figura:fondo_galaxia}).

To avoid contamination in the photometric measurements due to the light of the galaxies within the cluster, it is necessary to subtract the light profiles of the galaxies. We performed a series of proofs for obtaining the best modelled background-sky image (Figure~\ref{figura:fondo_galaxia}) using SExtractor, 
which has the ability of creating {\sc segmentation} and {\sc background} images, among others. Exploiting this capability, we created the background-sky image using different parameters  (e.g, {\sc back\_size}, {\sc back\_filter}, {\sc back\_filterthresh} etc.).
We performed various runs of SExtractor testing different parameters (e.g., {\sc detect\_minarea}=3, 5, 8, {\sc detect\_thresh}=1.5, 2.0, 2.5, {\sc back\_size}=16, 32, 64, {\sc back\_filter}=1, 3, 5 etc.). 
Finally, we selected the SExtractor parameters ({\sc detect\_minarea}=3, {\sc detect\_thresh}=1.5, {\sc back\_size}=16, {\sc back\_filter}=3) with which we obtained the best recovery of spectroscopically confirmed GCs and with the lowest scatter in magnitudes compared to the literature (Figure~\ref{figura:mag_comparison}).

Compared with the FDS survey data (\citealt{Iodice:2016}, 0.21~arcsec~pixel$^{-1}$ and \citealt{Cantiello:2018}, 0.26~arcsec~pixel$^{-1}$) the S-PLUS observations have lower-sampling (0.55~arcsec~pixel$^{-1}$) and are shallower (r$\sim$21.3~mag for S-PLUS \citealt{Buzzo:2022}; Paper II, and r$\sim$24.3~mag for FDS \citealt{Dabrusco:2016}).
However, the spatial coverage of the FDS survey is smaller than S-PLUS since it only covers the central FoVs (5 pointings) of S-PLUS in Fornax.
For testing the detections in this work, we compare the recovered sample with the spectroscopic GC (spec-GC) catalog of \citet{Avinash:2022}.
We recovered $\sim$1000 out of the total of 2341. In the first panel of Figure~\ref{figura:mag_comparison}, we plot the spectroscopic (blue histogram) and  S-PLUS recovered samples (black histogram). It is clear that we miss the faint part 
($i\gtrsim22.0$ mag) of the GCLF from the sample of \citet{Avinash:2022}. 

\subsection{PSF photometry}
\label{seccion:psf_photometry}

PSF photometry is a method to obtain photometry for unresolved or marginally resolved star clusters  \citep[e.g.,][]{Gallagher:2010, Fedotov:2011, Lomeli:2017}. At Fornax distance (19~Mpc) and considering the S-PLUS pixel scale (0\arcsec.55), GCs ($\sim$50~pc~pixel$^{-1}$) are unresolved sources, given the mean half mass radius, r$_{h}$, of a GCs is $\sim$4~pc \citep{vandenBergh:1991, Jordan:2005, Barmby:2006, Webb:2012}. Hence, we perform PSF photometry given the characteristics of the GC candidate selection.

The PSF photometry was obtained using a combination of SExtractor and PSFEx, in a similar manner to \citet{Lomeli:2017}, \citet{Lomeli:2022} and \citet{Lomeli:2022cfht}. A brief description of the procedure is given below: 
\begin{itemize}
\item[a)] We performed a first run of SExtractor for detection and selection of point sources based on their brightness versus compactness, as measured by SExtractor parameters \magauto\ (a Kron-like elliptical aperture magnitude; \citealt{Kron:1980}), \fluxradius\ (similar to the effective radius) and \class\ (discriminator between point sources and extended sources). For the PSF creation, in the 106 fields we selected a similar range in the space \magauto\ versus \fluxradius: 
\begin{itemize}
\item 12 $\lesssim$\magauto$\lesssim$21.5~mag; 
\item  1 $\lesssim$\fluxradius$.\lesssim$2.3~pixel 
\item  \class$\gtrsim$0.7 
\end{itemize}
\noindent In this range it was possible to select $\sim$1000 point sources for the PSF creation in each FoV. 
\item[b)] The PSF creation was preformed with PSFEx using the point sources selected in the last step. The spatial variations of the PSF were modeled with polynomials of a degree of 3. To create the PSF, the ﬂux of each star was measured in an aperture of 9~pixel of radius in all bands (equivalent to 4\arcsec.95$\times$4\arcsec.95); such an aperture, determined through the curve of growth method for each passband, is large enough to measure the total ﬂux of the stars, but small enough to reduce the likelihood of contamination by external sources. 
\item[c)] We performed a second run of SExtractor using the PSF created in the previous step, for measuring the PSF magnitude (\magpsf) and the point and extended sources discriminator, \spread.
\end{itemize}

To verify the plausibility of the PSF photometry, we compared the magnitudes measurements with those reported in \citet{Cantiello:2018} and confirmed spectroscopically by \citet{Avinash:2022}. In the last three panels of Figure \ref{figura:mag_comparison},
we compared the magnitude measurements in three bands: g, r, and i. 
It was possible to compare 523 (g-band), 579 (r-band) and 572 (i-band) GCs, which have magnitude estimations in both catalogs. In the bottom panels of the last three panels of Figure~\ref{figura:mag_comparison} we show the difference between this work and literature magnitudes, $\Delta$mag. The mean and sigma-dispersion for this differences are: $\Delta \overline{mag}=-$0.24, $\sigma=$0.45 in g, $\Delta \overline{mag}=-$0.014, $\sigma=$0.27 in r, and $\Delta \overline{mag}=$0.03, $\sigma=$0.24 in i band. Despite being different measurement methods, the estimates are within 1$\sigma$ of the error, so we can confirm that our PSF photometry is in agreement with the photometry reported in the literature.
The higher dispersion values observed  in the second panel of Figure~\ref{figura:mag_comparison}, i.e. in the g-band, is caused by the lower $S/N$ of GCs, intrinsically red objects, in bluer bands, which  generates an increase in the uncertainties especially for the faintest objects.

\section{Globular clusters selection. The sample of globular clusters candidates}
\label{seccion:sample}

Extracting a catalog of GCs is a challenging task due to the existence of various contaminants, such as foreground stars and background galaxies.
For the selection of new GC candidates in the 106 Fornax pointings of S-PLUS, we used SExtractor-derived structural and classifier parameters,  \fwhm, \class, \fluxradius\  and \spread, to define an initial cluster sample. This sample has been refined using  
color-magnitude and color-color diagrams that allowed us to separate, foreground stars, 
background galaxies and young stellar clusters (YSCs, e.g., \citealt{Whitmore:1999}, \citealt{Larsen:2002}, \citealt{Whitmore:2023}) from old GCs. In addition, we make use of GAIA\footnote{\url{https://www.cosmos.esa.int/web/gaia/dr3}} DATA RELEASE 3 (GDR3) using the tabulated proper motions coordinates to reject Galactic objects,
and finally we estimated the redshift (z) of the remaining objects applying Spectral Energy Distribution (SED) fitting techniques in order to reject background sources. 

\begin{figure} 
    \includegraphics[width=1.0\columnwidth]{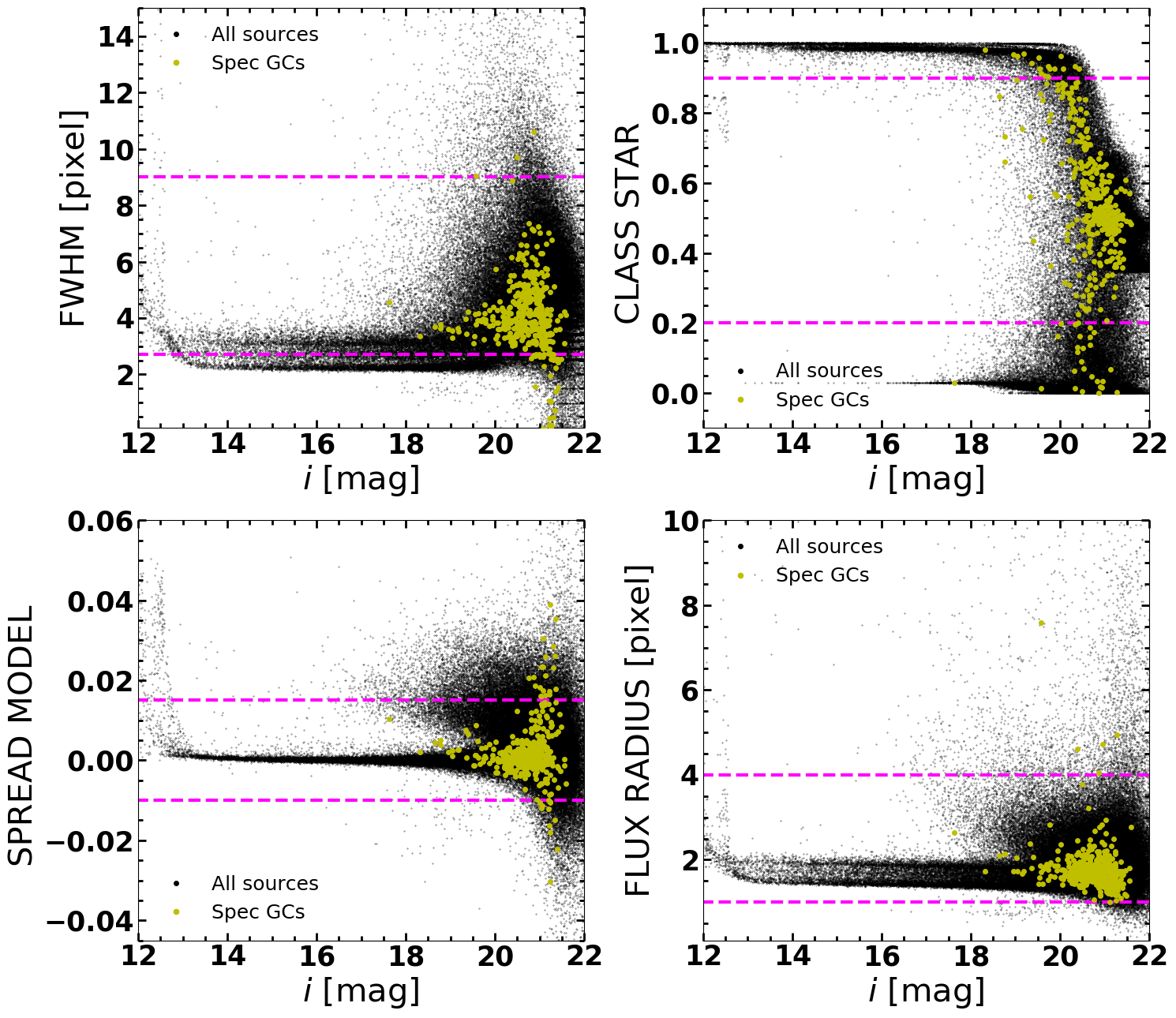} 
    \caption{Structural parameters used for the selection versus i-band magnitude. 
    In the four panels all detected sources  are shown as {\it small black dots} and spec-GC are depicted in {\it yellow}.
    In all panels the {\it horizontal magenta dashed lines} define the selection region of GC candidates.}
    \label{figura:structural_parameters} 
\end{figure}

\subsection{Structural parameters selection}
\label{seccion:structural_selection}

In the literature there are a variety of methods for GC candidates selection \citep[e.g.,][]{Mayya:2008, Fedotov:2011, Whitmore:2014, Munoz:2014, Lomeli:2017, Lomeli:2022}.
We used the structural and classifier parameters defined by the spec-GC \citep{Avinash:2022} to delimit the selection criteria. 
The parameters used for the selection are: 
\fwhm, as a discriminator between compact and marginally resolved sources; 
\class, as a stellarity classifier (compact sources are near 1);
\spread, as another stellarity classifier (compact sources are near 0); and 
\fluxradius, it is a proxy of the half-light radius (estimation of the size of objects in pixels). Since GCs are bright and well characterized in the i-band (see, \citealt{Lomeli:2022}), we set this band to perform the selection. In addition, we imposed that all the selected sources have a SExtractor {\sc flag}$_{i-band}$=0\footnote{
The {\sc flags} are warnings about the source extraction process. 
Different values of the {\sc flag} parameter indicate various problems with the photometry
\citep[see the SExtractor manual for further explanation,][]{Sextractor:1996}.}
and {\sc magerr\_psf} $\leq$ 0.2~mag in the broad bands (g, r, i, and z).

In Figure \ref{figura:structural_parameters}, we show the parameters \fwhm, \class, \spread\ and \fluxradius\ versus the i-band magnitude for all the detected sources (small black points) and the recovered spec-GC sample \citep[yellow circles,][]{Avinash:2022}.
By comparing the parameters of the spectroscopic GCs with those of the detections in the 106 S-PLUS Fornax fields, it is possible to obtain a refined sample of GC candidates. Below we describe the selection criteria:
\begin{itemize}
   \item \fwhm\ (top-left panel in Figure~\ref{figura:structural_parameters}): we considered all objects with 2.8 pixel $\leq$ {\sc fwhm} $\leq$ 9.0 pixel as GC candidates. At the distance of Fornax (19~Mpc, m-M=31.51; \citealt{Blakeslee:2009}) the pixel scale is 50.6~pc~pixel$^{-1}$.
   \item \class\ (top-right in Figure~\ref{figura:structural_parameters}): this is a discriminator between point sources (\class$\sim$1) and extended sources (\class$\sim$0). We chose objects with 0.2 $\leq$ class $\leq$ 0.90, where the bulk of the spectroscopic GCs are located.
   
   \item \spread\ (bottom-left panel in Figure~\ref{figura:structural_parameters}): this is the discriminator between marginally resolved point sources and extended sources provided by the combination of SExtractor with PSFex. From Figure~\ref{figura:structural_parameters}, we select the objects displaying \spread $\leq$ 0.015 (see, for example, \citealt[]{Lomeli:2017, Rosa:2019, Lomeli:2022b}). 
    
   \item \fluxradius\footnote{It is the radius of the circle centered on the light barycenter that encloses half of the total ﬂux.} (bottom-right panel in Figure~\ref{figura:structural_parameters}): it is a proxy of the half-light radius (r$_{e}$) estimated by SExtractor. From the comparison between spec-GC sample and all detections in Figure~\ref{figura:structural_parameters} we chose as GC candidates the sources with 1~pixel $\leq$ \fluxradius\ $\leq$ 4~pixel.  
\end{itemize}

In this work we used a combination of \class\ and \spread, which is not commonly used in the literature because both parameters are discriminators between point and resolved sources. \class\ is easier to obtain than \spread\ but the latter is more powerful though requires a higher computational investment. We selected this method because, by using the combination of both parameters, the number of contaminating sources decreases (see Section~\ref{seccion:contaminantes_2017}). Finally, after applying the above criteria we obtained a catalog of 597,634 sources. The structural parameters for each GC candidate are listed in columns 9-12 of Table~\ref{tabla:datos}.

\subsection{Magnitude selection}
\label{seccion:magnitude_selection}

All the data have been corrected for Galactic extinction ($A_{V}$=0.039~mag, $E(B-V)$=0.013~mag, in direction to NGC~1399) using the values from \citet{Schlafly:2011} provided by the NASA Extragalactic Database\footnote{The NASA/IPA Extragalactic Database (NED) is operated by the Jet Propulsion Laboratory, California Institute of Technology, under contract with the National Aeronautics and Space Administration.}.
The values are published in the S-PLUS Data Release 4 (DR4; \citealt{Herpich:2024}).

To avoid the contamination of bright point sources (e.g., galactic stars, ultra compact dwarf), we select objects displaying i~$\geq$~19.04~mag (M$_{V}$~$\sim-$11.50~mag, see transformation equation in Appendix~\ref{apendice_C}). The magnitude limit is $\sim$1 magnitude brighter than that of Omega Cen (M$_{V}=$~$\sim-$10.30~mag), however in agreement with the spectroscopic sample in which there are GCs with i~$\leq$~19.04~mag. Considering a distance modulus of (m-M)=31.51 mag, this magnitude selection corresponds to the 3$\sigma$ limit of the GC luminosity function assuming $\sigma=$1.40$\pm$0.06 for massive galaxies \citep[e.g.][]{Whitmore:1995} and galaxy clusters \citep[e.g.][]{Karla:2013}.
For comparison, the dispersion values found for the GCLF of NGC~1399 and the Milky Way (MW) are $\sigma=$1.23$\pm$0.03 \citep[][]{Villegas:2010} and $\sigma=$1.15$\pm$0.10 \citep[e.g.][]{Harris:1996, Jordan:2007}, respectively.

\begin{figure*} 
    \centering
    \includegraphics[width=0.65\columnwidth]{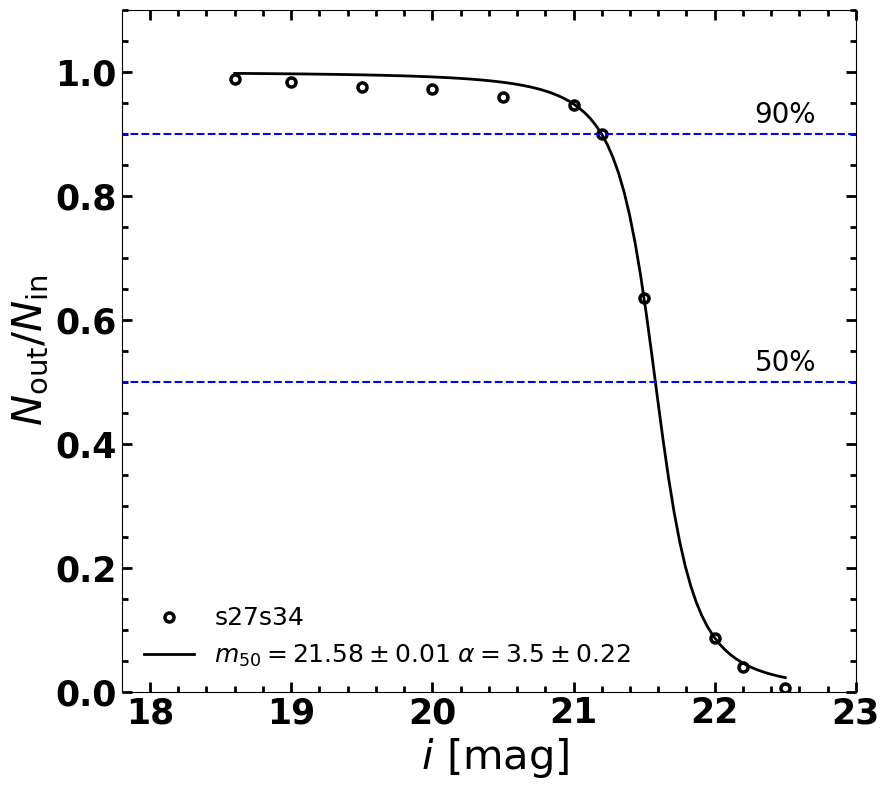}
    \includegraphics[width=0.65\columnwidth]{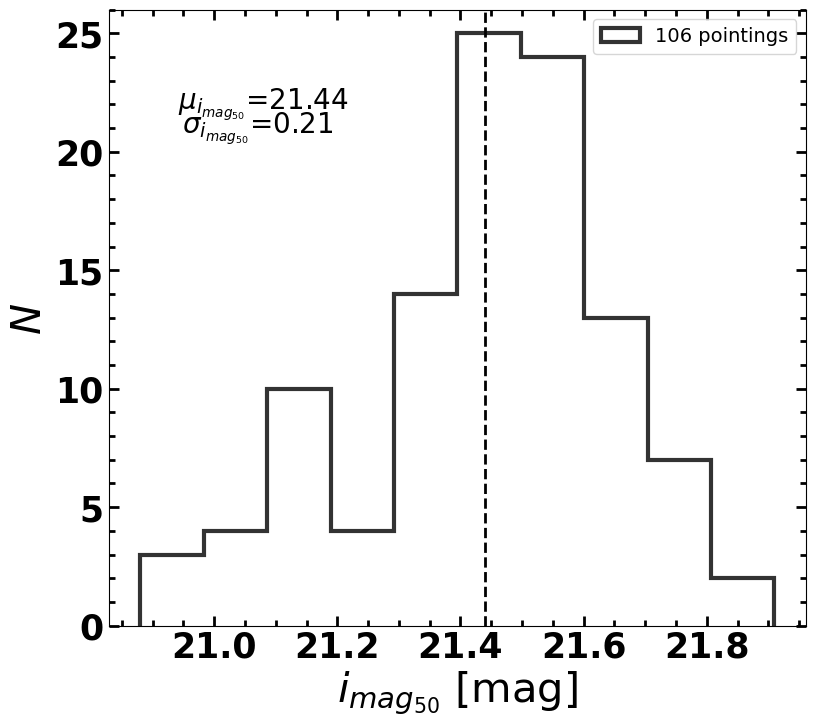} 
    \includegraphics[width=0.65\columnwidth]{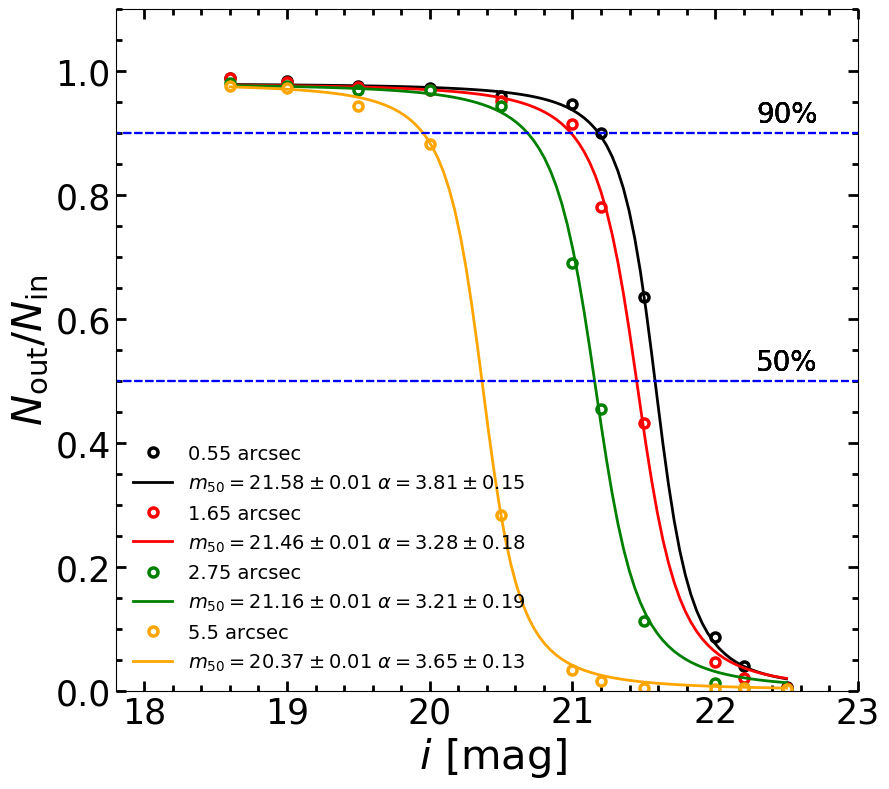}     
    \caption{{\it Left:} example of one completeness curve in the i-band for the s27s34 field. The {\it horizontal blue dashed line} represents the completeness at 50\% and 90\% levels.
    {\it Center:} distribution of i$_{mag_{50}}$ magnitudes recovered for all the Fornax pontings. The {\it vertical black dashed line} represents the mean value of the distribution.
    {\it Right:} completeness tests for mock sources with different Gaussian profile sizes: 0\arcsec.55 (black curve), 1\arcsec.65 (red curve), 2\arcsec.75 (green) and 5\arcsec.5 (yellow curve).
    } 
    \label{figura:completitud} 
\end{figure*}

\begin{figure} 
    \includegraphics[width=1.0\columnwidth]{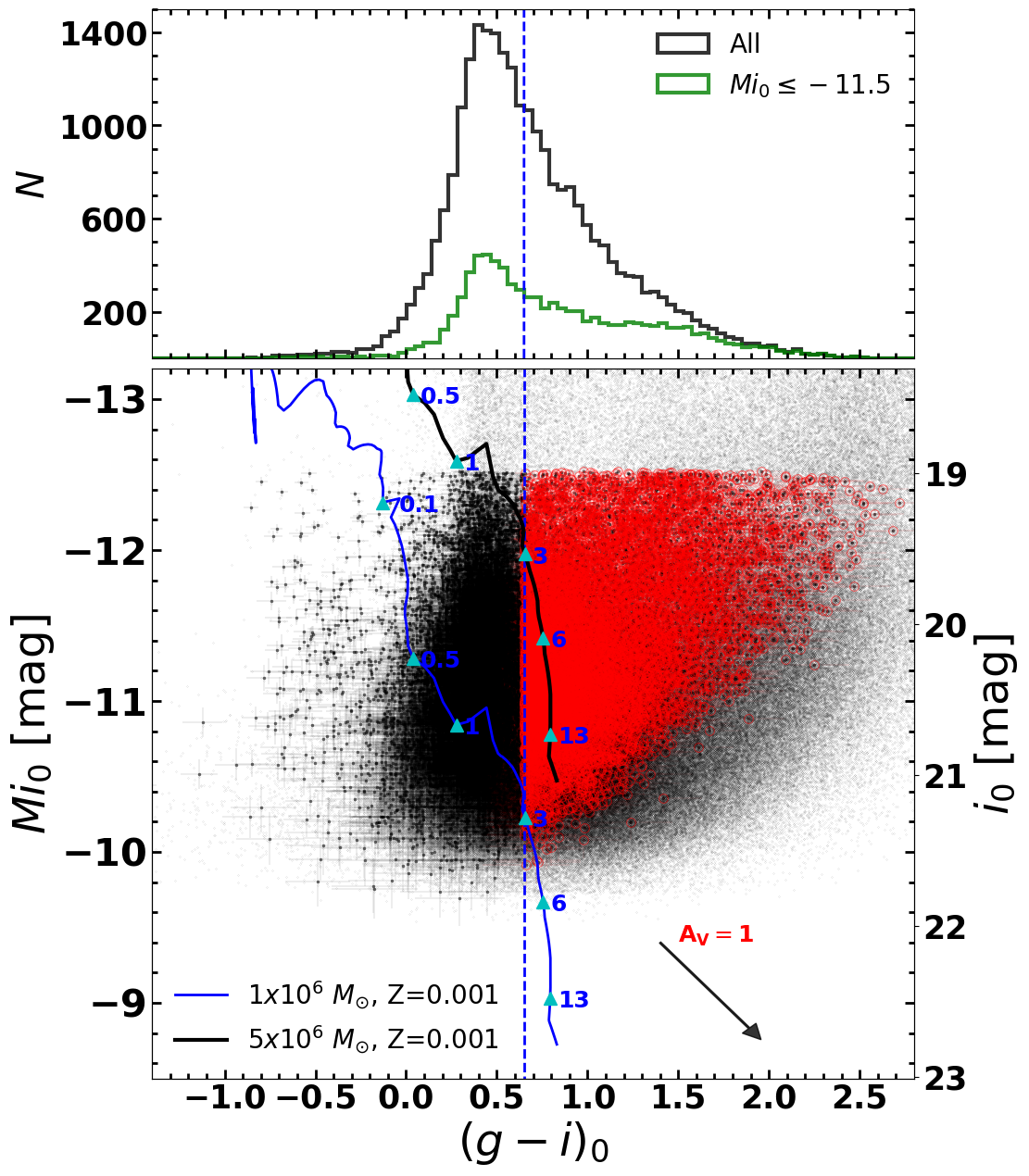}
    \caption{{\it Top:} color histogram of cluster candidates. The black line shows the color distribution over the entire range of magnitudes, whereas the green histogram shows the distribution for bright ($M_{i_0}\leq-$11.5 mag) clusters.
    The blue dashed vertical line at $(g-i)_{0}=0.65$ mag separates foreground stars, background galaxies and young cluster candidates from GCs. {\it Bottom:} $M_{i_{0}}$ versus $(g-i)_{0}$ CMD of all cluster candidates in the 106 Fornax pointings. Sources having $(g-i)_{0}\geq$ 0.65~mag are GC candidates (red small open circles), bluer objects are foreground stars, background galaxies and young cluster candidates (black  points), and  all the detections are represented by a cloud of grey dots. The evolutionary locus of the Single Stellar Populations (SSPs) models from \citet{Bruzual:2003} for a single metallicity Z=$0.001$, two values of masses of $1\times10^{6}$~\msun\  (blue solid line) and $5\times10^{6}$~\msun\  
    (black solid line), 
    and a Kroupa Initial Mass Function (IMF), are shown. Locations corresponding to ages of 0.1, 0.5, 1, 3, 6 and 13~Gyr are depicted with cyan triangles in each SSP. The chosen color cut separates clusters older than 3~Gyr from the younger ones for unreddened SSPs. The reddening vector with $A_{V}=1$~mag is represented by the black arrow. 
    } 
    \label{figura:color_magnitud} 
\end{figure}

On the other hand, in all the astronomical observations exists bias detection towards the faint magnitudes. To ensure the verisimilitude of the observations in the faint part of the GCLF it is therefore necessary to estimate the completeness magnitude limit.
To find out at which magnitude the sample of GC candidates is complete at a 50\% level, $m_{50}$, we carry out completeness tests. In this study we follow the completeness recipes presented in the works of \citet{Lomeli:2017, Lomeli:2022, Lomeli:2022b}. In the next subsections, we present a brief description of the followed procedure.

\subsection{Monte Carlo cluster simulations}
\label{seccion:monte_carlo_simulations}

We generated mock GCs using the {\sc iraf}/{\sc daophot} tasks {\sc addstars} and {\sc mkobjetcs}. A GC is defined by an intensity profile that follows the PSF obtained in this work and magnitudes in the range 18-23 i-mag considering intervals of 0.5~mag. Around 1300 clusters were generated for each FoV, 100 for each simulated magnitude. The coordinates of these sources were randomly generated and inserted on to an observed S-PLUS image. The same object detection criteria used for real objects were applied on the mock-object added frames.

In the left panel of Figure\,\ref{figura:completitud}, we show an example of one completeness curve in the i-band for one of the cental fields, s27s34.
In order to quantitatively obtain the magnitude
at which the sample is 50\% complete, we fitted the points with the Pritchet function \citep[e.g.,][]{McLaughlin:1994, Karla:2013, Lomeli:2017, Lomeli:2022, Lomeli:2022b} given by:
\begin{equation}
   \normalem{ f(m) = \frac{1}{2}\left[1 - \frac{\alpha(m-m_{\textrm{50}})}{\sqrt{1+\alpha^{2}(m-m_{\textrm{50}})^{2}}} \right] },
    \label{equ:prichet}
\end{equation}
\noindent where $\alpha$ is a fitting constant that determines the curve slope. The function fit values for the s27s34 field are $m_{50}$=21.58$\pm$0.01 and $\alpha$=3.50$\pm$0.22. We repeat the same process for the 106 pointings in the i-band. In the center panel of Figure~\ref{figura:completitud}, we show the magnitude at which the sample is 50\% complete in the i-band, i$_{mag_{50}}$, and the distribution of i$_{mag_{50}}$ for all the FoVs. We observe that the range of recovered magnitudes spans $\sim$1~mag around the mean of the distribution, ${\mu}_{i_{mag50}}=21.44\pm$0.21. The range of magnitudes between $\sim$20.90-21.80 that is recovered in the 106 fields is the range of magnitudes in which the sample is 50\% complete. The fact that there is a variation of i$_{mag_{50}}$ among different tiles may be attributable to different observational conditions such as air masses, exposure times, number of stacked images, etc. (see Appendix~\ref{apendice_A}). 

Even if GCs at Fornax distance using the S-PLUS data are unresolved objects, we can see in Figure \ref{figura:structural_parameters} that they can present a FWHM higher than the one of the PSF, $\simeq 3"$ ($2 \le FWHM\le8)$, since we are extracting objects at the S-PLUS detection limit \citep{Herpich:2024}. In order to understand the completeness of sources that do not follow the PSF profile, i.e. marginally resolved objects, we performed simulations, where a cluster is defined by an intensity profile that follows a Gaussian function of a given FWHM and a total magnitude (equal to the PSF simulations), with FWHM taking values of 0\arcsec.55 (black curve), 1\arcsec.65 (red curve), 2\arcsec.75 (green), see Figure~\ref{figura:completitud}. In the right panel of Figure~\ref{figura:completitud}, we observe that the number of sources detected decreases as we increase the FWHM size, obtaining brighter $m_{50}$ values (e.g., for FWHM=1\arcsec.65 and 2\arcsec.75,  i$_{mag_{50}}$=21.16$\pm$0.01 and 20.37$\pm$0.01 respectively).

\begin{figure*} 
    \includegraphics[width=1.0\columnwidth]{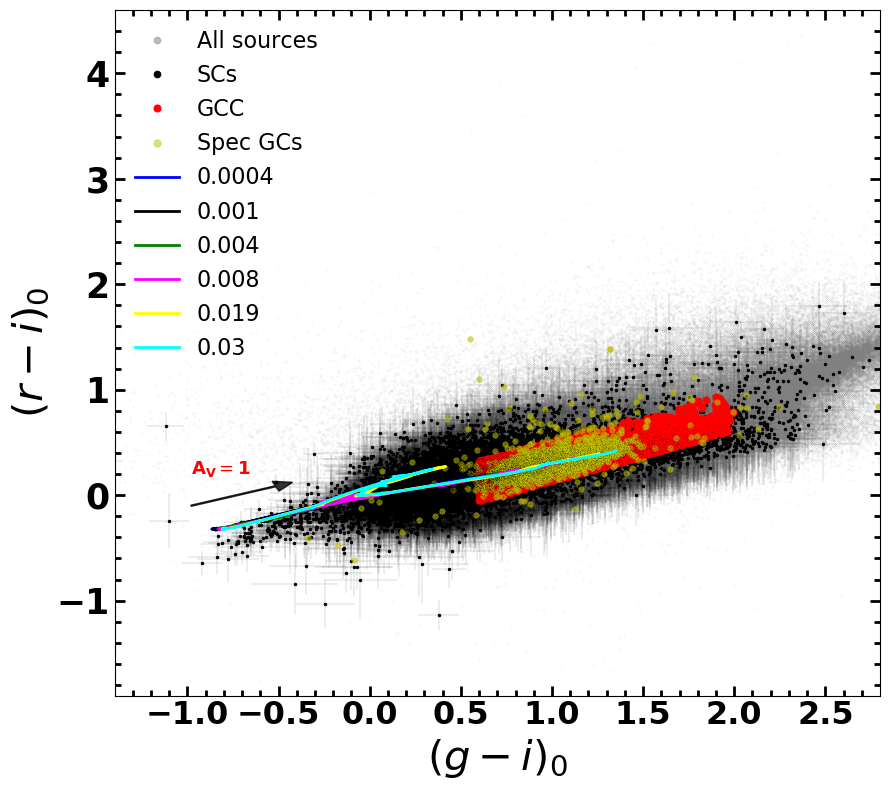}    
    \includegraphics[width=1.0\columnwidth]{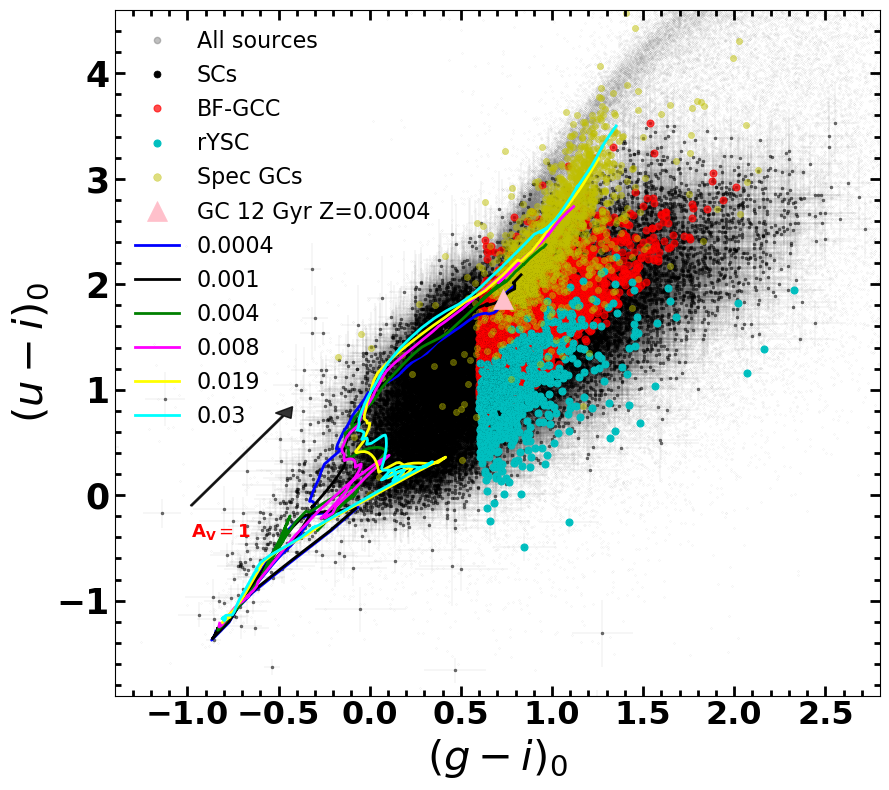}    
    \caption{{\it Left:} Color-color diagram $(r-i)_{0}$ versus $(g-i)_{0}$ of all cluster candidates (SCs,  black dots). 
    Sources having $(g-i)_{0}\geq$0.65~mag and -0.10~mag$\leq(r-i)_{0}\leq$0.70~mag are considered GC candidates (GCC, red dots).
    The chosen color cut separates clusters older than 3~Gyr from the younger ones for unreddened SSPs. 
    {\it Right:} Color-color diagram $(u-i)_0$ versus $(g-i)_{0}$ showing the $u$-selected cluster candidates (SCs). Objects displaying $(g-i)_{0}\geq$0.65~mag are GC candidates (BF-GCC, red dots), and bluer objects are possibles young star cluster candidates (rYSC, black dots).     
    The bluest color that a classical GC can have (corresponding to Z=0.0004 and to an age of 12~Gyr) is marked by a {\it rose solid triangle} ($(g-i)_{0}\sim$0.6 and $(u-i)_{0}\sim$1.9). 
    The reddened young star clusters that displays colors typical of GCs are contaminants and are identified by open cyan circles. 
    In both panels: The reddening vector with $A_{V}=1$~mag is represented by the black arrow.  The evolutionary loci of SSPs from \citet{Bruzual:2003} for different metallicities using a Kroupa IMF are shown by solid curves of different colors, following the color notation shown in each panel. The spectral GCs (spec-GC) are depicted in {\it yellow circles}.}
    \label{figura:color_color} 
\end{figure*}

\subsection{Color selection}
\label{seccion:color_selection}

In the bottom panel of Figure~\ref{figura:color_magnitud}, we show the $Mi_{0}$ versus $(g-i)_{0}$ color-magnitude diagram (CMD), corrected for Galactic extinction, for the sample of GC candidates selected with the structural parameters plus magnitude criteria (black solid points). 
The evolutionary loci of the SSPs models from \citet{Bruzual:2003} at typical metallicities of YSCs (Z=0.008$\sim$1/3 solar) and GCs (Z=0.001) are shown. 
These models correspond to synthetic clusters of mass $1-5\times10^{6}$~\msun\   
obeying a Kroupa IMF between masses 0.1 and 100~\msun. 
In the top panel of Figure~\ref{figura:color_magnitud}, we show the $(g-i)_{0}$ color histogram of all 
cluster candidates (black histogram) and that of the clusters brighter than $M_{i_0}\leq-$11.5~mag (green histogram).
For the bright cluster sample there seems to be a second distribution that peaks at $(g-i)_0\sim0.5$~mag. 
The color that separates the two distributions corresponds to the $(g-i)_{0}$=0.65~mag bin.
The blue and red distributions correspond to the unresolved foreground, background contaminants or YSCs in star-forming galaxies
and GCs, respectively. 
Based on the SSPs models and the color distribution, we use $(g-i)_{0}$=0.65~mag 
to separate GCs from other objects \citep[e.g.][]{Hwang:2008, Simanton:2015, Lomeli:2022, Whitmore:2023}. 
Different photometric studies used a selection cut in color $(g-i)_{0}\sim$0.4-0.5 mag \citep[e.g.,][]{Pota:2013, Cantiello:2018}. However, when compared with the SSP models used here (Z=0.001) such a color is equivalent to 1~Gyr, a very young age to be consistent with old classic GCs. On the other hand, in both panels of Figure~\ref{figura:color_color}, we plot the spectroscopically confirmed GCs (yellow circles)  by \citet{Avinash:2022}. 
The total number of GCs in this catalog is 2341. It was only possible to plot 1807 GCs
with good photometric estimations, and only approximately 2\% have a color $(g-i)_{0}\leq0.65$~mag.
Taking into account the colors of the spec-GCs, $(g-i)_{0}$, it is possible to infer that the bulk of the GCs displays $(g-i)_{0}>0.65$~mag (both panels of Figure~\ref{figura:color_color}). Therefore, we consider that a color cut $(g-i)_{0}\leq0.65$~mag is not suitable for the selection of classic old GCs since the number of contaminant sources can increase considerably and the loss of GC candidates is low. On the other hand, in right panel of Figure~\ref{figura:color_color}, we show a color-color diagram, $(r-i)_{0}$ versus $(g-i)_{0}$, used for the selection of GC candidates \citep[e.g.,][]{Pota:2013, Kartha:2014, Pota:2015}. 
The {\it red dots} are the GC candidates which meet the criteria described above and displaying $(r-i)_{0}\geq0.19$~mag $\pm~1\sigma$ ($\sim$3~Gyr, Z=0.001). 
However, in these kinds of color-color diagrams, there is a large number of objects that can
contaminate the GC candidates sample. As shown by the SSP (solid lines in colors), in these diagrams,
it is difficult to break the age-metallicity
degeneracy without using the u-band.
In Section~\ref{seccion:contaminantes_u}, we explain the GC candidate selection using a color-color diagram employing the S-PLUS u-band. 

Although using the structural parameters we have been able to reject most of the contaminating objects in our sample as Galactic stars or background galaxies, there is a possibility that the GC candidates sample still contains a number of contaminants. In next subsections we describe the processes used to obtain a sample much reduced in contaminants.

\subsection{Refining the GC selection through the comparison with DESI and GAIA and the application of SED fitting}
\label{seccion:distancias_selection}

We cross-matched the GC candidate sample with DESI Legacy Imaging Surveys\footnote{The DESI Legacy Surveys team is producing an inference model of the extragalactic sky in the optical and infrared. The original Legacy Surveys (MzLS, DECaLS and BASS) conducted dedicated observations of $\sim$14,000 square degrees of extragalactic sky visible from the northern hemisphere in three optical bands (g,r,z), which was augmented with four infrared bands from NEOWISE.} (hereafter, DESI) \citep{Desi:2023} which has covered the Fornax cluster, with a pixel scale sampling of 0.20 pix arcsec$^{-1}$ (twice of the S-PLUS resolution) and a photometric depth of 23.4~mag in r-band. 
With the cross-match we recovered $\sim$99\% of GC candidate, of which} we used the DESI object classification (“type”), which includes point sources (``PSF"), round exponential galaxies with a variable radius (``REX"), deVaucouleurs (``DEV") profiles (elliptical galaxies), exponential (``EXP") profiles (spiral galaxies) and Sérsic (``SER") profiles. Ultimately, we retain only the objects classified as “PSF” and “REX”  ($\sim$60\%). In fact, cross-matching  \citet{Avinash:2022} spectroscopic catalog with the light profile classification provided by DESI, we found that the GCs where in their majority classified as 
``PSF" ($\sim$68\%), 
then as ``SER" (15\%),  
``DEV" (7\%), and  
``EXP" (2\%).      
Therefore considering only objects classified as "PSF" might exclude real GC candidates.
At the same time, from a visual inspection and profile examination (using {\sc iraf}/{\sc imexam}), we concluded that the 'REX' objects on average had a similar profile to 'PSF' objects, while the "SER" objects presented extended profiles. 

\begin{figure} 
    \includegraphics[width=1.0\columnwidth]{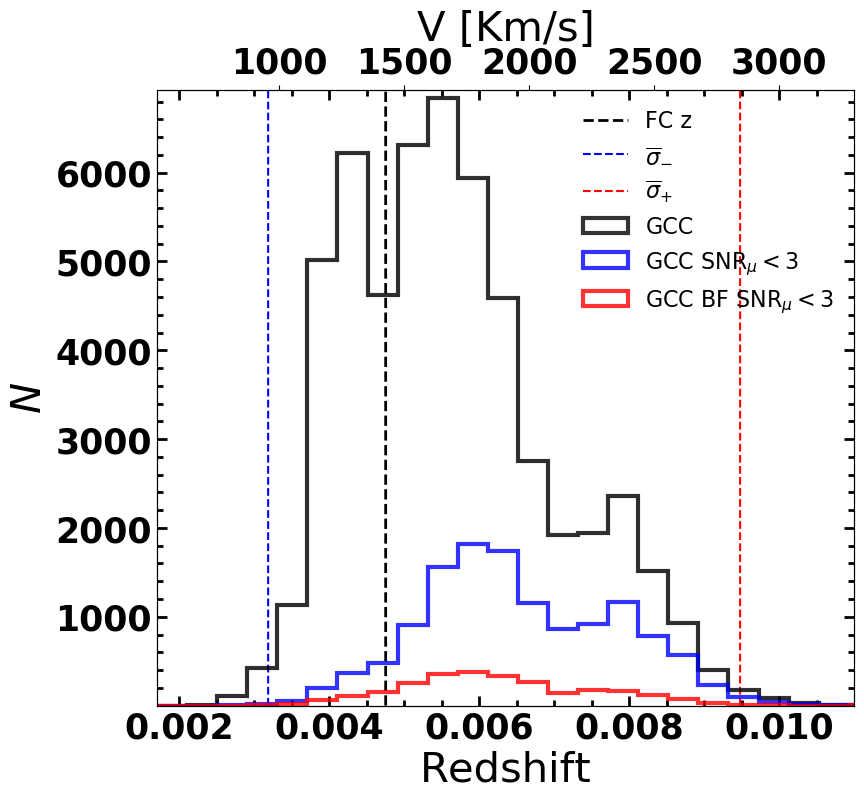}
    \caption{
    Redshifts estimated with LEPHARE.   
    {\it Black solid line:} sample of GC candidates before DESI and GAIA rejections.
    {\it Blue solid line:} sample after DESI and GAIA rejections.
    {\it Red solid line:} sample of bona-fide GC candidates.
    {\it Black dashed line:} the redshift of NGC~1399, the dominant galaxy of Fornax.
    {\it Blue and red dashed lines:} lower and upper range taking into account the mean errors for the redshift estimations.
    }
    \label{figura:redshift} 
\end{figure}

We also used the GAIA\footnote{\url{https://gea.esac.esa.int/archive/}} catalog \citep{GAIA:2016} to reject local contaminant sources. In order to separate Galactic stars from other objects, we used the proper motions from the catalog provided by Gaia Data Release 3 (Gaia DR3;  \citealt{GAIADR3:2021}) centered in NGC~1399 with a coverage of all the Fornax field of S-PLUS.
The number of sources of the Gaia catalog in the Fornax area is 1,129,284. We use the signal-to-noise ratio of the proper motion (SNR$_{\mu}$, \citealt{Voggel:2020}, \citealt{Buzzo:2022}),

\begin{equation}
   SNR_{\mu} =\sqrt{\mu_{RA}^{2}+\mu_{DEC}^{2}} /  \sqrt{\sigma\mu_{RA}^{2}+\sigma\mu_{DEC}^{2}},
    \label{equ:prichet}
\end{equation}

\noindent to select stars. Here, $\mu$ is the proper motion and $\sigma\mu$ the dispersion in each coordinate.
In fact, non-stellar objects are expected to have proper motions that are consistent with 0 at the 3$\sigma$ confidence level, while genuine stars are expected to have $SNR_{\mu}$ $>$ 3. 
We performed a cross-match between Gaia proper motions and our catalog, with which it was possible to reject $\sim$50\% of GC candidates. The error for the cross-match
was the mean of the ({\sc ra,dec}) coordinate errors from the Gaia catalog ($\sim$0.7~arcsec).

\begin{figure*} 
    \centering 
    \includegraphics[width=2.0\columnwidth]{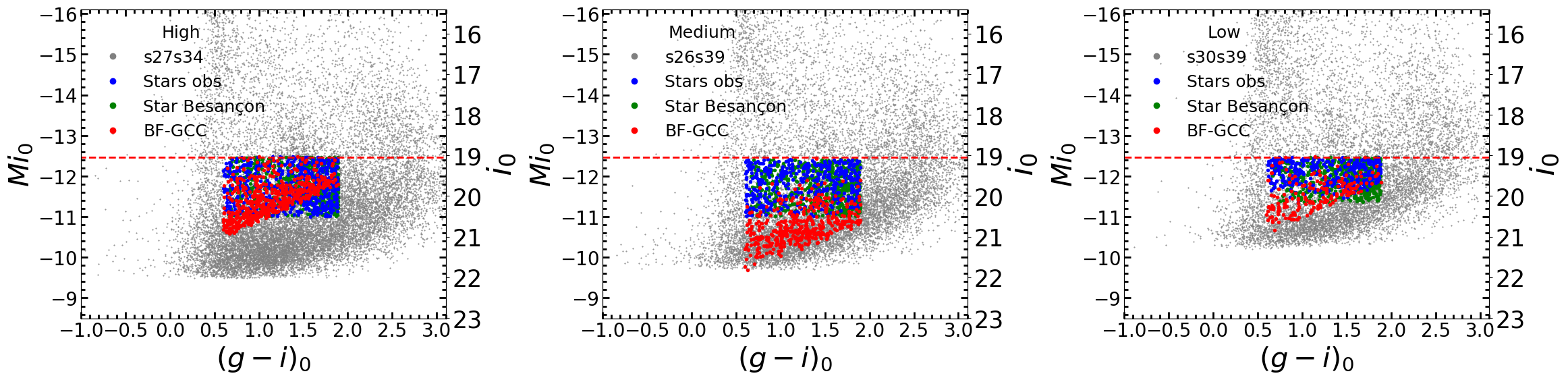}
    \caption{Contaminants estimation in the $i_{0}$ versus $(g-i)_{0}$ CMD in three selected fields, which have high (left), medium (center) and low (right) crowding.
    {\it Cloud of black small dots} are all the detections in each pointing.  
    {\it Blue circles} are the observed stars (sources with \class$>$0.90), 
    {\it green circles} are the expected stars from Besançon models \citep{Robin:2003} and 
    {\it red circles} are the BF-GCC.
    {\it Red dashed line} represents the the bright selection limit in magnitude which is estimated using the turn-over of the GCLF (see Section~\ref{seccion:magnitude_selection}).  
    } 
    \label{figura:cmd_contaminantes} 
\end{figure*}

\begin{figure*}
    \centering 
    \includegraphics[width=1.45\columnwidth]{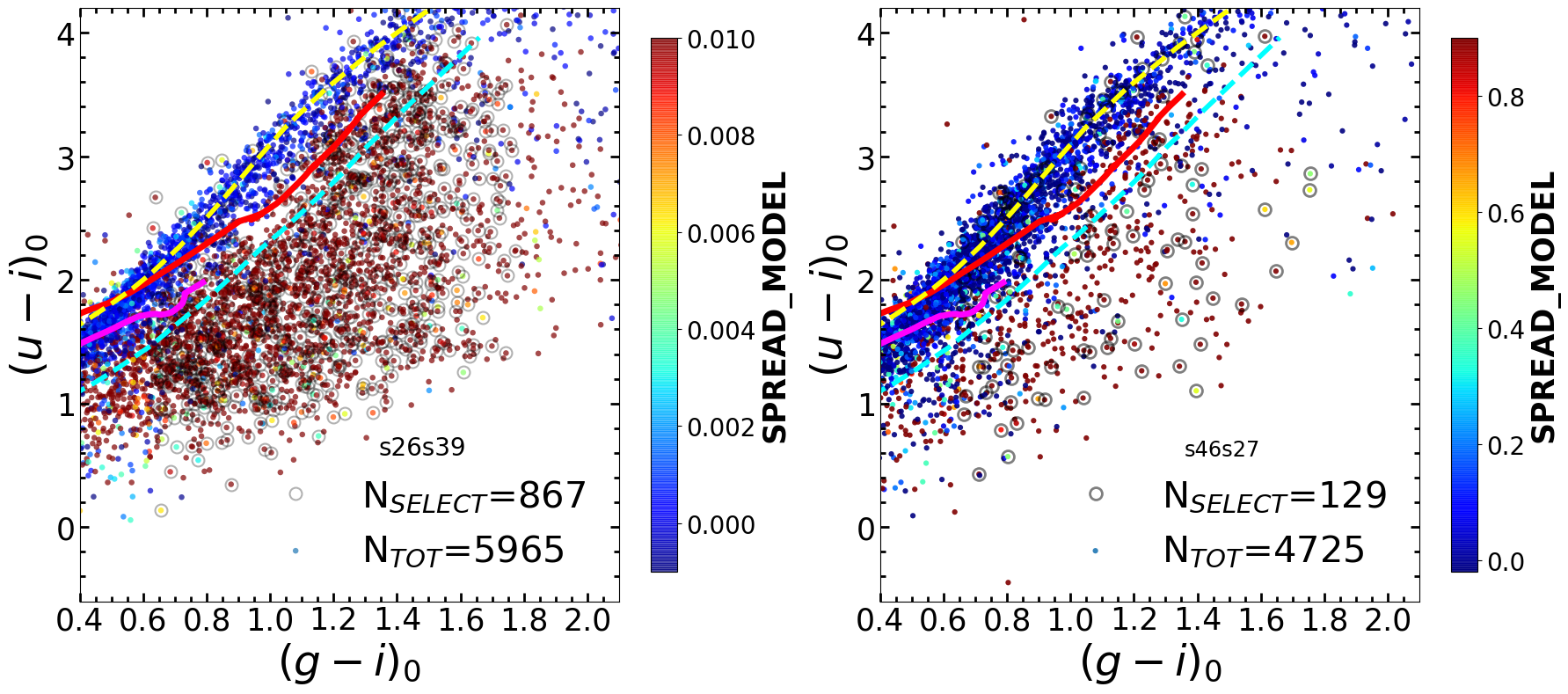}\vspace{0.3cm}       
    \includegraphics[width=1.45\columnwidth]{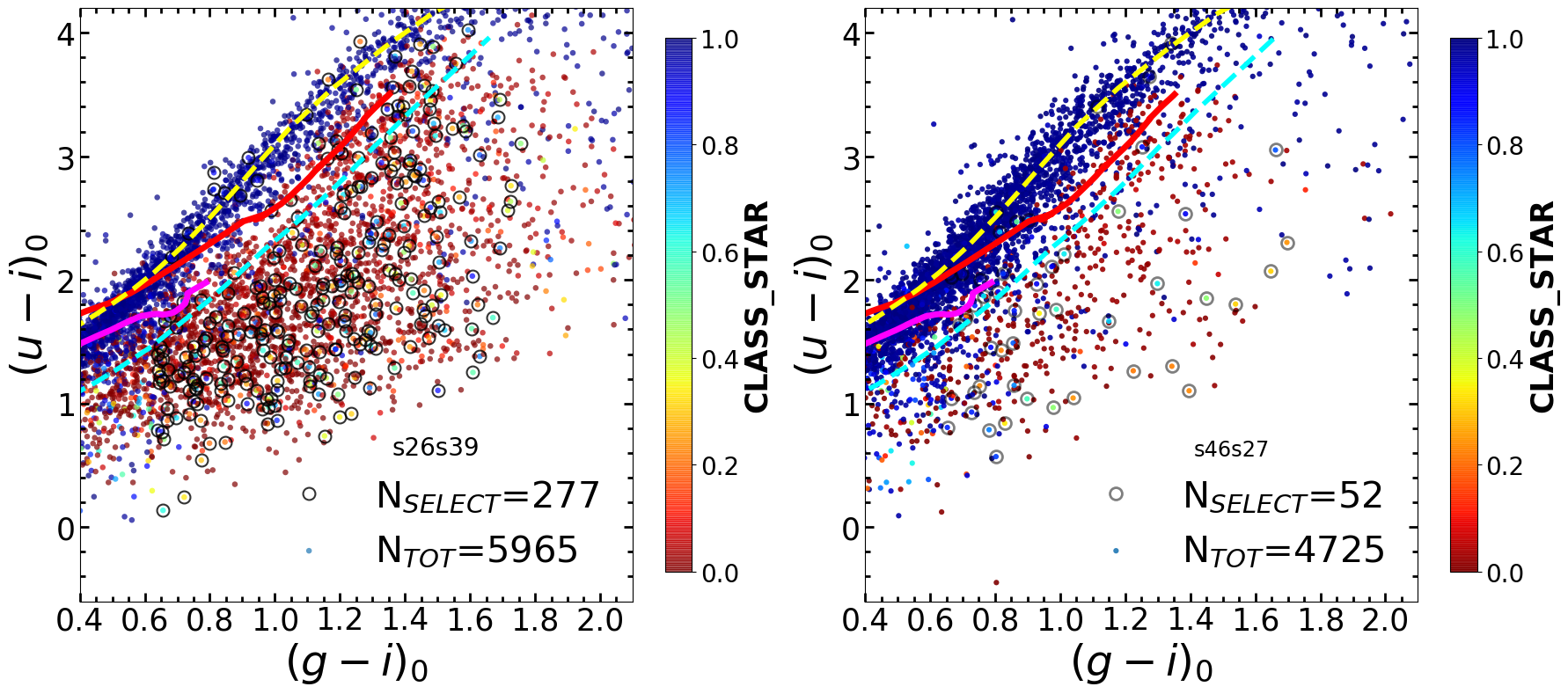}\vspace{0.3cm}
    \includegraphics[width=1.45\columnwidth]{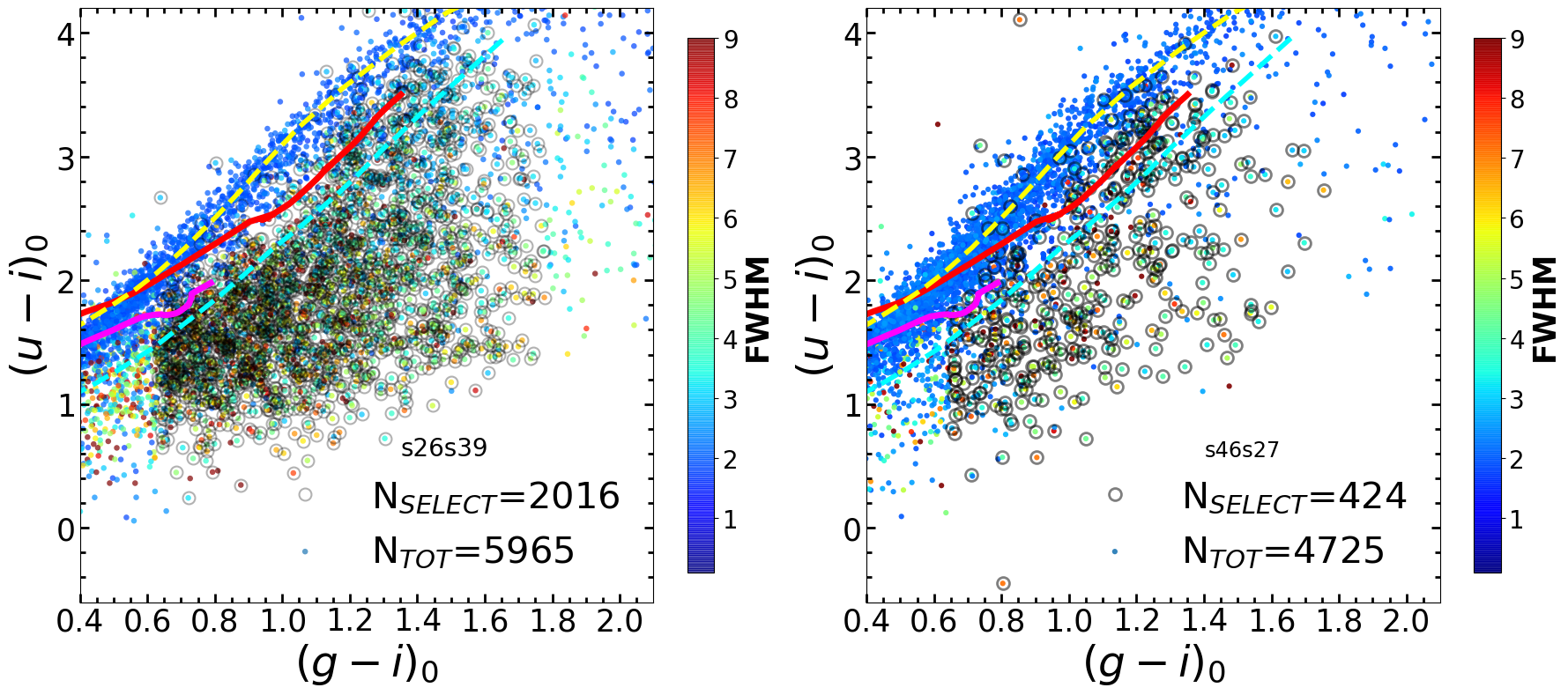}   
    \caption{Contaminants estimation in the $(u-i)_{0}$ versus $(g-i)_{0}$ color-color diagram.
    Comparison of structural parameters in the S-PLUS-s26s39 Fornax field {\it (left panels)} and 
    in the S-PLUS-s46s27 control field {\it (right panels)}. 
     From top to bottom we show \spread$\times$100, \class\ and \fwhm. 
    N$_{TOT}$ are all the objects 
    present  in each FoV ({\it small colored dots}) and 
    N$_{SELECT}$ are the objects that meet with the 
    selection criteria ({\it small colored dots into black open circles}),  respectively.   
    The {\it magenta} and {\it red lines} sketch the loci of SSPs models with Z=0.0004 and Z=0.03, respectively.
    The {\it cyan} and {\it yellow lines} sketch the loci of zero age main sequences (ZAMSs) stars 
    with Z=0.0001 and Z=0.04, respectively.   
    } 
    \label{figura:cmd_contaminantes_v2} 
\end{figure*}

We used LEPHARE\footnote{\url{http://www.cfht.hawaii.edu/~arnouts/ lephare.html}} \citep[][]{Arnouts:2002}, to reject extragalactic contaminant sources.
LEPHARE is a software of template-fitting (TF) based on a $\chi^{2}$ minimization, to fit data with both galactic and stellar templates. The templates used for this analysis are a set of galactic SEDs derived by the COSMOS survey collaboration \citep{Scoville:2007}, and the Pickles stellar spectra library \citep{Pickles:1998}. 
We used physically motivated priors to perform the LEPHARE fitting: 
$(g-i)$ color, absolute magnitude in i-band (M$_i$), extinction, 
and redshift z=0.4, 
considering that the photometric redshift distribution of S-PLUS sources with r$<$22~mag 
peaks at z=0.2 \citep[][]{Bom:2024}.
In Figure~\ref{figura:redshift}, we show the redshift results from the SEDs fitting for the sample before GAIA rejections (black solid line), after GAIA rejections (blue solid line) 
and the sample of bona-fide GC candidates (BF-GCC, see Section~\ref{seccion:contaminantes_u}, 
red solid line). We show the velocity of NGC~1399, 
$v_{*}$ = 1424.91$\pm$3.90, z$_{*}$ = 0.00475$\pm$0.00001 (black dashed line; \citealt[][]{Graham:1998}), and the range for the z estimations, z$-\sigma$ (blue dashed line) and z$+\sigma$ (red dashed line), where sigma is the mean error. We observed that a $\sim$99\% of the distributions fall within the estimated range. 
The final catalog of GC candidates consists of 12,999 objects.
In Table~\ref{tabla:selection} we list all the selection criteria.

\begin{table}
    \setlength\tabcolsep{2.2pt}
    \begin{small}
	\centering
    \begin{center}
	\caption{GC candidates selection resume. }
	\label{tabla:selection}
	\begin{tabular}{c c c} 
        \hline
        Parameter  &  Value  & $N_{GCC}$  \\ 
        (1)        & (2)     & (3) \\
       \hline  
        All        &  --                              & $\sim$3~million   \\
       \fwhm       &  2.8$\leq$\fwhm$\leq$9.0~[pixel] &  --       \\
       \class      &  0.2$\leq$\class$\leq$0.9        &  --     \\ 
       \spread     &  \spread$\leq$0.15               &  --     \\ 
       \fluxradius &  1$<$\fluxradius$\leq$4~[pixel]  &  $\sim$500,000       \\
       \magpsf     &  19.5$-1.0\sigma$ $\leq$ i $\leq$ mag$_{50}$~[mag]  &    \\
       \magerrpsf  &  g,r,i,z-bands$\leq$0.2~[mag]    & -- \\ 
       $(g-i)_{0}$ &  $(g-i)_{0}$$\geq$0.65$\pm1.0\sigma$~[mag] & -- \\
       $(r-i)_{0}$ &  $(r-i)_{0}$$\geq$0.19$\pm1.0\sigma$~[mag] & -- \\
       $(u-g)_{0}$ &  $(u-g)_{0}$$\geq$1.05$\pm1.0\sigma$~[mag] & $\sim$50,000  \\
       Gaia pm     &  SNR$_{\mu}$<3  & -- \\
       TF$_{crit}$ & $\chi_{red}^{2}$(galactic)$<\chi_{red} ^{2}$(stellar)  &  $\sim$25,000 \\
       DESI   &  “type” classification & 12,999   \\
       BF-GCC          &  --       & 2643     \\
       \hline
	\end{tabular}
    \end{center}    
        \end{small}
        \begin{scriptsize}
            \tablecomments{
                (1) Selection parameter. 
                (2) Selection cut.
                (3) Number of objects selected for cut.
                Not all rows have a number specified, as certain cuts are made simultaneously.}
                \end{scriptsize}
\end{table}

\subsection{Contamination of the GC sample from reddened young star clusters, stars and galaxies}
\label{seccion:contaminantes_u}

GCs' color is the most useful discriminator between young and old GC populations. For example, metal-poor SSPs (Z$\leq$0.001) predict $(g-i)\geq$ 0.65~mag for populations older than $\sim$3~Gyr \citep{Bruzual:2003}. 
The use of color–color diagrams involving ultraviolet and optical filters is known to break the age-metallicity degeneracy (e.g., \citealt{Georgiev:2006}; \citealt{Bastian:2011}; \citealt{Fedotov:2011}; \citealt{Lomeli:2022}, \citealt{Whitmore:2023}). The evolutionary loci of clusters in such a color-color diagram for theoretical SSPs from \citet{Bruzual:2003} of different metallicities, have been shown in Figure~\ref{figura:color_color}. There, we plot all the sources selected with the structural parameters and magnitude cuts (see Sections\,\ref{seccion:structural_selection} and \ref{seccion:magnitude_selection}) as {\it black dots}, while sources that also meet with the distance selection criteria (GAIA-LEPHARE selection) and DESI classification (see Section~\ref{seccion:distancias_selection}) are shown as {\it red} and {\it cyan open circles}.

The $(u-i)$ colors of reddened young ($<10$~Myr) clusters are distinctly different from that of clusters older than $\sim$3~Gyr, which allows us to break the age-reddening degeneracy. Thus, possible background
galaxies and reddened YSCs (contaminants) would lie below the SSP locus for age$>$3~Gyr having a bluer $(u-i)$ color for a given $(g-i)$.
In other words, for a redder, $(g-i) > 0.65$ mag, cluster to be considered as a genuine GC, its  $(u-i)$ color, after taking into account photometric errors, should correspond to a location above the SSP locus in Figure~\ref{figura:color_color}. As illustrated in Figure~6 of \citet{Lomeli:2022}, the real errors in photometry are larger than the formal error bars, which limits the use of the colors for a precise determination of age. Nevertheless, the photometric quality is good enough to separate background galaxies and reddened YSCs from GCs. Finally, from Figure~\ref{figura:color_color}, we can select the sample of bona-fide GC candidates (BF-GCC) 
as all sources {\it (red empty circles)} which meet all the selection criteria and it is also possible to obtain a sample of possible reddened YSCs ({\it cyan open circles)}. 

We obtain a mean estimation of stars and background galaxies contaminants in the final sample. 
The size of the region that we are analyzing is large, so we randomly selected 12 regions from the center to the outermost zones, 
in which there is a high, medium and low density of detected sources. 
We used the Besançon model of the Galaxy \citep{Robin:2003} to estimate which fraction of sources corresponds to foreground stars in the Milky Way. 
In the panels of Figure~\ref{figura:cmd_contaminantes}, we show the $i_{0}$ versus $(g-i)_{0}$ CMD in three selected fields, which have high (left), medium (center) and low (right) crowding, where the {\it clouds of black dots} are all the detections in each pointing. The {\it blue circles} are the observed stars, {\it green circles} are the expected stars from Besançon models and {\it red circles} are the BF-GCC. The observed stars were selected using a typical value of \class$>$0.90 and $SNR_{\mu}$, and for contamination estimation we used the stars that meet the selection criteria (colors and magnitudes) for GC candidates. The modeled stars are restricted to the selection parameter space, which is the reason they are overlapping with the observed stars. We estimate the percentage of contaminating objects as the difference between the recovered stellar objects and the objects predicted by the Besançon model divided by the number corrected for incompleteness of candidate GCs. 
We obtained a mean value of point sources contaminants of $\sim$29\%. 
We repeat the same procedure in the $(u-i)_{0}$ versus $(g-i)_{0}$ color-color diagram, where the mean value of contaminants is $\sim$20\%.

\subsection{Second contaminants estimation test}
\label{seccion:contaminantes_2017}

As a final test, in order to estimate the number of contaminants (foreground stars and background galaxies) 
in our sample, we used a control ﬁeld (CF) from the S-PLUS data at high Galactic latitudes, to reduce the contribution from  Milky Way sources, avoiding lines of sight 
with known nearby galaxies. 
The CF is SPLUS-s46s27 with central coordinates RA$_{J2000}$=68.32, DEC$_{J2000}$=-59.53.
The Galactic extinction value in that direction is $A_{V} = $0.033 \citep{Schlafly:2011}.

In Figure~\ref{figura:cmd_contaminantes_v2} we show the comparison between three structural
parameters in the $(g-i)_{0}$ versus $(u-i)_{0}$ color-color diagram for one Fornax field (FF, left panels) and the CF (right panels). In all the panels the sources depicted by black open circles are the objects that meet the GC candidates selection criteria in the parameters \spread\, \class\ and \fwhm.
In this example, the number of GC candidates in the CF is considerably smaller than in the Fornax field
for each of those structural parameters. 
The objects that meet the selection criteria in the CF 
are considered possible contaminants (column 3 in Table~\ref{tabla:spread_class_fwhm}). The percentage of contaminants in each structural parameter will 
be the ratio of the number of objects in the CF divided by the number of objects in the FF (column 4 in Table~\ref{tabla:spread_class_fwhm}). 
Thus, the percentage of possible contaminants per structural parameter is 15\% (\spread), 19\% (\class) and 21\% (\fwhm).
We emphasize that in the selection of GCs we are using the combination of structural parameters, so when using the combination of these parameters, the percentage of contaminants in this field is reduced by up to 11\% (11/103).
We follow the same procedure in 12 randomly selected fields finding the mean values of contaminants per structural parameter of 27\% (\spread), 27\% (\class) and 30\% (\fwhm), while the mean value of contaminants using the combination of all the structural parameters was 18\%. 
In Table~\ref{tabla:spread_class_fwhm}, we show the numbers (columns 2 and 3) and percents (column 4) in all the FoVs where we estimated the contamination.

\begin{table}
    \setlength\tabcolsep{10.0pt}
    \begin{small}
    \begin{center}
    \centering
    \caption{Number of contaminants per structural parameter.}
	\label{tabla:spread_class_fwhm}
	\begin{tabular}{c c r r r r} 
        \hline
        Parameter & FF       &  CF   & CF/FF  \\ 
        (1)       & (2)    & (3)   & (4)   \\
        \hline  
        \spread   & 5667    &  1400  & 24\%  \\
        \class    & 2445    &   572  & 23\% \\
        \fwhm     & 17036   &  4664  & 27\% \\  
        All       & --      &  --    & 18\% \\
	\hline
	\end{tabular}      
\end{center}    
 \begin{scriptsize}
            \tablecomments{
                (1) Parameter of selection.
                (2) FF, Fornax field.  
                (3) CF, control field.
                (4) Ratio between CF and FF. All is the percent of contaminants using all the selection criteria in all the fields.
                }
                \end{scriptsize}
        \end{small}
\end{table}

To ensure the genuineness of our BF-GCC sample, we performed two additional tests: a comparison of our sample with galaxy catalogs from the literature, and a visual inspection. In order to discard contamination for galaxies, we compared our BF-GCC sample with a compilation of 1005 Fornax galaxies reported in the literature (see Paper I). When performing a coordinate cross-match with an error of up to 10 arcsec, we obtained 7 matches with our catalog. We also compared the BF-GCC sample with the compilation of ultra diffuse galaxies (UDGs) from \cite{Zaritsky:2023}, obtaining 4 matches, three of which are among the 7 matches with the Fornax galaxies compilation. Finally, we compared  the BF-GCC sample with the catalog of 61 nucleated dwarf galaxies from \citet{Yasna:2018}, with a similar coordinate error, 
resulting in no match in this case. As next step, we performed a visual inspection of each one of the objects. With this last inspection, we rejected $\sim$50 objects that had a large extended emission and objects contaminated by other sources. The total number of objects in the BF-GCC sample is 2643. After the entire selection process we have created a BF-GCC sample with a low number of contaminants
($\sim$20\%), in which we perform further analysis in the following sections. 

\begin{table*}
    \setlength\tabcolsep{2.8pt}
    \begin{scriptsize}
	\centering
	\caption{Broad band colors and observational properties of the 10 brightest GC candidates.}
	\label{tabla:datos}
	\begin{tabular}{l c c c c c c c c c c c c c r} 
\hline 
Name    & RA     & DEC  & i  & $(u-i)_{0}$ & $(g-r)_{0}$  & $(g-i)_{0}$ & $(g-z)_{0}$ & \fwhm  & {\sc class}    & {\sc spread}     & {\sc flux}  & M$_{i}$     & flag  \\
        &        &      &    &             &              &             &   &    &  {\sc star}   &   {\sc model} &{\sc radius}   &   &  \\
(1)    & (2)    &   (3)  & (4)   &   (5)             & (6)             &    (7)         & (8) & (9)   &  (10)  & (11)   & (12)  & (13)   & (14)   \\
      
\hline
GC\_SPLUS-s30s31\_1 &   51.91487  &   -39.14339   &   19.02$\pm$0.03   &    1.46$\pm$0.14  &  0.48$\pm$0.06  &  0.63$\pm$0.06   &  0.60$\pm$0.08    &    4.96  &  0.89 &  0.001  &  2.05  &   -12.38 &  02 \\ 
GC\_SPLUS-s26s42\_2 &   66.72938  &   -34.23268   &   19.02$\pm$0.03   &   --  &  1.24$\pm$0.17  &  2.20$\pm$0.17   &  2.54$\pm$0.17    &    4.29  &  0.60 &  0.002  &  1.90  &   -12.38 &  00 \\ 
GC\_SPLUS-s31s38\_3 &   64.86757  &   -39.47520   &   19.02$\pm$0.02     &   --  &  1.20$\pm$0.10  &  1.94$\pm$0.09   &  2.34$\pm$0.09    &    3.14  &  0.69 &  0.000  &  1.77  &   -12.38 &  00 \\ 
GC\_SPLUS-s25s38\_4 &   58.42777  &   -32.43845   &   19.03$\pm$0.02    &   2.50$\pm$0.21  &  1.35$\pm$0.09  &  2.24$\pm$0.09   &  2.74$\pm$0.09    &    6.56  &  0.30 &  0.004  &  2.56  &   -12.37 &  01 \\ 
GC\_SPLUS-s31s37\_5 &   64.07645  &   -39.44286   &   19.03$\pm$0.02      &   3.26$\pm$0.40  &  1.00$\pm$0.12  &  2.30$\pm$0.11   &  2.85$\pm$0.11    &    7.83  &  0.72 &  0.004  &  3.03  &   -12.37 &  01 \\ 
GC\_SPLUS-s27s32\_6 &   50.17298  &   -35.15112   &   19.04$\pm$0.03     &   --  &  1.16$\pm$0.18  &  2.30$\pm$0.17   &  2.89$\pm$0.17    &    5.99  &  0.82 &  0.004  &  2.73  &   -12.36 &  00 \\ 
GC\_SPLUS-s29s40\_7 &   66.96018  &   -37.06000   &   19.04$\pm$0.03   &   2.62$\pm$0.21  &  0.64$\pm$0.05  &  1.04$\pm$0.05   &  1.10$\pm$0.06    &    3.61  &  0.82 &  0.003  &  1.75  &   -12.36 &  02 \\ 
GC\_SPLUS-s31s38\_8 &   65.78262  &   -40.69203   &   19.05$\pm$0.02     &   -- &  1.23$\pm$0.09  &  2.10$\pm$0.09   &  2.51$\pm$0.09    &    4.22  &  0.88 &  0.009  &  1.82  &   -12.35 &  00 \\ 
GC\_SPLUS-s24s28\_9 &   42.22746  &   -31.50407   &   19.05$\pm$0.02    &   --  &  1.23$\pm$0.10  &  2.08$\pm$0.10   &  2.45$\pm$0.10    &    2.97  &  0.83 &  0.001  &  1.58  &   -12.35 &  00 \\ 
GC\_SPLUS-s32s28\_10 &   49.21475  &   -41.01128   &   19.05$\pm$0.03    &   --  &  1.02$\pm$0.11  &  1.79$\pm$0.11   &  2.05$\pm$0.11    &    4.04  &  0.67 &  0.001  &  1.90  &   -12.35 &  00 \\ 
	\hline
	\end{tabular}
                \begin{scriptsize}
                \justify
                \tablecomments{
                (1) Assigned name, which follows the convention of FGC\_field\_n, 
                where F is for Fornax and GC for globular cluster, 
                field corresponds to the name of the pointing in which the GC candidate is located,
                and n is 1 for the brightest object in the i-band, and increases sequentially as 
                the magnitude increases.
                (2,3) Right ascension, Declination, in J2000.
                (4) Magnitude of PSF (\magpsf) in the i-band and magnitude error from SExtractor. 
                (5,8) $(u-i)_{0}$, $(g-r)_{0}$, $(g-i)_{0}$, and $(g-z)_{0}$  colors; 
                the error is the quadrature sum of the error in each band from SExtractor.
                (9,12) The structural parameters for each GC candidate: \fwhm, \class, \spread\ and \fluxradius.
                (13) M$_{i}$ magnitude.
                (14) FLAG classification: 00 clusters without u-band photometry; 01 determined as a reddened young
                 cluster from u-band photometry; 02--determined as BF-GCC from u-band photometry; and 03--error
                 in (u-i) is likely to be larger than the indicated formal error.}
                \end{scriptsize}
        \end{scriptsize}
\end{table*}

In summary, we used SExtractor for detecting $\sim$3 million sources and we used structural and classifier parameters (\fwhm, \class, etc.) as a first discriminating step, obtaining a sample of $\sim$500,000 sources. The next discriminator was the i-band magnitude and colors $(g-i)_{0}$ and $(r-i)_{0}$, from which we obtained a sub-sample of $\sim$50,000. sources. 
Then we used DESI classification, GAIA and LEPHARE, to reject Galactic and extragalactic contaminants from which we obtained a sub-sample of $\sim$13000 sources (flag=00 in Table~\ref{tabla:datos}).  
Finally, using the $(u-i)_{0}$ color we separated the young reddened stellar clusters (flag=01 in Table~\ref{tabla:datos}) and selected the BF-GCC sample 
(flag=02 in Table~\ref{tabla:datos}) resulting in 2643 sources. 
In Table~\ref{tabla:datos} we list the 10 brightest BF-GCC presenting their names (column 1), coordinates (columns 2-3), i-magnitudes (column 4) broad-bands colors (columns 5-8), structural parameters (columns 9-12), absolute i-magnitude (column 13) and flag classification (column 14), as an example of the published table.  
In the next section we present the results of the analysis of the properties for the BF-GCC sample.

\section{Properties of the Globular cluster system: Results and discussion}
\label{seccion:analysis}

We emphasize that one of the great strengths of S-PLUS is the large spatial coverage combined with its 12-band filter system, which allows to perform statistical estimations over large sky areas. 
Having obtained a sample of BF-GCs in the 106 Fornax pointings, the next natural step is analyzing their photometric properties. In the following sections, we present the analysis of their color distributions, the GCLFs and the spatial distribution, respectively. 
All magnitudes and colors have been corrected for the foreground
Galactic extinction using the A$_{V}$ values given in Table~\ref{tabla:splus}, and the \citet{Cardelli:1989} reddening curve. 

\subsection{Color distributions}
\label{seccion:colores}

\begin{figure} 
\begin{center}
    \includegraphics[width=1.0\columnwidth]{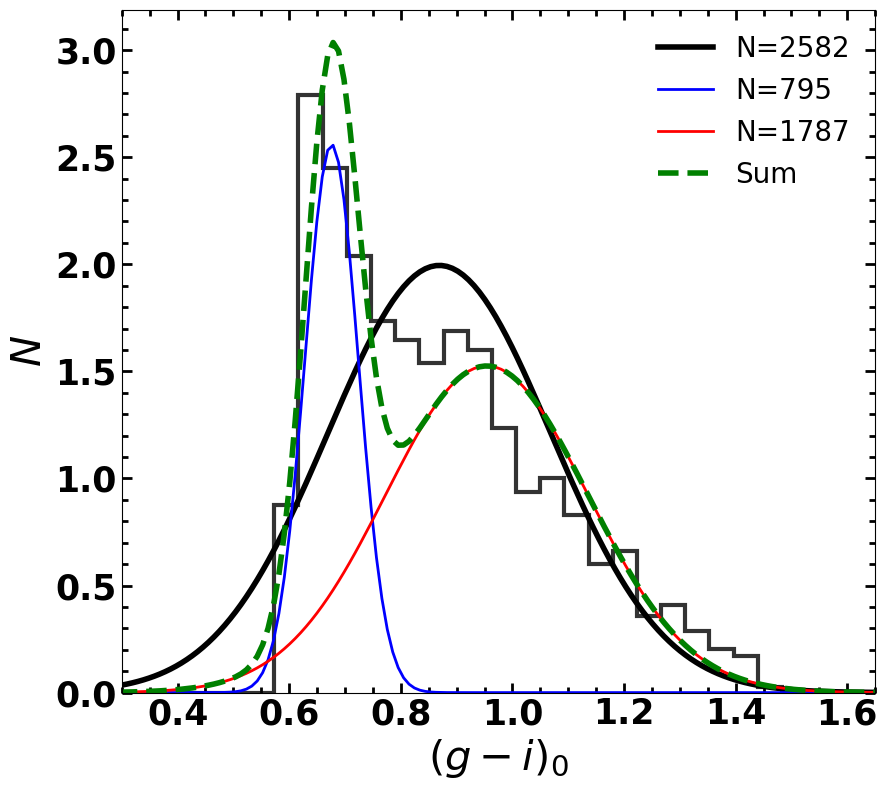}    
    \caption{$g-i)_{0}$ color distribution for the BF-GCC sample. {\it Black solid line} is the unimodal fit returned by the GMM analysis and {\it blue} and {\it red solid line} is the bimodal fit returned by the GMM analysis. The {\it green dashed line} is the sum of the red and blue fits.}  
    \label{figura:color_ancho_narrow} 
\end{center}    
\end{figure}

\begin{figure}
    \includegraphics[width=1.\columnwidth]{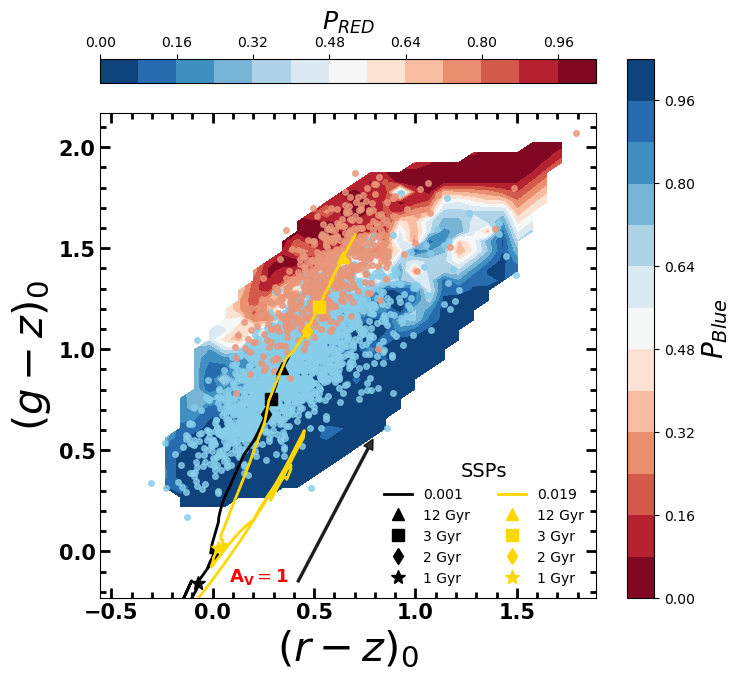}  \caption{ $(g-z)_{0}$ vs $(r-z)_{0}$ diagram with color bars indicating the GMM probability distribution to be red ({\it top color bar}) and blue ({\it right color bar}) GC candidates. 
    The evolutionary loci of SSPs from \citet{Bruzual:2003} corresponding to a Kroupa IMF with Z=0.001 (black solid curve) and Z=0.019 (yellow solid curve) are shown. Locations corresponding to 1, 2, 3, and 12 Gyr in each locus are indicated with different symbols (see the symbol code in the plot). The reddening vector with $A_{V}=1$~mag is represented by the black arrow.  
    } 
    \label{figura:colores_bi_analisis} 
\end{figure}

Previous studies (e.g., \citealt{Bassino:2006}, \citealt{Kim:2013}, \citealt{Cantiello:2018}) have shown that the color distribution of the GC system in Fornax is bimodal. However, they have been focused on the central region of the Fornax cluster, where NGC~1399 is located, and they cover, at most, one virial radius and they are much deeper, sampling GCs of typical GC masses, i.e. 10$^5$~\msun. Here we show that the color bimodality of the GC system is evident up to, at least, 3~\rv\ (see Section~\ref{seccion:color_spacial}).
A common result found in early-type galaxies is a bimodal distribution of GCs colors (\citealt{Zepf:1993}, \citealt{Gebhardt:1999}, \citealt{Larsen:2001}). 
A correlation between color and metallicity was also observed in different galaxies \citep[e.g.][]{Peng:2006, Brito:2011}. Given this correlation, the bimodal color distribution has been attributed to a bimodality in the abundance of metals \citep[see,][]{Brodie:2006}.
The bimodality would indicate the presence of two different populations of GCs: the metal-poor (commonly referred to as blue) and the metal-rich (usually designated as red).
The two kinds of GCs also seem to be spatially segregated, with the distribution of metal-poor GCs having a larger scale length as compared to the metal-rich GCs \citep[e.g.][]{Larsen:2003, Webb:2012, Schuberth:2010, Kartha:2014}.

In Figure~\ref{figura:color_ancho_narrow} we show the $(g-i)_{0}$ color distributions for the BF-GCC sample where the sub-index 0 stands for Galactic reddening-corrected colors, using the $A_V$ values in Table~\ref{tabla:splus}.
In total we  analyzed 10 colors with the broad-bands: 
$(u-g)_{0}$, $(u-r)_{0}$, $(u-i)_{0}$, $(u-z)_{0}$, $(g-r)_{0}$, $(g-i)_{0}$, $(g-z)_{0}$, $(r-i)_{0}$, $(r-z)_{0}$, $(i-z)_{0}$
and 15 colors with the narrow-bands: 
$(J0378-J0410)_{0}$, $(J0378-J0430)_{0}$, $(J0378-J0515)_{0}$, $(J0378-J0660)_{0}$, $(J0378-J0861)_{0}$, $(J0410-J0430)_{0}$, $(J0410-J0515)_{0}$, $(J0410-J0660)_{0}$, $(J0410-J0861)_{0}$, $(J0430-J0515)_{0}$, $(J0430-J0660)_{0}$, $(J0430-J0861)_{0}$, $(J0515-J0660)_{0}$, $(J0515-J0861)_{0}$, $(J0660-J0861)_{0}$.
In Section~\ref{seccion:distribucion_espacial}, we analyzed the $(g-i)_{0}$ and $(g-z)_{0}$ colors distribution at differents \rv\ from the center of Fornax.

\begin{table} 
\begin{center}
\setlength\tabcolsep{7.5pt} 
\caption{Total number of GC candidates at differents \rv.}
\label{tabla:total_number} 
\begin{tabular}{l r c c c c c c c c l r l}
\hline
Sample  &   Obs   & Blue$_{gi}$  &  Red$_{gi}$   & Blue$_{gz}$  & Red$_{gz}$ \\
(1)     &  (2)    &      (3)     & (4)           & (5)          & (6) \\       
\hline

0.5\rv &   99     & 48   & 51   & 76   & 23 \\
1\rv   &  315     & 139  & 176  & 233  & 82 \\
2\rv   &  863     & 432  & 431  & 658  & 205 \\
3\rv   & 1597     & 816  & 781  & 1212 & 385 \\
BF-GCC & 2653     & 1390 & 1263 & 2034 & 619 \\
\hline
\end{tabular}
\end{center} 
             \tablecomments{(1) Subsamples. 
                (2) Number of BF-GCC in each \rv.
                (3,4) Number of BF-GCC in each \rv, divided by color, $(g-i)_{0}$.
                (5,6) Number of BF-GCC in each \rv, divided by color, $(g-z)_{0}$.}
\end{table}

To confirm the presence of bimodality in the color distributions we used the Gaussian Mixture Modeling (GMM) code \citep{Muratov:2010} which carries out a robust statistical test for evaluating bimodality, and uses a likelihood-ratio to compare the goodness of the fit for double-Gaussian versus a single-Gaussian. This method is independent of the binning of the sample. The results from the GMM test in the color distributions are shown in Table \ref{tabla:colores}. 
For a distribution to be considered bimodal, the Kurtosis must be negative, the distance between the peaks of the distributions (D) must be D$>2$ and $p$-values must be small \citep[e.g.][]{Muratov:2010}. According to GMM results and visual inspection, we find evidence of color bimodality in the colors $(g-i)_{0}$ and $(g-z)_{0}$, while we did not find strong statistical evidence to confirm bimodality in any of the narrow-band colors. A possible explanation for that might be linked to the fact that the narrow-band filters sample a short spectral range. In addition, the larger errors in the narrow-bands magnitude estimations might extend the color distribution, fading the bimodality. Moreover, we are looking at only very bright (massive) GCs.  The blue tilt found for bright (massive) GCs a washes out a color bimodality for massive GCs \citep[e.g.,][]{Mieske:2010, Fensch:2014}.

In Table~\ref{tabla:total_number} we present the BF-GCC candidates divided in blue and red sub-popultions using $(g-i)_{0}$=0.86 (columns 3 and 4) and $(g-z)_{0}$=1.24 (columns 5 and 6) according to GMM results which are in agreement with the values found in \citet[][]{DAbrusco:2022} for NGC~1399. We will use this classification in order to analyse the projected distribution of the BF-GCC in the rest of the colors. In Figure~\ref{figura:colores_bi_analisis}, we show one example of these novel color-color diagrams using colors different to those used for the GC candidates selection (see Section~\ref{seccion:color_selection}).
From the SSPs metal-poor model shown in Figure~\ref{figura:color_color}, we set the color cut selection of GC candidates with a minimum age of $\sim$3~Gyr.
When compared with Figure~\ref{figura:colores_bi_analisis}, it is observed that the sample of BF-GCC remains older than $>$1~Gyr (Z=0.001, black star). In the 
$(g-z)_{0}$ versus $(r-z)_{0}$ diagram,  $\sim$20\% (453 objects) of the blue sub-sample displays an age  
$\lesssim3$~Gyr (Z=0.001, black square), while $\sim$53\% (1086 objects) is older than 12~Gyr (Z=0.001, black triangle). 
On the contrary, 99\% of the red sub-sample seems to be older than 12~Gyr (Z=0.001, black triangle). 

There are two possible explanations: the projected color shift in the BF-GCC towards ages younger than 3~Gyr can be attributed to the fact that in the color selection we accept objects with 1$\sigma$  photometric errors; or according to this brief analysis, independently of selection cuts, the blue subsample appeared to have more than one formation peak, a portion older than $>$12~Gyr, formed in a first burst of star formation \citep{Ashman:1992}, or invoking the mass-metallicity relationship \citep[][]{Cote:1998, Strader:2005, Forbes:2018_metallicity}, formed in less-massive satellite galaxies that are subsequently acquired by giant galaxies during the accretion process.
While the other portion of GCs probably was formed in more recent burst $\sim$2-3~Gyr, which would then not be old classical GCs, but rather intermediate-age clusters \citep[e.g.,][]{Simanton:2015}. 
Meanwhile, without taking into account intrinsic extinction the red subsample is entirely older than $>$12~Gyr.

In Figure~\ref{figura:colores_bi_analisis}, we show the contours that represent the probability (estimated with GMM) for a BF-GCC to be blue ({\it left color-bar}) or red ({\it top color bar}). The locus of both populations together with the probability bars reinforces the hypothesis of the existence of two independent subpopulations.

\subsection{GCLFs}
\label{seccion:funcion_luminosidad}

Studies have shown that the GCLF for early-type galaxies is Universal \citep[e.g.,][]{Whitmore:1995, Brodie:2006, Jordan:2007, Lomeli:2022}, which implies that the peak of the distribution, or turn-over (TO), is the same in all galaxies. This has led many authors \citep[e.g.,][]{Richtler:2003, Forbes:2018} to assume that GC systems are fundamental pieces to understand the formation and evolution processes of their parent galaxies. By understanding the formation processes of individual galaxies, we could understand the formation and evolution processes of galaxy clusters. However, the universality of the GCLF for late-type galaxies has not been fully tested. Examples of studies where universality has been proven are found in the MW \citep[e.g.,][]{Harris:1996, Bica:2003}, M31 \citep[e.g.,][ suggest a second peak in the GCLF]{Peacock:2010, Wang:2019} and a few nearby galaxies \citep[see][]{Lomeli:2022}. Here we show the global GCLF in all observed fields, which may include GC systems in all kind of galaxies (elliptical, spiral, dwarf, etc.). Subsequent studies will be concentrated in the analysis of individual galaxies in Fornax.

\begin{figure} 
    \includegraphics[width=1.\columnwidth]{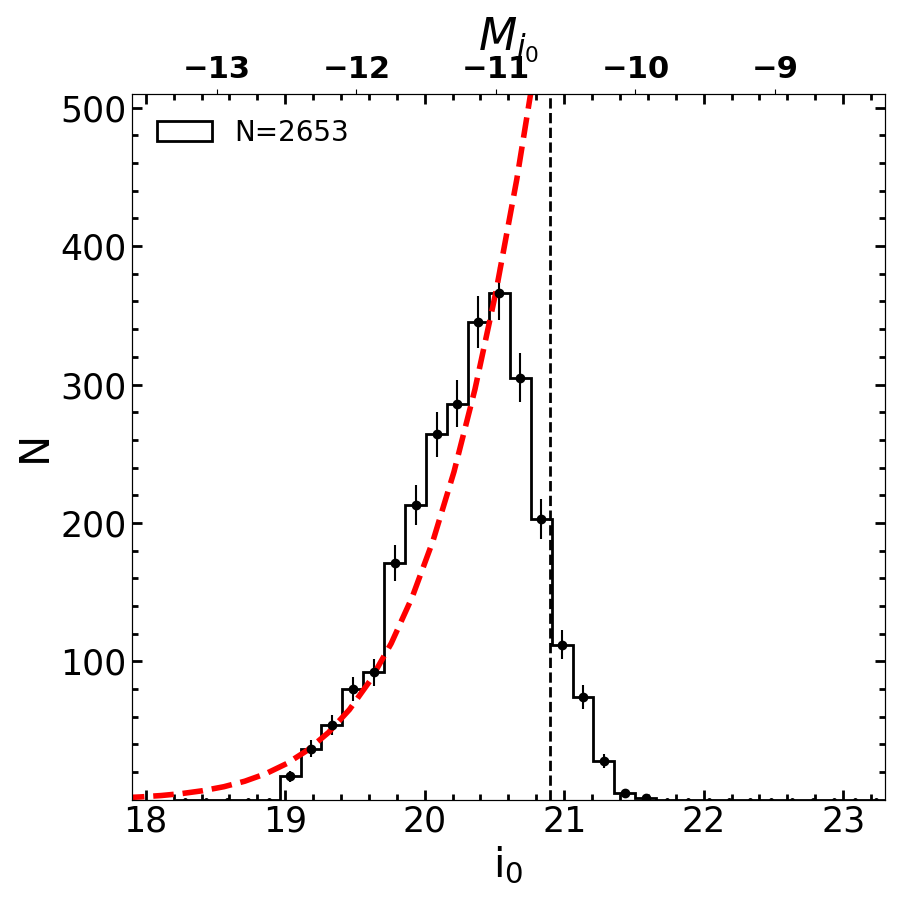}      
    \caption{i-band GCLF distribution (histogram). {\it Red dashed line:} is the expected GCLF using a log-normal distribution corrected for incompleteness in magnitude. The {\it vertical dashed lines} indicate the magnitude at which the detection is 50 per cent is complete. Poisson error bars ($\sqrt{N}$) are indicated.    
    } 
    \label{figura:gclf} 
\end{figure}

\begin{figure*} 
\begin{center}    \includegraphics[height=9cm,width=15cm]{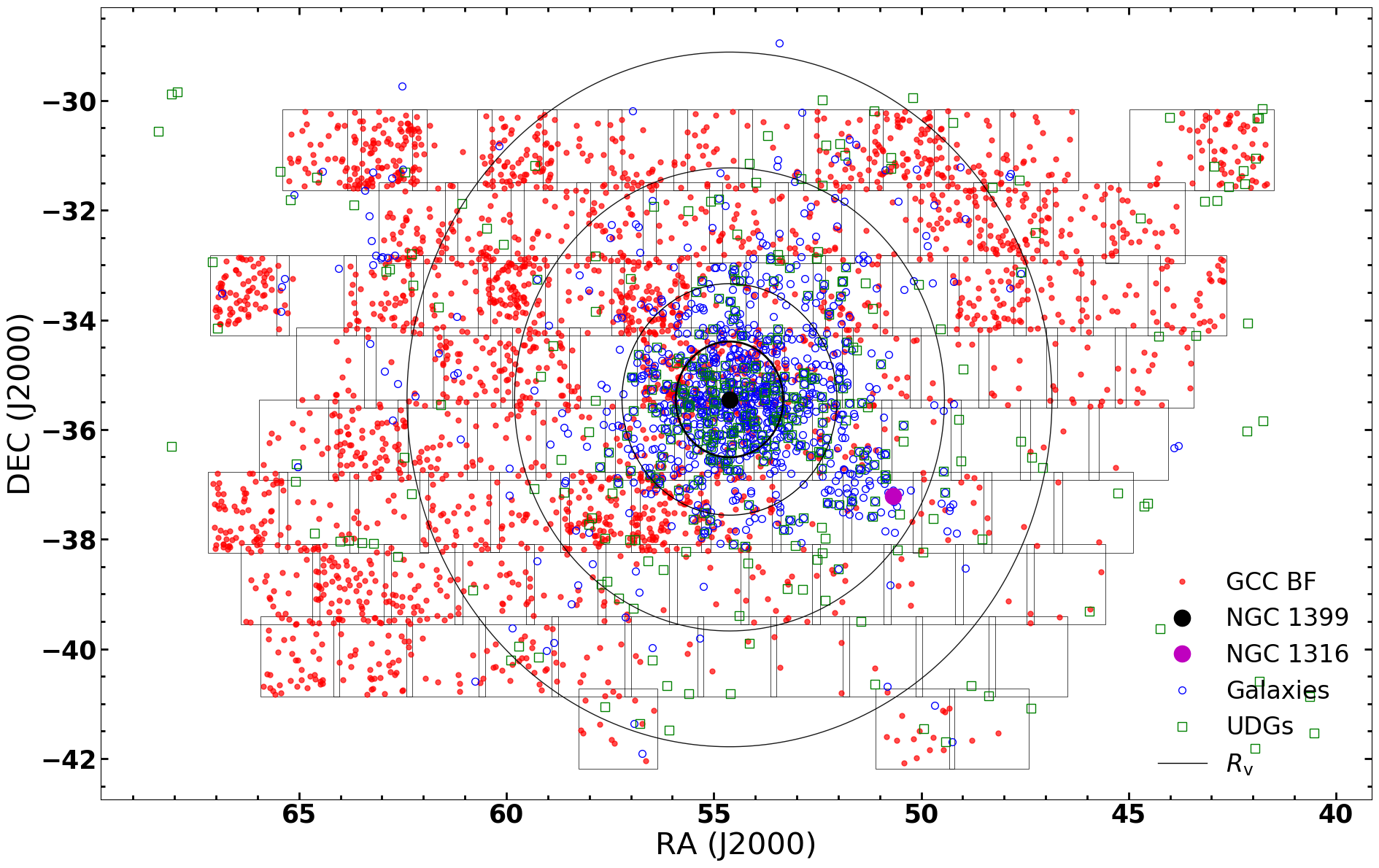}    
    \includegraphics[height=5cm,width=7.5cm]{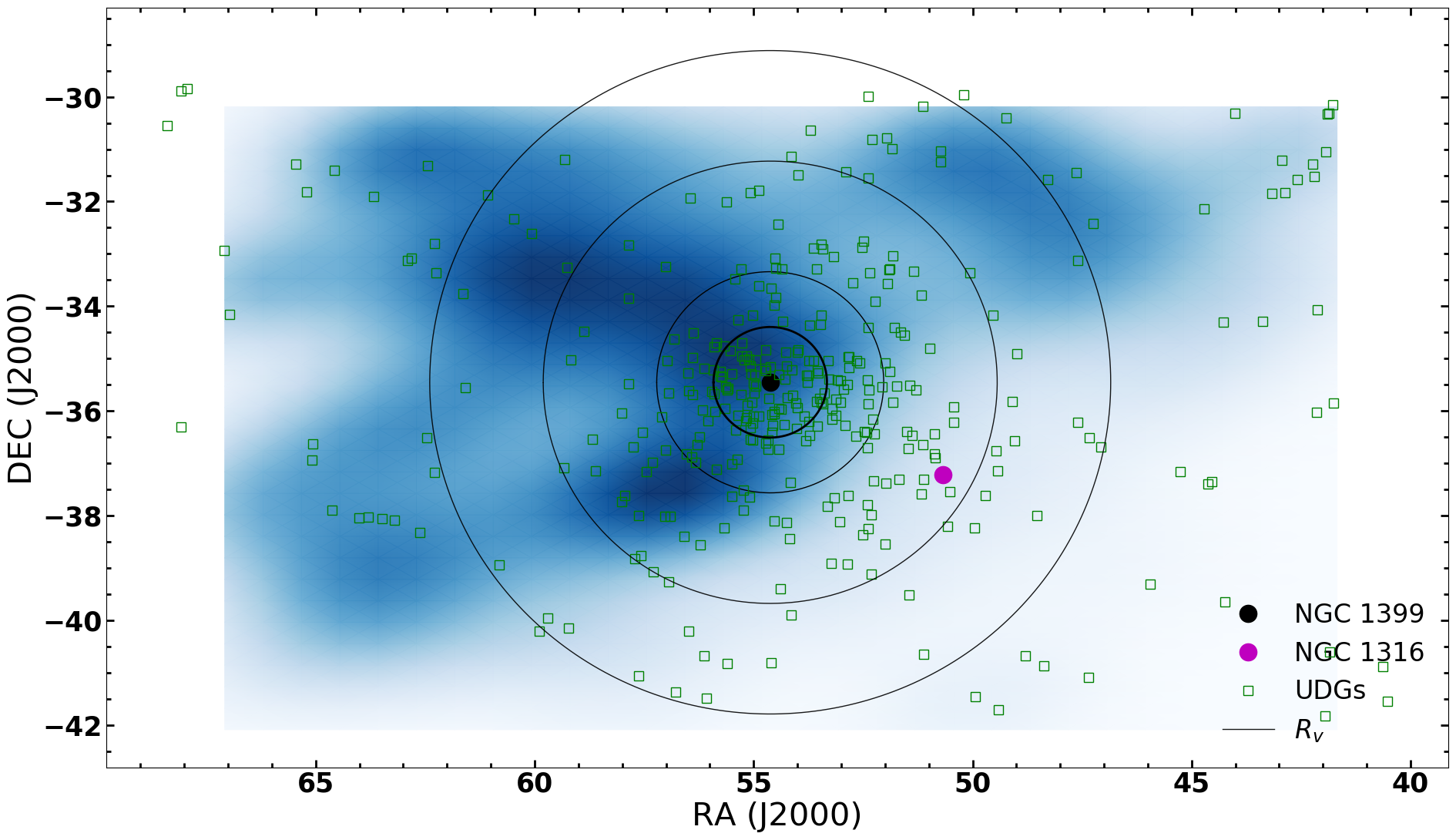}    
    \includegraphics[height=5cm,width=7.5cm]{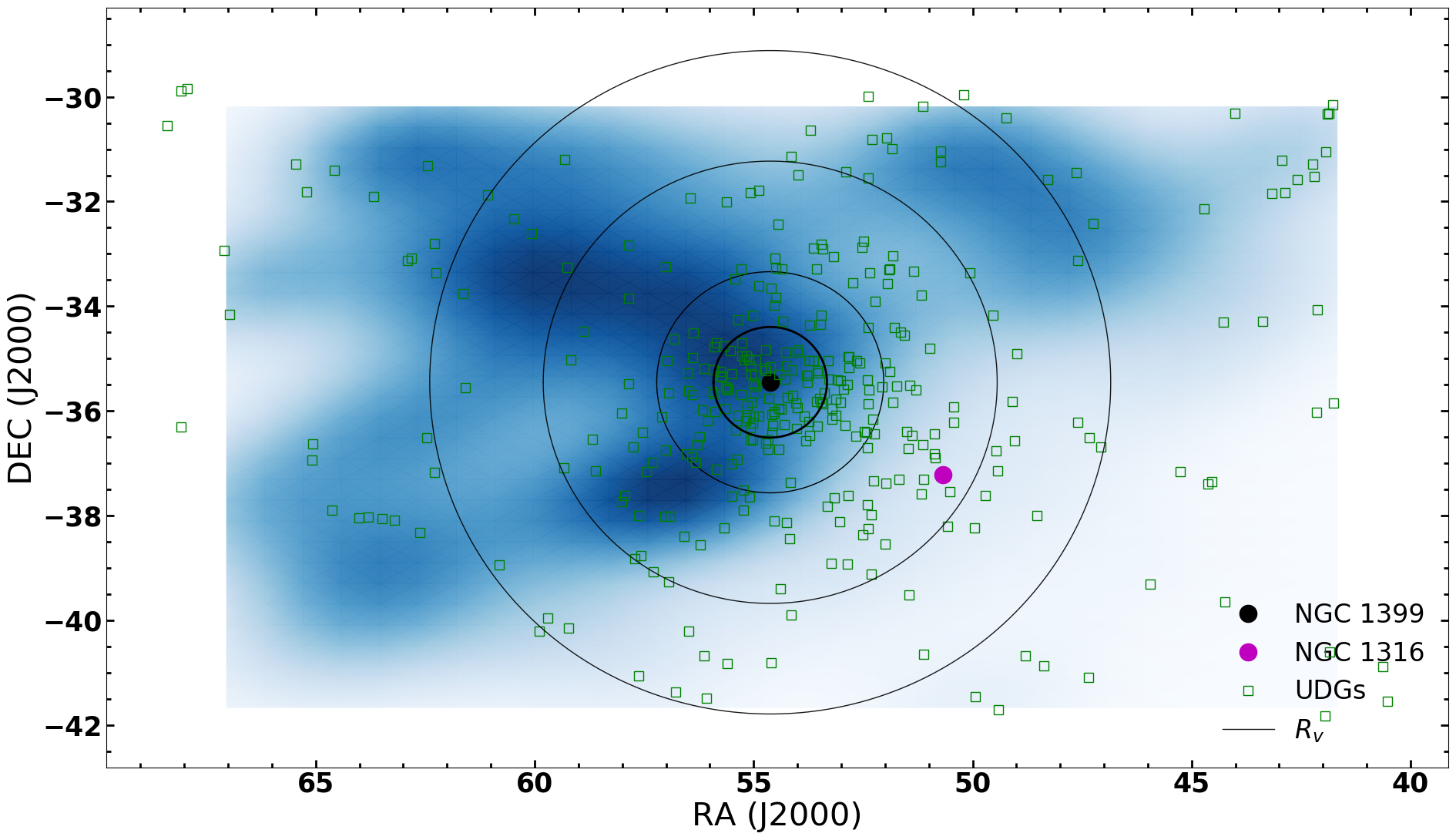}     
    \caption{Spatial distribution in \ra,\dec\ (J2000) coordinates for the BF-GCC sample ({\it red dots}). Galaxies from the literature are indicated with {\it blue circles} and {\it green empty squares}.
    The {\it black empty circles} are the multiples (0.5, 1, 2, 3) of the \rv\ centered in NGC~1399. {\it Black solid point:} represent the center of NGC~1399 galaxy.
    In the bottom panels we show the GCs  distribution smoothed with a Gaussian kernel ({\it blue distribution}) of the BF-GCC on the left, and including only objects identified as ``PSF'' by DESI on the right, see text for more details. 
    }     \label{figura:spatial_distribution} 
\end{center}    
\end{figure*}

\begin{figure*} 
    \centering
    \includegraphics[height=9cm,width=13cm]{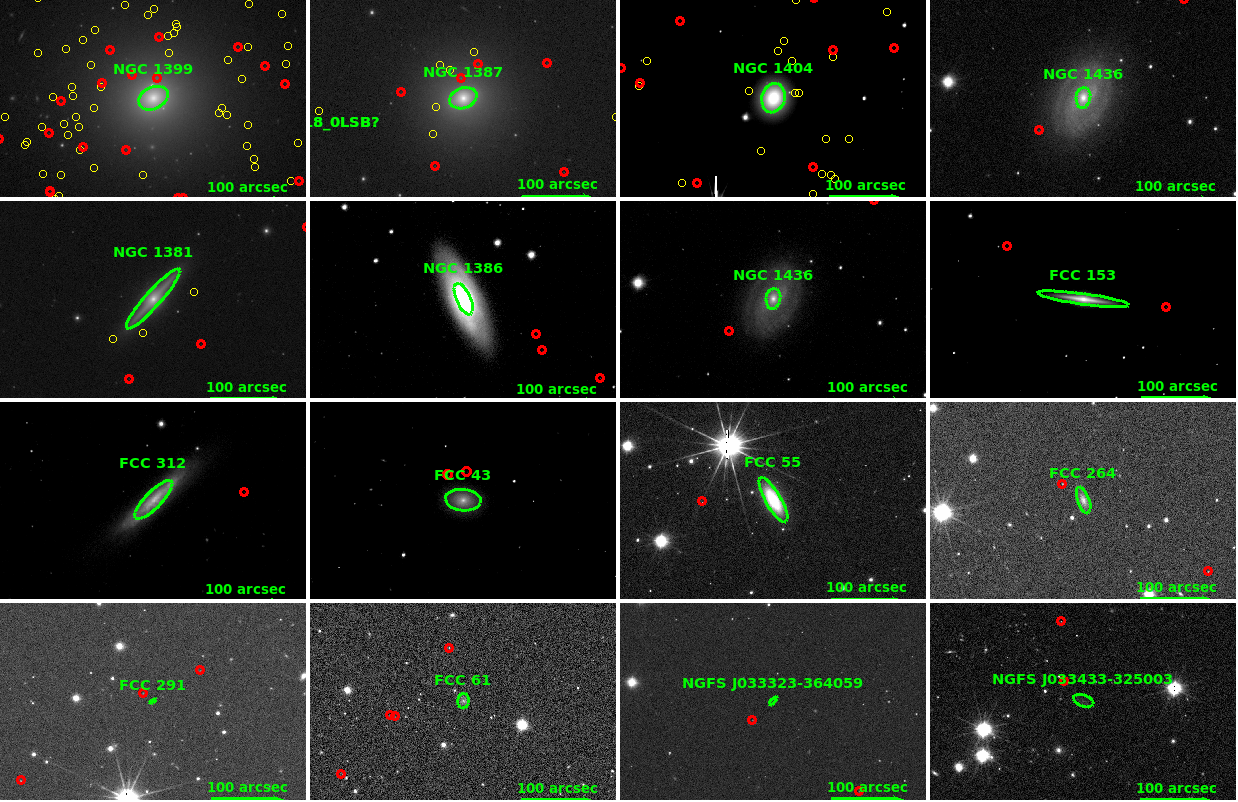}     
    \caption{Sample of galaxies ({\it green ellipses}), GC candidates ({\it red circles}) and spec-GCs recovered in S-PLUS i-band (\it yellow circles). All images are snaps in i-band from S-PLUS and are aligned such that north is up and east to the left and the angular size equivalent to 100 arcsec (green arrows) are shown.
    } 
    \label{figura:gcs_galaxies} 
\end{figure*}

In Figure~\ref{figura:gclf}, we show the GCLF (black solid histogram) in the i-band for the BF-GCC sample. For completeness we show the 12-bands GCLFs in  Figure~\ref{figura:gclf_apendice}. 
We emphasize that it is only possible to observe the brightest part of the GCLF because the S-PLUS images are not deep enough to detect, in a confident manner, objects fainter than $i\sim21.44$ mag. Thus it is not possible to observe the TO generally found in early-type galaxies \citep[e.g.,][]{Whitmore:1995}. However, it is possible to estimate the expected TO 
in the i-band according to the SSPs models from \citet{Bruzual:2003},
for an old stellar population (12~Gyr) with a low-metal content (Z=0.0004), for which $(g-i)_{AB}=0.7307$ mag. From the transformation equation for metal-poor stars by \citet{Jordi:2006} (see Appendix~\ref{apendice_C}),
the TO at M$_{{\text V}}=-7.4\pm0.10$ mag \citep[e.g.][]{Harris:1996,Jordan:2007},  translates into a TO at M$_{{\text i}}=-8.3\pm0.10$~mag. Considering that (m-M)=31.51 mag for Fornax, the TO is expected to occur at i=23.21~mag. The red dashed line in Figure~\ref{figura:gclf} is the expected GCLF using a log-normal distribution corrected for incompleteness in magnitude, 
\begin{equation}\label{eqn:GCLF}
dN/dM= N_{0}e^{-( M-M_{0})^{2}/2\sigma_{\rm M}^{2}}, 
\end{equation} where $N_{0}$ is a normalization factor, $M$ is the absolute magnitude of the fitted bin, $M_{0}$ is the absolute magnitude of TO and $\sigma_M$ is the dispersion. The GCLF is corrected for incompleteness using the expected TO (described above) and the 3$\sigma$ limit of the GCLF assuming $\sigma=$1.40$\pm$0.06 \citep[e.g.][]{Whitmore:1995, Karla:2013}, and for a factor correction caused by selection criteria in the colors, eg., $(u-i)_{0}$. This factor correction was estimated by the number of non detections in u-band (u-band is shallower) respect to the detections in i-band.

\subsection{Spatial distribution}
\label{seccion:distribucion_espacial}

Taking into advantage the large spatial coverage of S-PLUS, in this subsection we present the analysis of the projected spatial distribution of the identified BF-GCC, as well as that of their colors and of the GCLF at different radius from the center of Fornax. 

\subsubsection{GCs spatial distribution}
\label{seccion:distribucion_espacial2}
In Figure~\ref{figura:spatial_distribution}, we show the BF-GCC (red dots) spatial distribution in \ra,\dec\ (J2000) coordinates in the 106-FoVs ($\sim$200~deg$^{2}$) analysed here. Different multiples (0.5, 1, 2, 3) of the virial radius, \rv\ (black empty circles), centred on NGC~1399 (\ra=54.620941, \dec=-35.450657) are displayed.
A compilation of spectroscopically confirmed galaxies from \citet{Analia:2024} ({\it blue empty circles}) and ultra diffuse galaxies ({\it green empty squares}) from \citet{Zaritsky:2023} are also shown. 

In Figure~\ref{figura:spatial_distribution}, we noted that the highest concentration
of GCs is towards the center where NGC~1399 is located. 
One \rv\ is equivalent to $\sim$2~degrees, 
which at the Fornax distance is equivalent to $\sim$720~kpc.
The number of BF-GCC inside 1~\rv\ centred on NGC~1399 is 315. 
We note that, although all the spectroscopically confirmed galaxies from the literature are shown, there are areas where a concentration of BF-GCC is displayed, lacking their host galaxies (for example from 61\grad\ to 66\grad\ in {\ra} and -41\grad\ to -37\grad\ in {\dec}).
On the other hand, in the fields covering from 45\grad\ to 49\grad\ in {\ra} and -40\grad\ to -36\grad\ in {\dec}, there seems to be a scarcity in the detections of GC candidates. In addition, we also note that in some 
FoVs for example in s24s41 
(62\grad\ to 64\grad\ -32\grad\ to-30\grad\ ), the distribution of GC candidates is homogeneous thorough the field. 
The difference in number of recovered BF-GCC could be real or the results of a variation of the observing conditions.
In appendix~\ref{apendice_A}, we provide an analysis of different parameters such as: air-mass, exposure time and background. We find that, even if a non linear combination of these parameters is affecting our ability of recovering GCs, they are not sufficient to explain the variation of detections. To further investigate this behaviour, we calculated the reason between the number of objects detected in a tile and the number of BF-GCC extracted in the same tile. In fact, the observing conditions would affect in the same manner both detections. In the last row of Figure~\ref{apendice_A} we show the result of this experiment, which suggests  that, particularly in the South-West, the lower number of detections is probably related to observational conditions of each tile. Such result is consistent with the maps of the sky RMS and median value, which are also lower in the South-West region, reflecting worse observing conditions.

GCs literature studies in Fornax  are mostly focused on the central part, where the galaxy NGC~1399 is located.
For instance, \citet{Bassino:2006} studied the GCs distribution in a limited central area up to a radius of 275~kpc. 
Another example is the study of  \citet{Dabrusco:2016} where all GCs are distributed within ~210~kpc, whereas in \citet{Cantiello:2018}  86\% of their sample is concentrated within ~0.5~\rv, equivalent to 360~kpc (using our distance convention). Our sample extends up to 3~\rv\ with a complete spatial coverage, although it can reach up to 5~\rv\ (in East-West direction) with an incomplete spatial distribution (in North-South direction), see Figure~\ref{figura:spatial_distribution}. 
With the spatial coverage reached in this study, we are able for the first time to explore the large scale distribution of GCs within a galaxy cluster. 
As previously mentioned, GCs could be associated to the BCG \citep[e.g.,][]{Dirsch:2003, Bassino:2006, Blakeslee:2012}, to other galaxies \citep[e.g.,][]{Villegas:2010, Kim:2013} or to the intracluster light \citep[e.g.,][]{Schuberth:2008, Kaviraj:2012, Reina:2022, Kluge:2024, Saifollahi:2024}. 
The GCs distribution also highlights the interactions between galaxies within a cluster \citep[][]{Federle:2024}. In the bottom panels of Figure~\ref{figura:spatial_distribution} we show the distribution smoothed with a Gaussian kernel ({\it blue distribution}) of the BF-GCC on the left, and on the right the same distribution including only objects identified as  ``PSF'' by DESI (see Section~\ref{seccion:distancias_selection}), therefore leaving only objects consistent with a PSF like profile
to have even a higher purity. 
The bottom left and right panels of Fig~\ref{figura:spatial_distribution} presents similar features: with the clustering of GCs toward N-E and a lower density of GCs (caused by worse observing conditions, see Figure~\ref{figura:propiedades_apuntados}) towards the S-W. In the N-E area the darkest smoothed areas, which correspond to the highest density of GCs distribution, share the same spatial distribution as UDGs.
Previous studies in Virgo \citep[e.g.,][]{Powalka:2018} and Fornax \citep[e.g.,][]{DAbrusco:2022} have observed substructures in the GC population around large galaxies. The substructures are
expected in galaxy formation scenarios that involve accretion or merger events. 
Here we noted that the GCs might be clustered along substructures, which might trace back to the cluster build up, in both panels, with the right panel only presenting a lower number of objects.

\begin{figure*}
\centering
    \includegraphics[width=0.5\columnwidth]{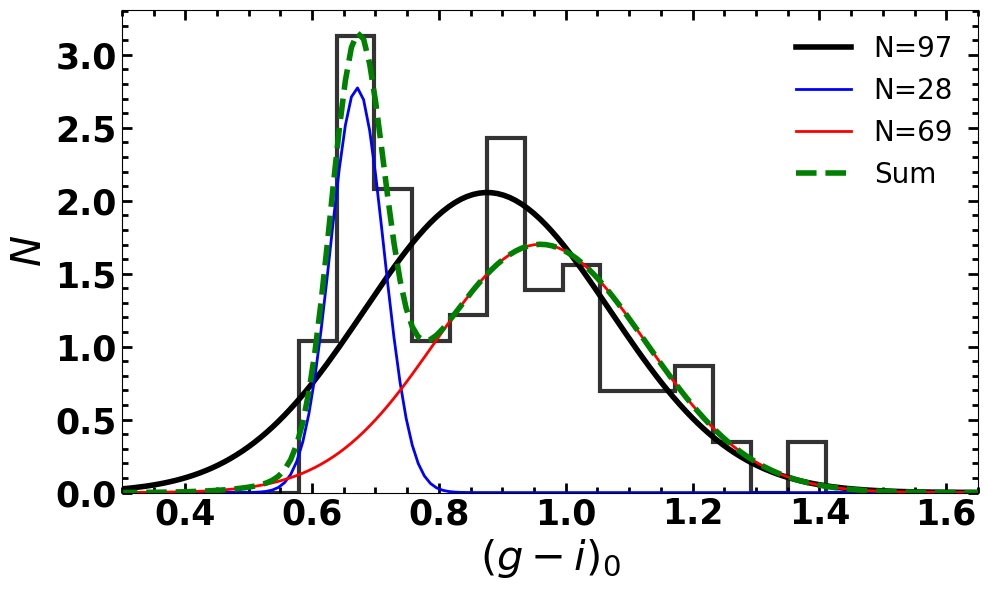}
    \includegraphics[width=0.5\columnwidth]{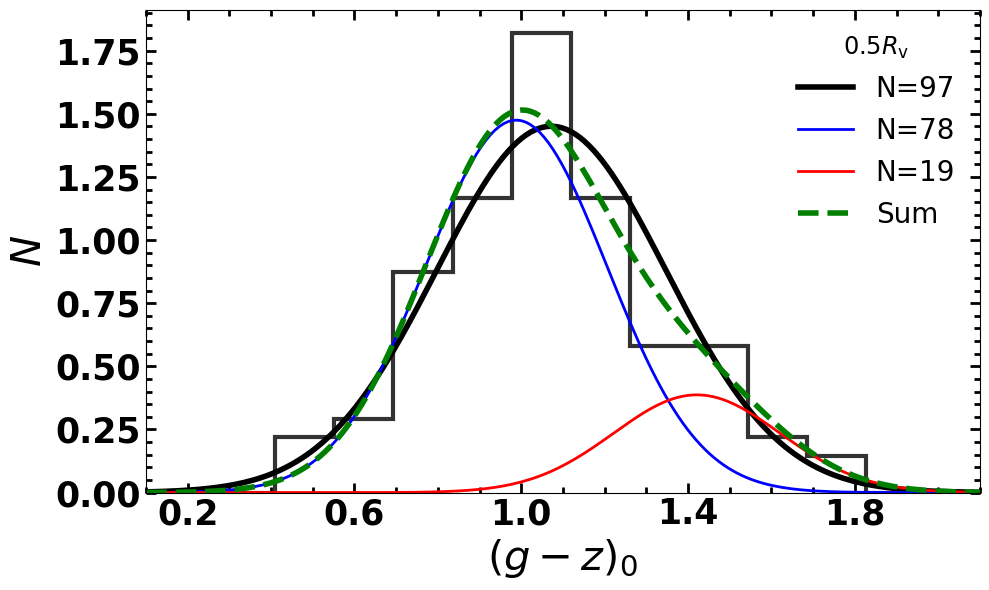}
    \includegraphics[width=0.5\columnwidth]{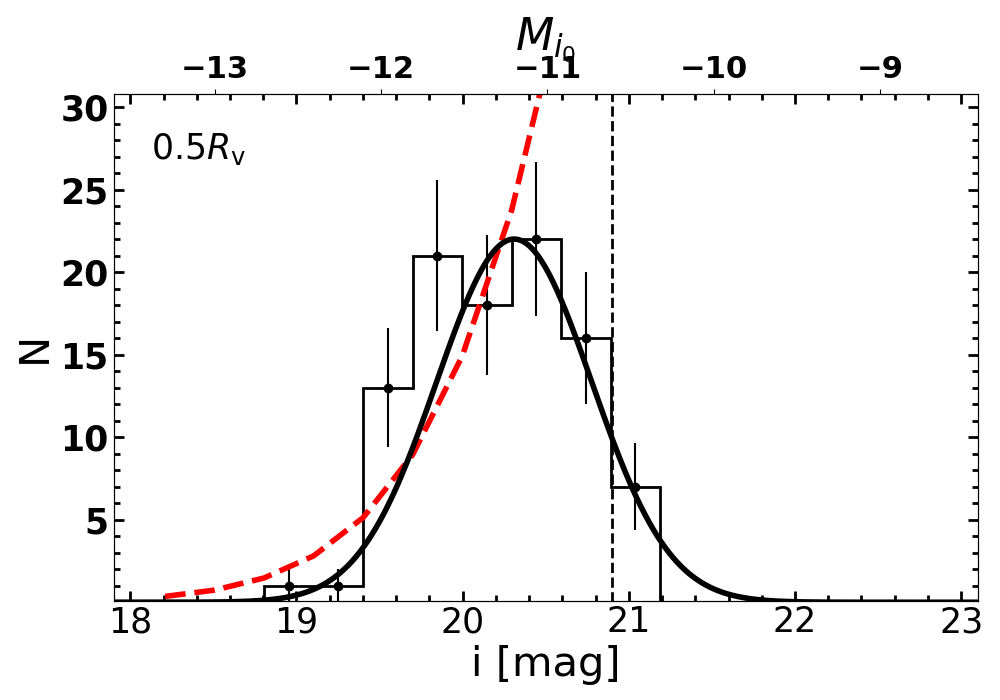}  
    
   \includegraphics[width=0.5\columnwidth]{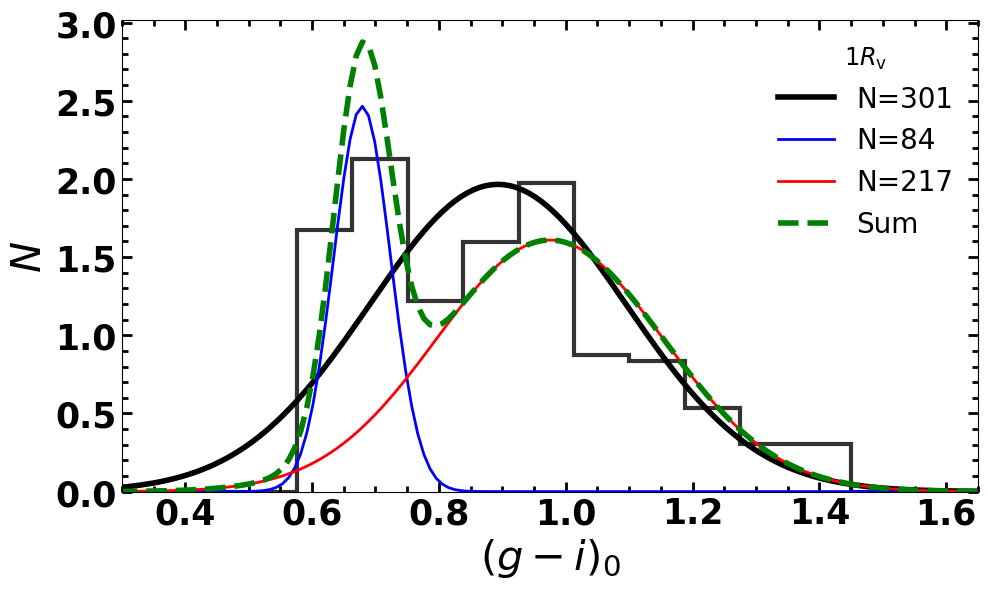}
    \includegraphics[width=0.5\columnwidth]{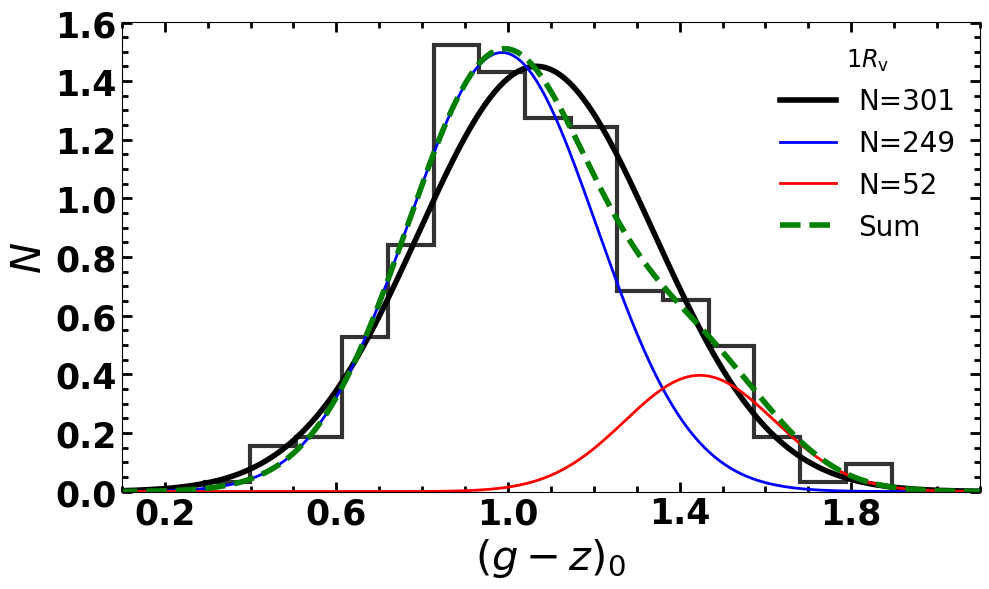}
    \includegraphics[width=0.5\columnwidth]{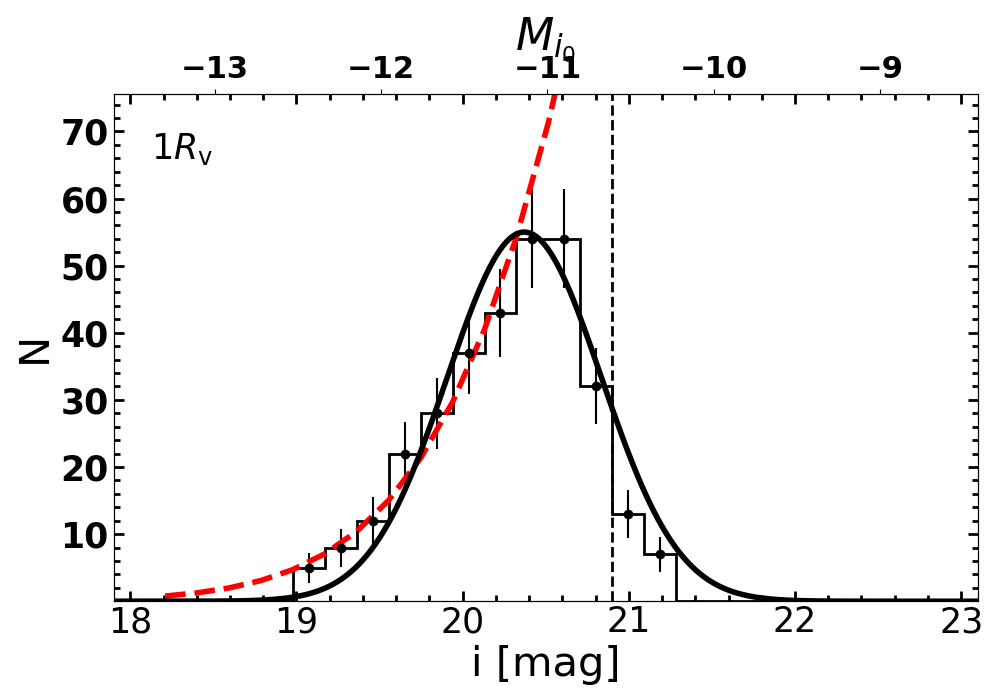} 

    \includegraphics[width=0.5\columnwidth]{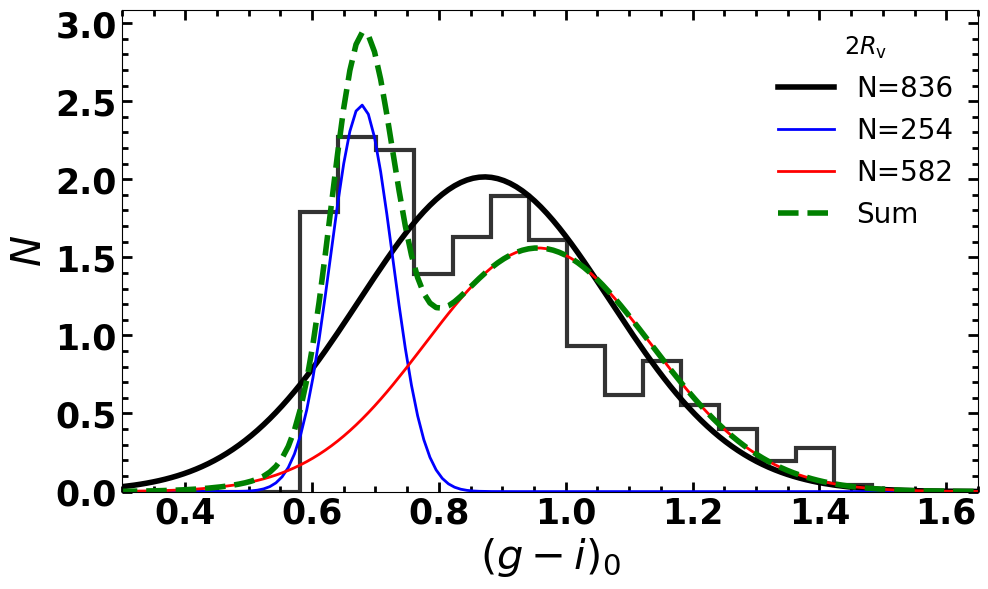}
    \includegraphics[width=0.5\columnwidth]{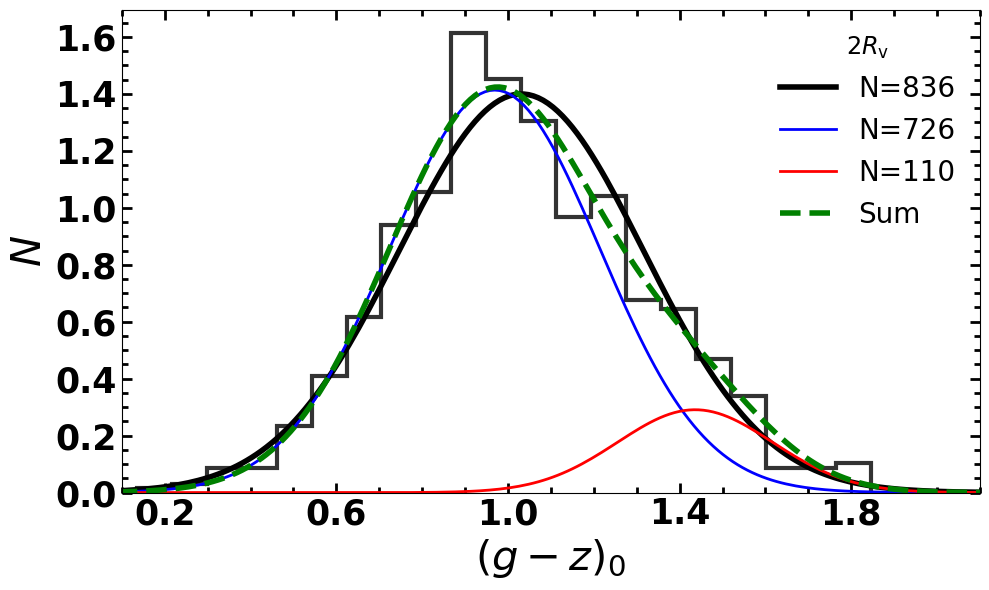}
    \includegraphics[width=0.5\columnwidth]{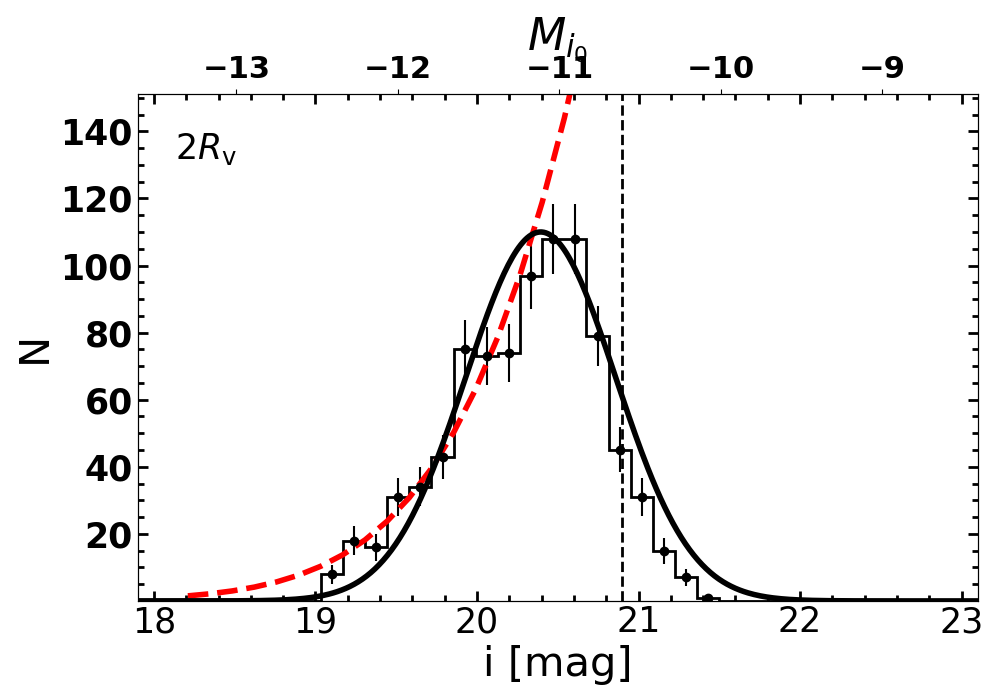} 
    
    \includegraphics[width=0.5\columnwidth]{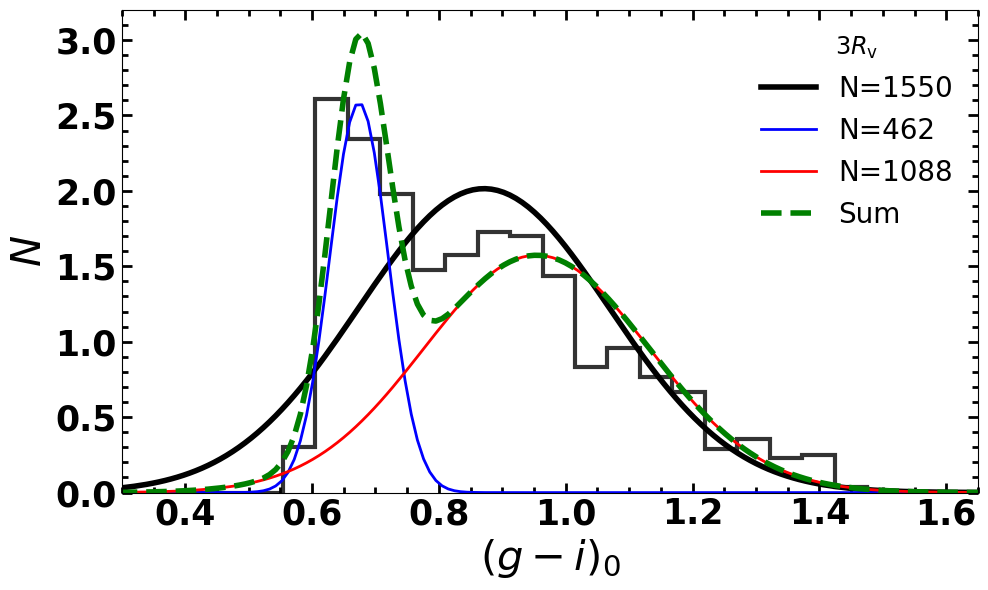}
    \includegraphics[width=0.5\columnwidth]{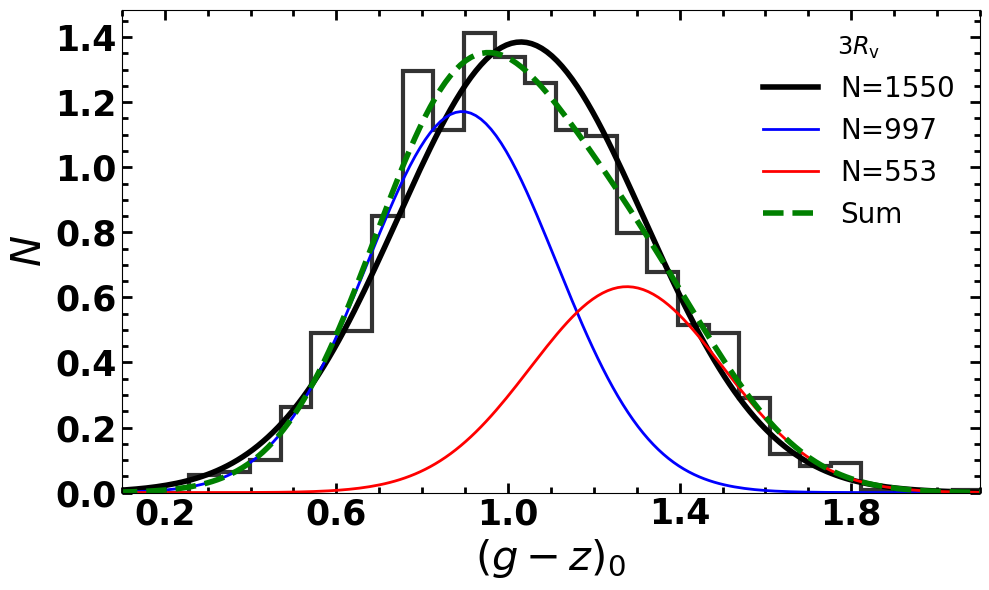}
    \includegraphics[width=0.5\columnwidth]{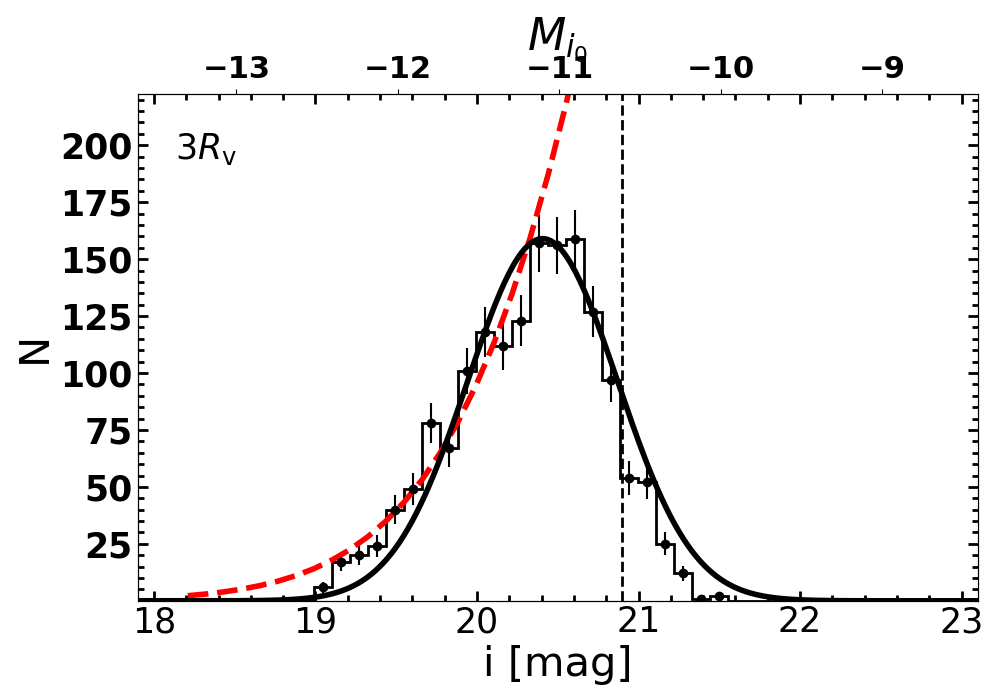}  
    
    \caption{Colors and GCLF at different \rv\, 0.5, 1, 2, and 3 (from top, to bottom panels). 
    First and second columns, are the $(g-i)_{0}$ and $(g-z)_{0}$ colors; color code is the same as in Figure~\ref{figura:color_ancho_narrow}. 
    Third column shows the i-GCLF; color code is the same as in Figure~\ref{figura:gclf}. 
    } 
    \label{figura:color_ancho_radios} 
\end{figure*}

As illustrative images in Figure~\ref{figura:gcs_galaxies} we show snaps of 20 galaxies with different Hubble morphological types (E, S, S0, and dwarfs). Each stamp shows the detected GC candidates (red circles) and spec-GCs recovered in S-PLUS i-band ({\it yellow circles}) close to each galaxy. The green ellipses plotted are the semi-axes \aimage\ and \bimage\ multiplied by the {\sc kron\_radius} obtained in our photometry analysis. Each image has a dimension of 390$\times$318~arcsec. In the literature there are $\sim$1000 spectroscopically confirmed galaxies belonging to the Fornax, most of which are also located in the inner parts of the cluster. A large portion of GC candidates is located in the outer parts of Fornax, where there are no classified galaxies yet. 
Thus with our GC candidates sample it is possible to detect galaxies with an indirect method, increasing the number of galaxies classified in Fornax and also study disrupted field GCs which do not have a host galaxy. In the next subsections
we discuss more about the GC candidates spatial distribution.

\subsubsection{Colors and GCLF at different \rv}
\label{seccion:color_spacial}

In various studies it has been found that the spatial distribution of GCs is bimodal \citep[e.g.,][]{Zepf:1993}, in which the most metal-poor clusters are distributed in the outer parts of their parent galaxies, while the metal-rich ones present a more homogeneous distribution with a concentration peak towards the inner parts of the host galaxies \citep[e.g.,][]{Hargis:2014, Kartha:2014}. 
In this scenario a color gradient is likely, with the redder
GCs concentrated towards the center and the bluer GCs populating the outermost parts of the system \citep{Hargis:2014}. 
With S-PLUS data it is possible to expand this gradient to a large number of colors at radii of up to 5~\rv\ in \ra.

In the first and second columns of Figure~\ref{figura:color_ancho_radios}, we show the $(g-i)_{0}$ and $(g-z)_{0}$ color distributions at differents \rv\  (0.5, 1, 2, and 3). Near NGC 1399 (i.e. the Fornax center), it can be seen a possible trace of bimodality in the color $(g-i)_{0}$. As we move away from the cluster center, this bimodality fades and becomes a long tail of red clusters. 
On the contrary, in $(g-z)_{0}$ color, a bimodality distribution is preserved at large distances from the center, similarly to the observations in Virgo cluster \citep[e.g.][]{Durrell:2014, Zhang:2015}. 
In both colors, it is observed that the GC candidates with a bluer color predominate with respect to the GC candidates with redder colors. A larger population of blue GC candidates may be due to the existence of dwarf galaxies (not yet catalogued), that host bluer GCs populations due to their low mass (mass-metallicity ratio, \citealt{Cote:1998}).
According to the fitting results from GMM (Table~\ref{tabla:colores}) in each subsample, we concluded that the bimodality in both colors is preserved as we move away from the center.

In the third column of Figure~\ref{figura:color_ancho_radios}, we show the GCLF distribution in i-band at different \rv\ (0.5, 1, 2, and 3). We obtained the peak magnitude from the fit of the GCLF made with GMM, using a single Gaussian {\it (black solid line)}. The GCLF corrected for incompleteness is the {\it red dashed line}. 
It is noted that the mean value found with GMM of each distribution observed increases slightly at differents \rv\ 
(20.21$\pm$0.05, 20.27$\pm$0.03, 20.29$\pm$0.02 and 20.31$\pm$0.01), which implies a mass decrease of $\sim$10\%  between the central and the outermost peak distribution assuming the proportionality between mass to light. The increase in the peak mass of the GCLF should be related to the concentration of massive galaxies, towards the center \citep{Jordan:2007}; 
the mean GCs mass is greater in massive galaxies compared to less massive ones \citep[e.g.,][]{Harris:2013, Lomeli:2022cfht}, while the decrease in mass is related to the concentration of dwarf galaxies towards the outer parts of the cluster. However, this result must be taken with care, because we are not looking at individual galaxies, and the differences in the estimated mass are within the errors (2$\sigma$). 

\subsection{Missing galaxies}
\label{seccion:missing_galaxies}

The number of galaxies spectroscopically confirmed or considered likely members on morphological basis in Fornax, is $\sim$1000 \citep[see][]{Analia:2024}. On the other hand, reviewing the literature \citep[eg.,][]{Jordan:2007fornax, Villegas:2010, DAbrusco:2022, Saifollahi:2024}, and rejecting repeated galaxies, there are 75 galaxies in Fornax in which $\sim$10,000 GCs have been observed. If we assume that the BF-GCC sample extracted in this work, is also associated to galaxies, the total number of GCs associated to the Fornax cluster is $\sim$13000. 
If, in a first order approximation, we assume that all galaxies have the same number of GCs and that the number of GCs in the Fornax cluster should be similar for the rest of the 925 galaxies, the total number of GCs would be $\sim$173,000 ($N\_{TOT}=N_{GAL,SPEC}/N_{GAL,OBS}\times N_{GC,OBS}$), of the same order of magnitude of GCs associated to Abell~1689 \citep{Karla:2013} cluster. 
Yet, the number of GCs observed in a galaxy depends on the galaxy mass and morphology, with irregular and dwarf galaxies having values from zero to tens \citep[e.g.,][]{Prole:2019, Annibali:2018, Karim:2024}, while galaxies similar in mass to MW have around tens to hundreds
 \citep[eg.,][$\sim$160]{Harris:1996, Harris:2010} GCs, and for massive early type galaxies is of the order of hundreds to thousands
\citep[eg.,][]{Harris:2014}.  According with
\citet{Villegas:2010} the galaxy (FCC\_335) has the minimum number (14 or 7, taking into account contaminants) of GCs, and the results obtained from simulations usually assign a number of $\sim$10 GCs to low or intermediate mass galaxies \citep[e.g.,][]{Reina:2022}.
Considering that the Fornax cluster has a large population of dwarf galaxies, 
it is possible to make two statements for the BF-GCC sample:
a) a large percentage of GCs are not really bound to galaxies and belongs to the intracluster-medium, indeed \citet{Saifollahi:2024} found that a percent of the Fornax GCs are associated to the intracluster-medium and 
b) the number of galaxies belonging to Fornax is greatly underestimated outside 1~\rv, as explained in Section~\ref{seccion:distribucion_espacial2} the GC distribution can be used to identify the location of galaxies belonging to the Fornax cluster and trace back the process of clustering, where the passages of newly acquired members within the cluster potential, or galaxy-galaxy interactions,  may leave behind a tale of stripped GCs. 
Further work, using S-PLUS photometric data, (e.g., Haack et al., in prep.) is currently underway with the purpose of identifying new galaxy candidates. In addition, spectroscopic confirmation using data from the Gemini South telescope for a sample of GCs and missing galaxies is currently ongoing (e.g., Lomel\'i-N\'u\~nez et al., in prep.). Finally, the CHANCES/4MOST survey \citep{Sifon:2024}, which will recover the spectroscopic redshift of galaxies in galaxies clusters out to 5 effective radii (including the Fornax cluster), will be a breakthrough in our understanding of cluster formation, and solve the issue of missing galaxies ($m_r<20.5$).

\section{Conclusions}
\label{seccion:conclusions}
We studied the GC system in Fornax over $\sim$200 square degrees, using homogeneous data taken through the 12 optical bands of S-PLUS. We used SExtractor plus PSFEx to perform PSF photometry and developed a method of selection of GCs, using structural, evolutionary and distance (GAIA and SED fitting template) parameters. Detection of simulated clusters was carried out to obtain the incompleteness as a function of magnitude. We used the u-band photometry to evaluate the contamination of our sample of GC candidates by stars, background
galaxies or YSCs reddened. The contaminating fraction was found to be $\sim$20\%. In a following paper, the GC systems associated to individual galaxies will be studied, as well as we are obtaining spectroscopic data for a sub-sample of objects to obtain spectroscopic confirmation of their association to the Fornax cluster.

We used our data set to construct 10 and 15 colors in broad and narrow bands respectively.  We performed statistical tests to evaluate the bimodality of colors, finding that globally, according to GMM results and visual inspection, there is evidence of color bimodality in 2 colors, namely, $(g-i)_{0}$ and $(g-z)_{0}$ 
in the broad bands. On the contrary, in the narrow bands we did not find strong statistical evidence to confirm bimodality in any color. 
A possible explanation for not finding bimodality in the narrow band colors may be related to the nature itself of the narrow band filters, since they only sample a small spectral range. At the same time, the larger error on the magnitude estimations in the narrow bands might extend the color distribution, fading the bimodality.
Also, we studied the $(g-i)_{0}$ and $(g-z)_{0}$ color distributions at differents virial radius (0.5, 1, 2, and 3). We found that near the Fornax center there is a clear trace of bimodality in the color $(g-i)_{0}$. As we move away from the center of the cluster, the bimodality fades and becomes a long tail of red clusters. Dissimilarly,  in $(g-z)_{0}$ color, a bimodality distribution is preserved at large distances from the center.

We construct the GCLF in the 12-bands highlighting two points: 
a) in all bands the log-normal distribution typical for GC systems can be estimated and it is found to increase smoothly up to reaching a peak value, and then again to decrease smoothly. However, 
b) with the S-PLUS i-band 50\% completeness magnitude of 21.44, we are unable to reach the TO generally observed in early-type galaxies. 
Thus, we are only sampling the bright end of the GCLF.
Also, we studied the GCLF at differents virial radius (0.5, 1, 2, and 3), it is noted that the peak of each distribution observed increases slightly at differents \rv\, which implies a mass increment assuming the proportionality between mass to light, maybe resulting from the infall of group and filaments into the cluster.

\section{Acknowledgments}
LLN thanks Funda\c{c}\~ao de Amparo \`a Pesquisa do Estado do Rio de Janeiro (FAPERJ) for granting the postdoctoral research fellowship E-40/2021(280692). AC acknowledge Funda\c{c}\~ao de Amparo \`a Pesquisa do Estado do Rio de Janeiro (FAPERJ) for granting the research fellowship 'Apoio a jovem pesquisador fluminense' E-40/2021(270993) and the CNPQ fellowship 'Produtividade em Pesquisa do CNPq - Nível 2'. AVSC acknowledge financial support from CONICET (PIP 1504), Agencia I+D+i (PICT 2019-03299) and Universidad Nacional de La Plata (Argentina).
SW is supported by the United Kingdom Research and Innovation (UKRI) Future Leaders Fellowship `Using Cosmic Beasts to uncover the Nature of Dark Matter' (grant number MR/X006069/1). PKH gratefully acknowledges the Fundação de Amparo à Pesquisa do Estado de São Paulo (FAPESP) for the support grant 2023/14272-4. AAC acknowledges financial support from the Severo Ochoa grant CEX2021-001131-S funded by MCIN/AEI/10.13039/501100011033. CL-D acknowledges financial support from the ESO Comite Mixto 2022.
Swayamtrupta Panda acknowledge CNPq Fellow, Gemini Science Fellow. TSG would like to thank CNPq (PQ grant 314449/2023-0) and FAPERJ (JCNE grant 201.309/2021). MG acknowledges support from FAPERJ grant E-26/211.370/2021. KMD thanks the support of the Serrapilheira Institute (grant Serra-1709-17357) as well as that of the Brazilian National Research Council (CNPq grant 308584/2022-8) and of the Rio de Janeiro Research Foundation (FAPERJ grant E-32/200.952/2022), Brazil. VF thanks Fundação Coordenação de Aperfeiçoamento de Pessoal de N\'ivel Superior (CAPES) for granting the postdoctoral research fellowship Programa de Desenvolvimento da Pós-Graduação (PDPG)-P\'os Doutorado Estrat\'egico, Edital nº16/2022 and the Finance Code  001 (88887.838401/2023-00) funding. 

The S-PLUS project, including the T80S robotic telescope and the S-PLUS scientific survey, was founded as a partnership between the Sao Paulo Research Foundation (FAPESP), the Observatório Nacional (ON), the Federal University of Sergipe (UFS), and the Federal University of Santa Catarina (UFSC), with important financial and practical contributions from other collaborating institutes in Brazil, Chile (Universidad de La Serena), and Spain (Centro de Estudios de Fisica del Cosmos de Aragon, CEFCA). 

This research has made use of the NASA/IPAC Extragalactic Database (NED) which is operated by the California Institute of
Technology, under contract with the National Aeronautics and Space Administration.

This work has made use of data from the European Space Agency (ESA) mission {\it Gaia} (\url{https://www.cosmos.esa.int/gaia}), processed by the {\it Gaia} Data Processing and Analysis Consortium (DPAC, \url{https://www.cosmos.esa.int/web/gaia/dpac/consortium}). Funding for the DPAC has been provided by national institutions, in particular the institutions participating in the {\it Gaia} Multilateral Agreement.

The DESI Legacy Imaging Surveys consist of three individual and complementary projects: the Dark Energy Camera Legacy Survey (DECaLS), the Beijing-Arizona Sky Survey (BASS), and the Mayall z-band Legacy Survey (MzLS). DECaLS, BASS and MzLS together include data obtained, respectively, at the Blanco telescope, Cerro Tololo Inter-American Observatory, NSF’s NOIRLab; the Bok telescope, Steward Observatory, University of Arizona; and the Mayall telescope, Kitt Peak National Observatory, NOIRLab. NOIRLab is operated by the Association of Universities for Research in Astronomy (AURA) under a cooperative agreement with the National Science Foundation. Pipeline processing and analyses of the data were supported by NOIRLab and the Lawrence Berkeley National Laboratory (LBNL). Legacy Surveys also uses data products from the Near-Earth Object Wide-field Infrared Survey Explorer (NEOWISE), a project of the Jet Propulsion Laboratory/California Institute of Technology, funded by the National Aeronautics and Space Administration. Legacy Surveys was supported by: the Director, Office of Science, Office of High Energy Physics of the U.S. Department of Energy; the National Energy Research Scientific Computing Center, a DOE Office of Science User Facility; the U.S. National Science Foundation, Division of Astronomical Sciences; the National Astronomical Observatories of China, the Chinese Academy of Sciences and the Chinese National Natural Science Foundation. LBNL is managed by the Regents of the University of California under contract to the U.S. Department of Energy. The complete acknowledgments can be found at \url{https://www.legacysurvey.org/acknowledgment/.}

%

\vspace{5mm}
\facilities{S-PLUS}


\software{Source-Extractor \citep{Sextractor:1996},    PSFEx \citep{Psfex:2011}, GALAXEV \citep{Bruzual:2003},  GMM \citep{Muratov:2010}, IRAF \citep{Iraf:1986}. } 

\section*{Data availability}
The paper is based on publicly available archival data. The complete version of Table~\ref{tabla:datos} is included as a Machine Readable Table. All the fits files are storage in the S-PLUS cloud (\url{https://splus.cloud}).



\appendix

\section{Image features}
\label{apendice_A}

\begin{figure*} 
\centering
    \includegraphics[width=0.24\columnwidth]{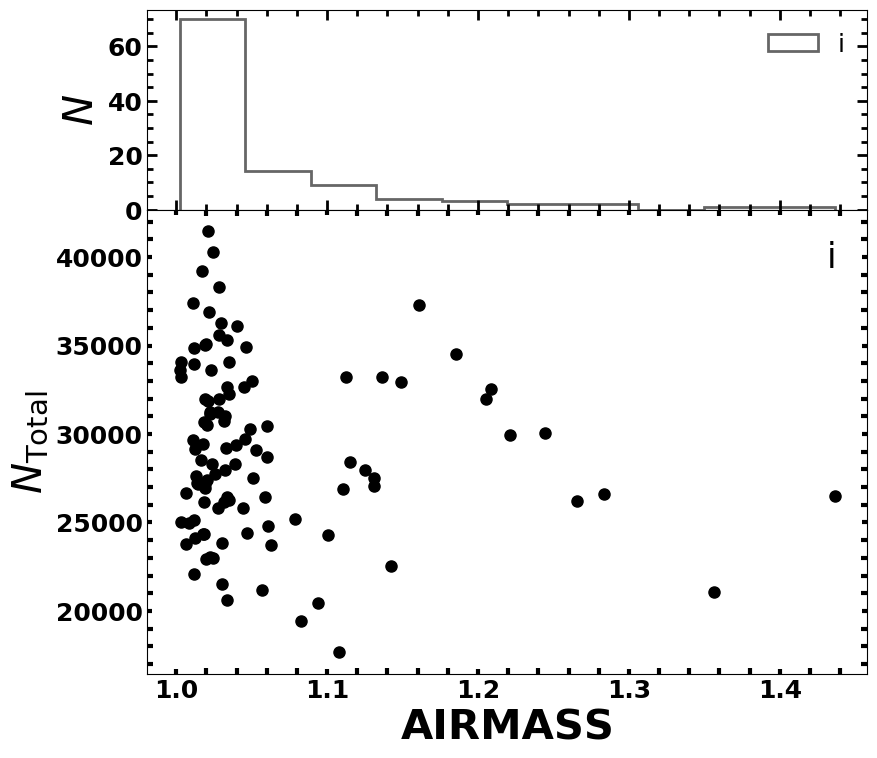}
    \includegraphics[width=0.24\columnwidth]{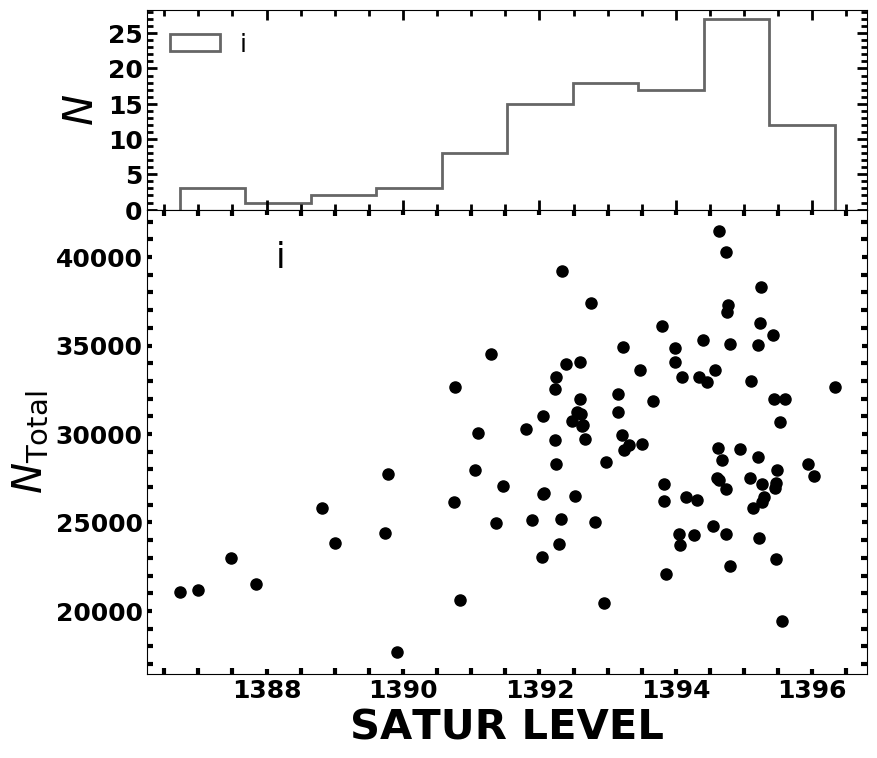}
    \includegraphics[width=0.24\columnwidth]{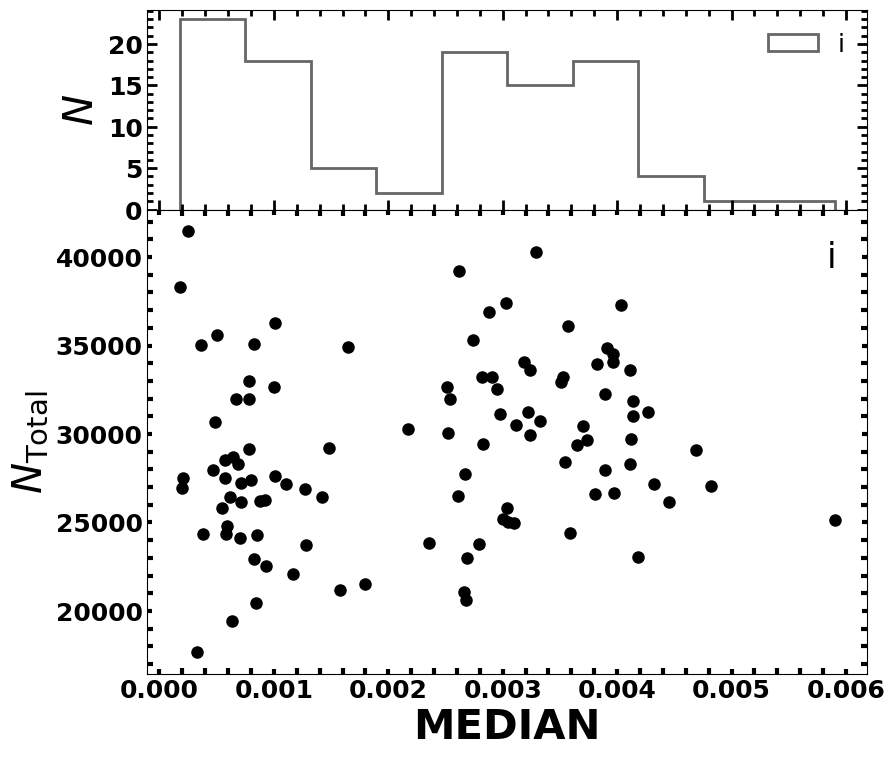}
    \includegraphics[width=0.24\columnwidth]{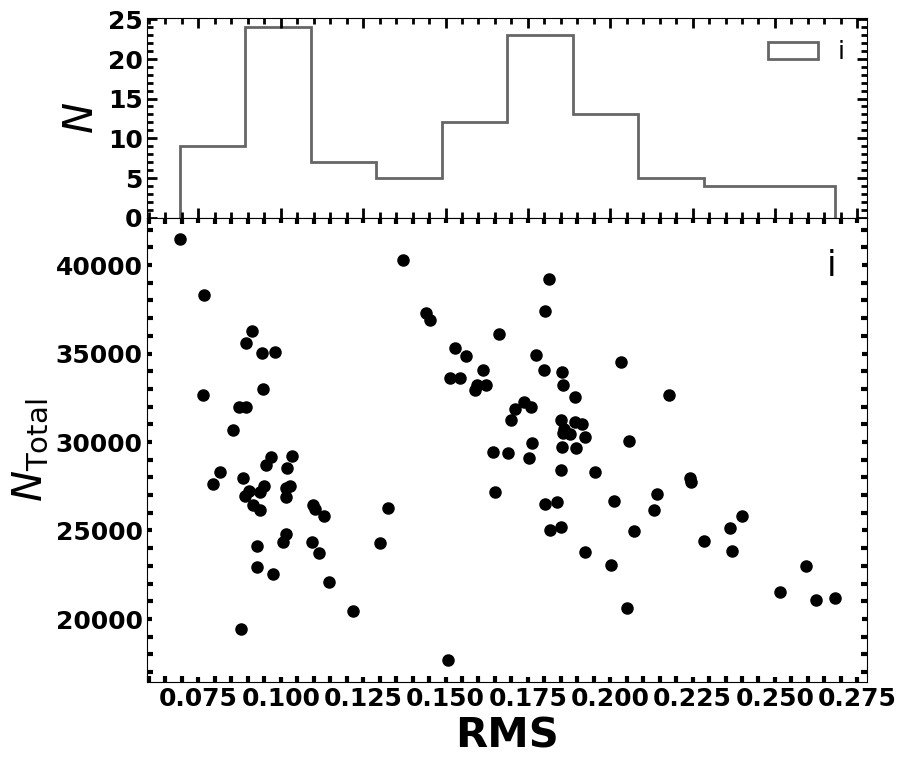}    
    
\vspace{0.3cm}    

    \includegraphics[width=0.24\columnwidth]{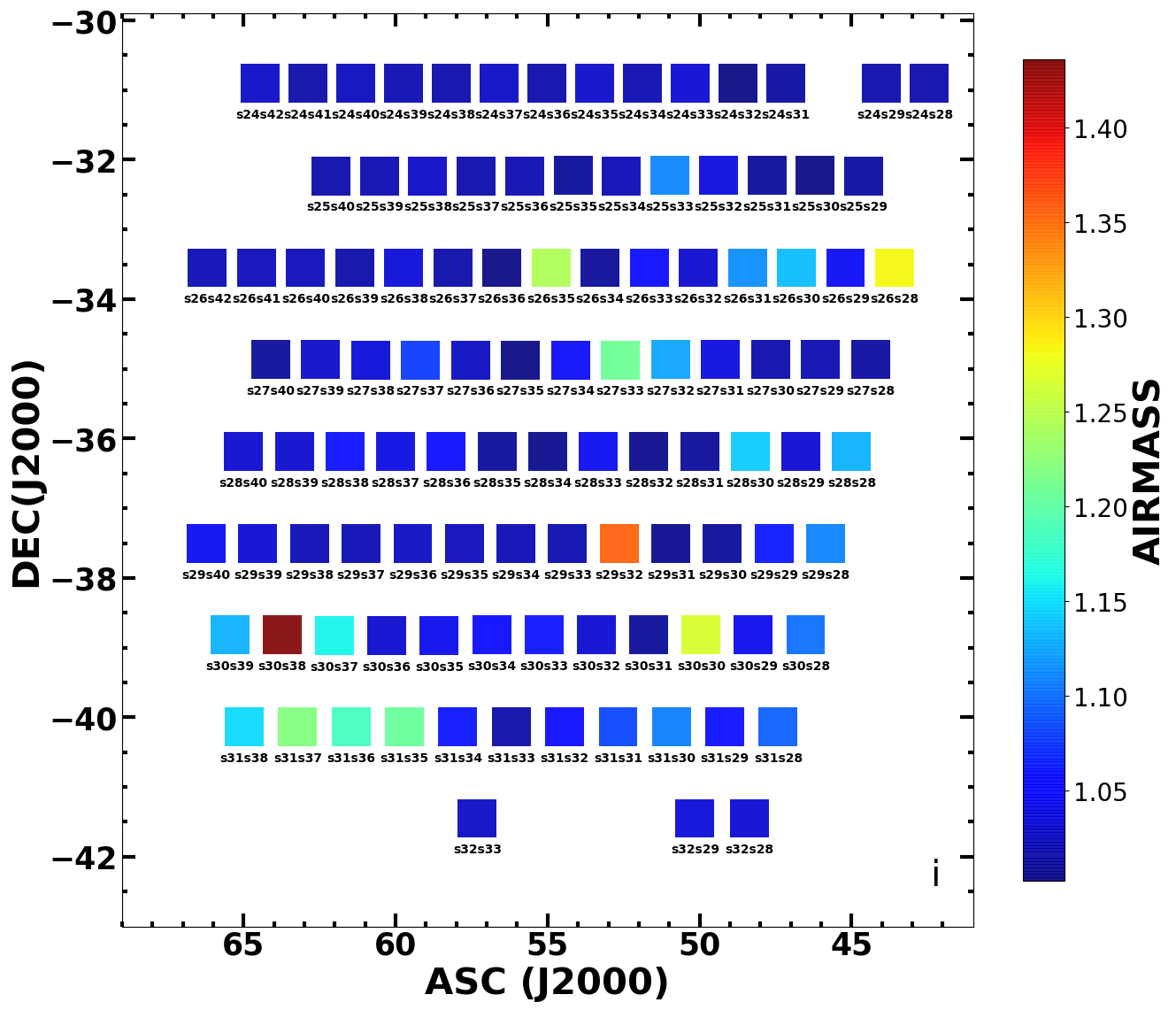}
    \includegraphics[width=0.24\columnwidth]{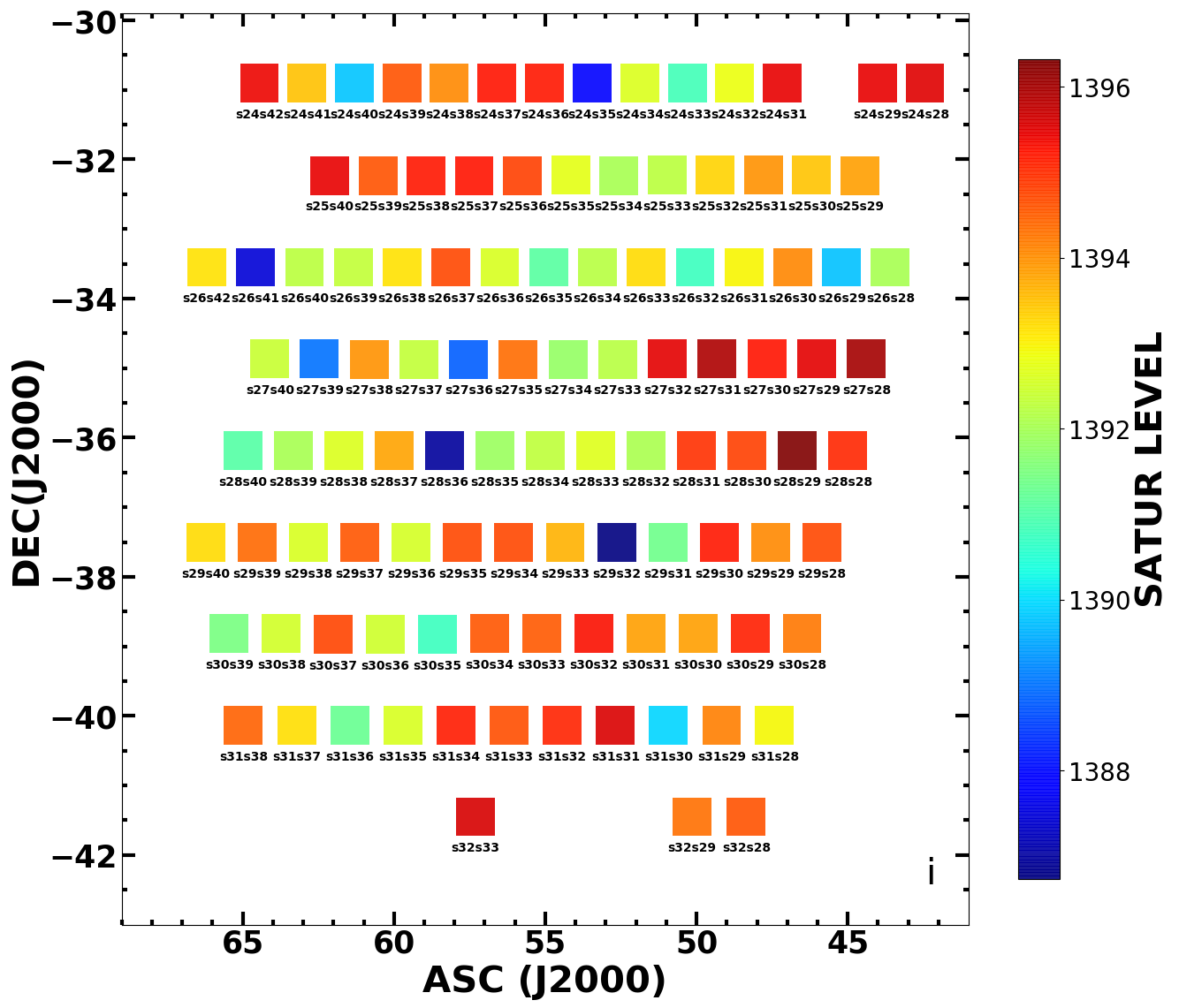}
    \includegraphics[width=0.24\columnwidth]{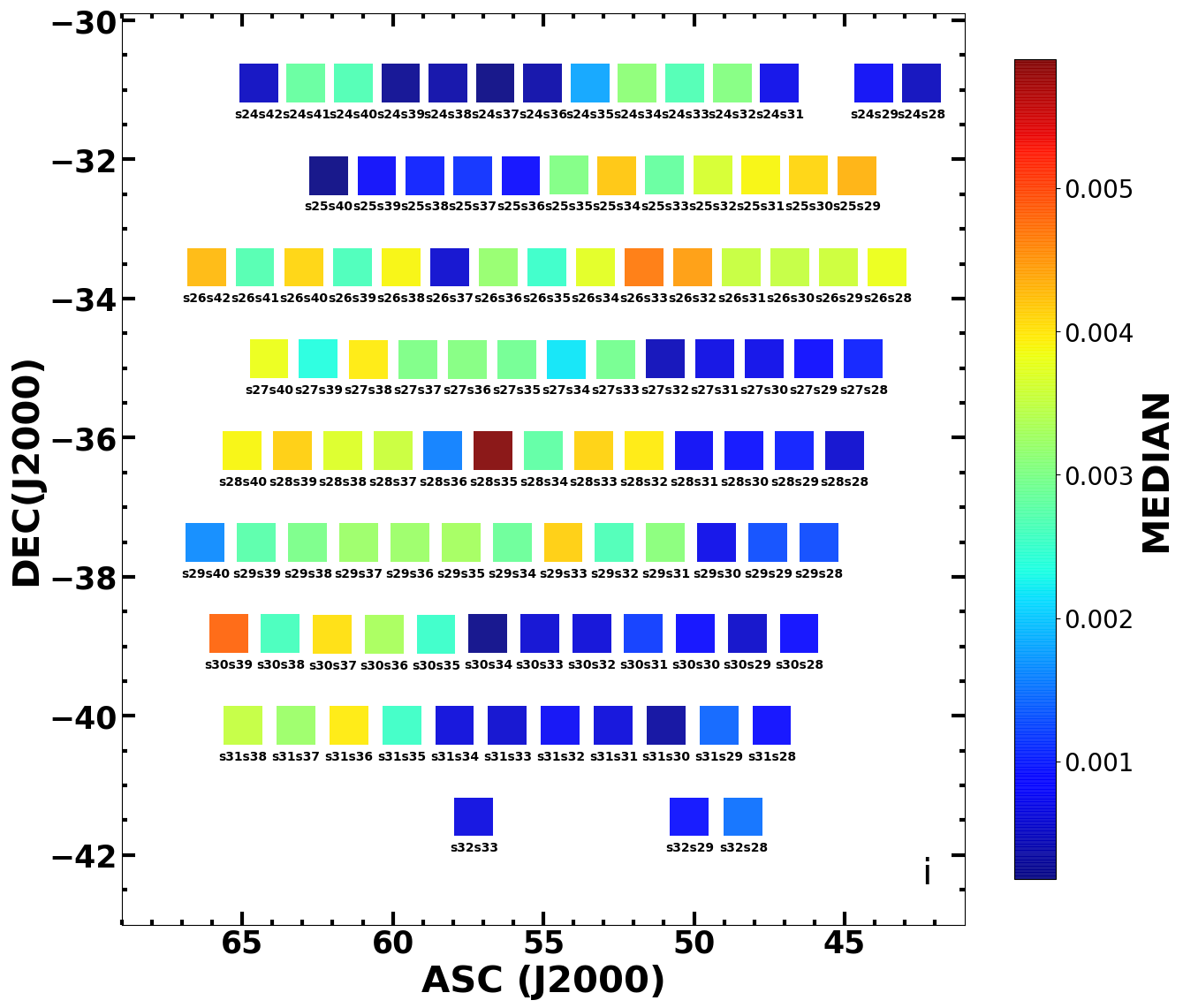}
    \includegraphics[width=0.24\columnwidth]{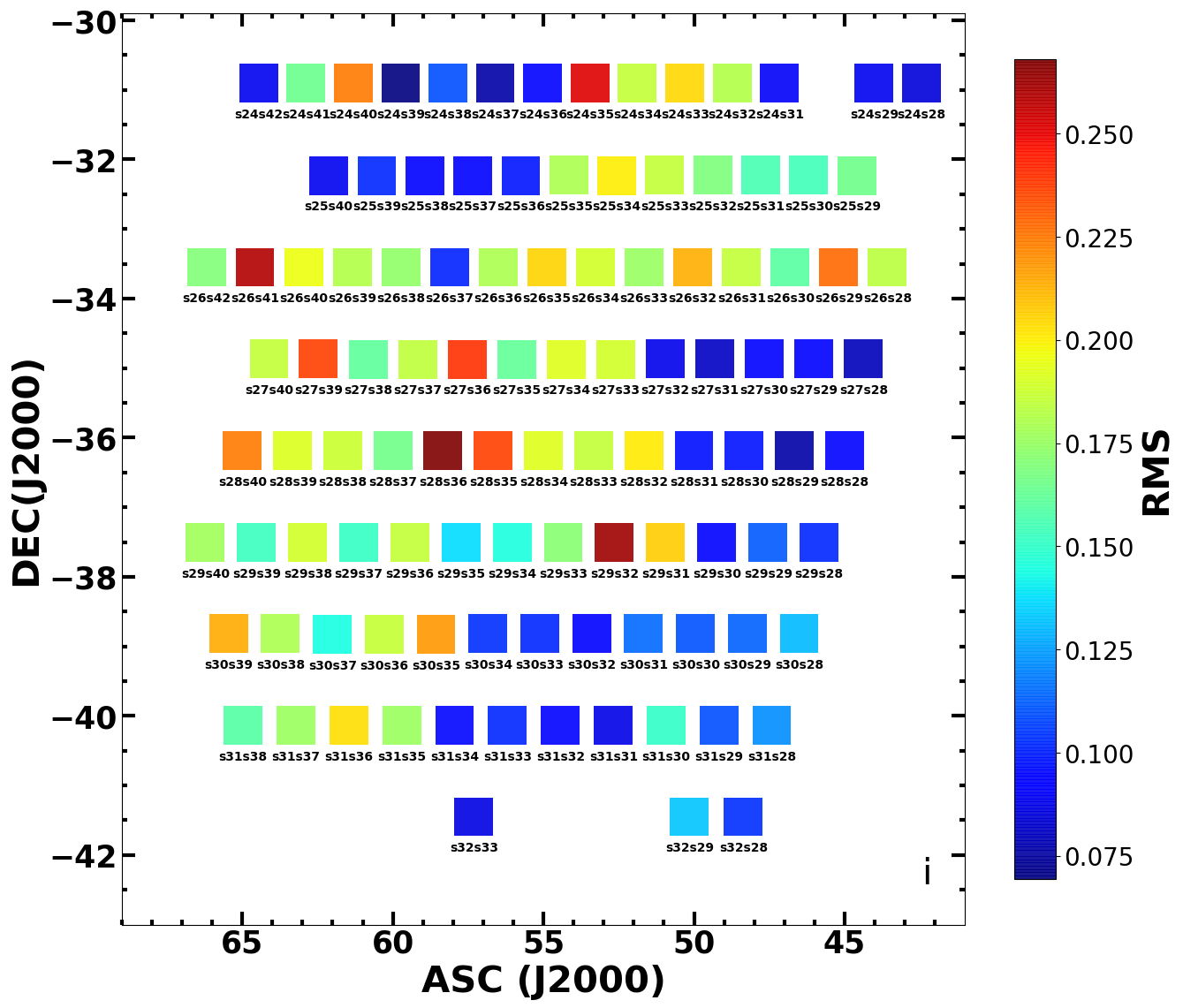}    

\vspace{0.5cm}
    \begin{flushleft} 
    \includegraphics[width=0.25\columnwidth]{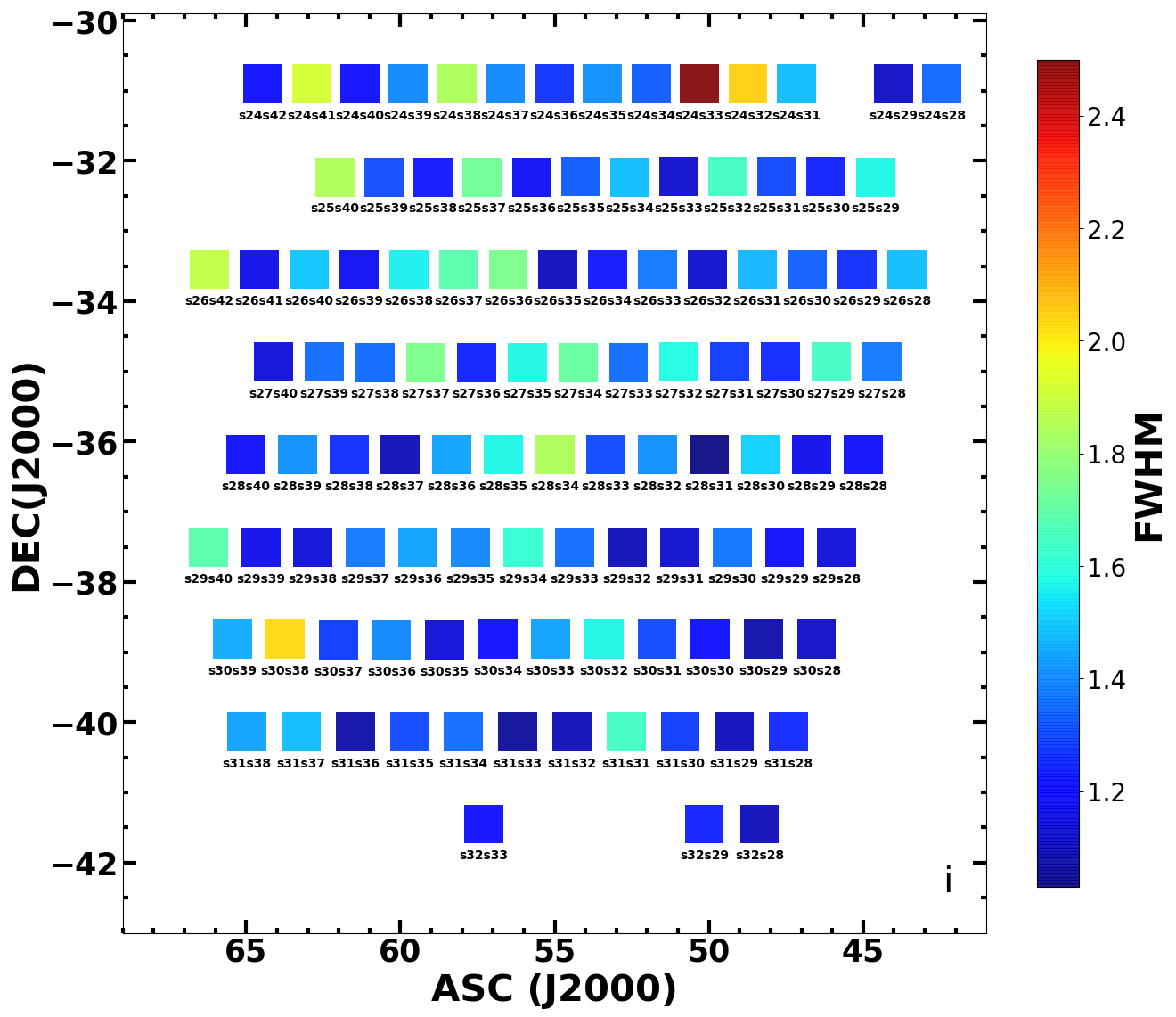}      
    \includegraphics[width=0.25\columnwidth]{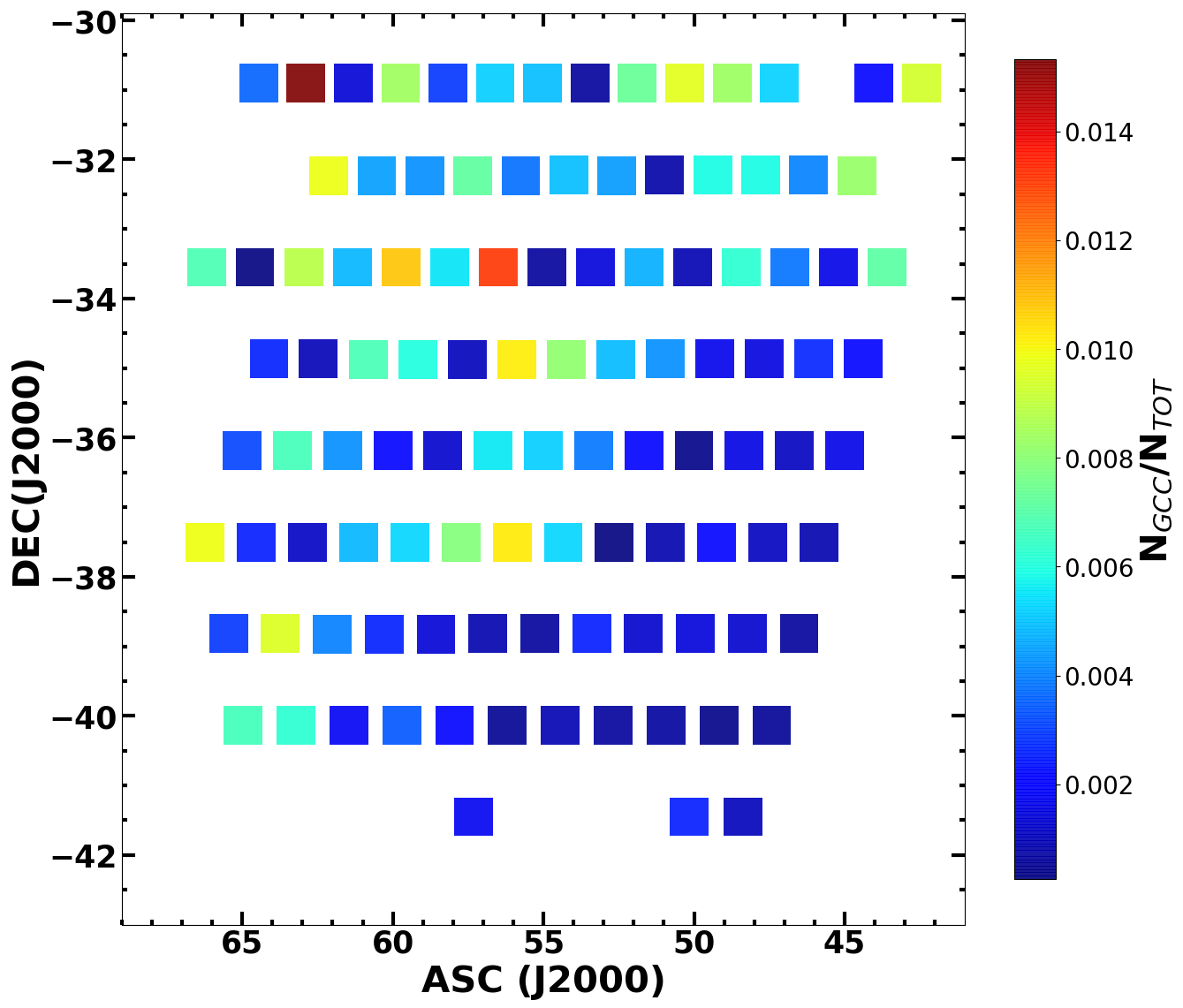}
    \end{flushleft} 

    \caption{Observational properties in the 106-FoVs in the i-band. In the first row panels we show the parameters: \airmas, \satur, \median\ and \rms\ versus the total number of sources detected in each FoV. In the second row panels we show all the FoVs in the (\ra,\dec) space with a color coded for each parameter: \airmas, \satur, \median\, \rms\, and in the third row panels the FWHM and ratio of the GCC divided by the total sources detected in each field.
    }    
    \label{figura:propiedades_apuntados} 
\end{figure*}

All astronomical observations are subject to weather. Depending on the conditions imposed by these 2 major factors is the image quality and the final detection and measurements that we can achieve. In Figure~\ref{figura:propiedades_apuntados}, we show four examples of the parameters that predetermine the final results, named: \airmas, \satur, and the \median\ and \rms\ of the background. In the left panels of Figure~\ref{figura:propiedades_apuntados} the parameters are plotted versus the total number ($N_{Total}$) of sources detected in each FoV. In the right panels of Figure~\ref{figura:propiedades_apuntados} the parameters are the color coded in the (\ra,\dec) space. In the case of a higher value of \airmas\ reported a lower number objects is recovered, while for the \satur\ the trend is inverse, a higher value of saturation, a greater number objects is recovered. In the cases of \median\ and \rms\, a bimodal distribution appears, where the trend is not clear. However in the bottom panel, it is observed that panels with the lower values of \median\ and \rms\ are in the region south-west. It is in these FoVs where our selection of GC candidates is smaller in proportion to the rest of the pointings. While in the saturation color space, it is observed that in the same region (south-west) the values are highest in comparison with the rest of FoVs. After this brief analysis it is possible to conclude that the parameters that determine the detection and selection of GC candidates have a non-linear relationship between them.

\section{Statistical colors results}
\label{apendice_B}

We used Gaussian Mixture Modelling (GMM) code \citep{Muratov:2010} to confirm the presence of bimodality in the color distributions
according to the analysis presented in Section~\ref{seccion:colores}. In Tables~\ref{tabla:colores} and~\ref{tabla:colores} we presents the GMM statistics results for Figure~\ref{figura:color_ancho_narrow} and left and central panels of Figure~\ref{figura:color_ancho_radios} respectively.

\begin{table*}
    \setlength\tabcolsep{3.pt}
	\centering
    \begin{center}
	\caption{GMM fitting values for colors distribution.}
	\label{tabla:colores}
    \begin{scriptsize}
	\begin{tabular}{l c c c c c c c c c r c c c c c c} 
\hline
\multicolumn{4}{c}{Unimodal results} & \multicolumn{7}{c} {Bimodal results} \\
Color  &     Peak&   $\sigma$&   Peak1&   Peak2 &  $\sigma1$ & $\sigma2$&   \ngc & $f_{2}$ &    D &  Kurtosis & \multicolumn{3}{c}{$p$-values}  &   Bi  \\ 
(1)&        (2)  &   (3)  &   (4)   &      (5) &    (6)&     (7)&   (8) & (9) & (10) &   (11)  &   \multicolumn{3}{c}{(12)}  &   (13) \\
\hline
\multicolumn{13}{c}{Width bands} \\
\hline
$(u-g)_{0}$  & 0.74$\pm$0.01  & 0.30$\pm$0.01 & 0.59$\pm$0.03  & 0.9$\pm$0.03 & 0.2$\pm$0.01 & 0.3$\pm$0.01 & 2582 & 0.47 & 1.21$\pm$0.08  & 0.47 & 0.01 & 0.6  & 1.00  & N  \\  
$(u-r)_{0}$  & 1.32$\pm$0.01  & 0.37$\pm$0.01 & 1.26$\pm$0.02  & 1.46$\pm$0.24 & 0.32$\pm$0.02 & 0.43$\pm$0.09 & 2582 & 0.31 & 0.53$\pm$0.82  & 0.38 & 0.01 & 0.81  & 1.00  & N  \\  
$(u-i)_{0}$  & 1.60$\pm$0.01  & 0.38$\pm$0.01 & 1.47$\pm$0.04  & 1.88$\pm$0.14 & 0.29$\pm$0.02 & 0.39$\pm$0.05 & 2582 & 0.32 & 1.19$\pm$0.48  & 0.30 & 0.01 & 0.82  & 0.99  & N  \\  
$(u-z)_{0}$  & 1.76$\pm$0.01  & 0.41$\pm$0.01 & 1.62$\pm$0.04  & 2.04$\pm$0.12 & 0.32$\pm$0.02 & 0.42$\pm$0.03 & 2582 & 0.33 & 1.14$\pm$0.35  & 0.20 & 0.01 & 0.24  & 0.99  & N  \\  
$(g-r)_{0}$  & 0.59$\pm$0.01  & 0.19$\pm$0.01 & 0.52$\pm$0.24  & 0.6$\pm$0.07 & 0.09$\pm$0.06 & 0.20$\pm$0.05 & 2582 & 0.87 & 0.48$\pm$1.58  & 0.29 & 0.01 & 0.64  & 0.99  & N  \\  
$(g-i)_{0}$  & 0.87$\pm$0.01  & 0.2$\pm$0.01 & 0.68$\pm$0.01  & 0.95$\pm$0.01 & 0.05$\pm$0.01 & 0.18$\pm$0.01 & 2582 & 0.69 & 2.10$\pm$0.04  & -0.31 & 0.01 & 0.23  & 0.01  & Y  \\  
$(g-z)_{0}$  & 1.02$\pm$0.01  & 0.29$\pm$0.01 & 0.92$\pm$0.03  & 1.35$\pm$0.08 & 0.22$\pm$0.01 & 0.20$\pm$0.03 & 2582 & 0.24 & 2.02$\pm$0.28  & -0.28 & 0.01 & 0.09  & 0.01  & Y  \\  
$(r-i)_{0}$  & 0.28$\pm$0.01  & 0.15$\pm$0.01 & 0.27$\pm$0.0  & 0.37$\pm$0.1 & 0.13$\pm$0.01 & 0.24$\pm$0.03 & 2582 & 0.11 & 0.51$\pm$0.8  & 1.47 & 0.01 & 0.63  & 1.00  & N  \\  
$(r-z)_{0}$  & 0.43$\pm$0.01  & 0.23$\pm$0.01 & 0.42$\pm$0.01  & 0.88$\pm$0.2 & 0.21$\pm$0.01 & 0.35$\pm$0.07 & 2582 & 0.02 & 1.60$\pm$1.08  & 1.19 & 0.01 & 0.11  & 1.00  & N  \\  
$(i-z)_{0}$  & 0.15$\pm$0.01  & 0.18$\pm$0.01 & 0.14$\pm$0.01  & 0.17$\pm$0.05 & 0.15$\pm$0.02 & 0.23$\pm$0.02 & 2582 & 0.37 & 0.17$\pm$0.4  & 0.62 & 0.01 & 0.82  & 1.00  & N  \\  

\hline
\multicolumn{13}{c}{Narrow bands} \\
\hline
$(J0378-J0410)_{0}$  & 0.27$\pm$0.02  & 0.49$\pm$0.02 & 0.26$\pm$0.04  & 0.33$\pm$0.13 & 0.40$\pm$0.08 & 0.82$\pm$0.18 & 772 & 0.15 & 0.11$\pm$0.16  & 1.95 & 0.01 & 0.93  & 1.00  & N  \\  
$(J0378-J0430)_{0}$  & 0.30$\pm$0.02  & 0.49$\pm$0.01 & 0.29$\pm$0.12  & 2.14$\pm$0.74 & 0.47$\pm$0.05 & 0.12$\pm$0.19 & 565 & 0.01 & 5.37$\pm$2.04  & 0.52 & 0.01 & 0.01  & 1.00  & N  \\  
$(J0378-J0515)_{0}$  & 0.69$\pm$0.02  & 0.47$\pm$0.01 & 0.62$\pm$0.34  & 0.74$\pm$0.57 & 0.55$\pm$0.12 & 0.38$\pm$0.19 & 748 & 0.54 & 0.25$\pm$1.81  & 0.47 & 0.15 & 0.86  & 1.00  & N  \\  
$(J0378-J0660)_{0}$  & 1.11$\pm$0.02  & 0.49$\pm$0.01 & 1.10$\pm$0.43  & 1.19$\pm$0.53 & 0.46$\pm$0.21 & 0.88$\pm$0.29 & 771 & 0.05 & 0.12$\pm$1.73  & 0.72 & 0.02 & 0.93  & 1.00  & N  \\  
$(J0378-J0861)_{0}$  & 1.41$\pm$0.02  & 0.54$\pm$0.02 & 1.41$\pm$0.55  & 1.41$\pm$0.50 & 0.28$\pm$0.24 & 0.58$\pm$0.20 & 762 & 0.83 & 0.01$\pm$1.75  & 0.45 & 0.23 & 1.00  & 0.99  & N  \\  
$(J0410-J0430)_{0}$  & 0.06$\pm$0.02  & 0.46$\pm$0.02 & 0.07$\pm$0.05  & 0.04$\pm$0.22 & 0.36$\pm$0.09 & 0.73$\pm$0.18 & 565 & 0.22 & 0.05$\pm$0.57  & 1.75 & 0.01 & 0.98  & 1.00  & N  \\  
$(J0410-J0515)_{0}$  & 0.41$\pm$0.02  & 0.48$\pm$0.02 & 0.27$\pm$0.33  & 0.45$\pm$0.05 & 0.70$\pm$0.15 & 0.38$\pm$0.06 & 748 & 0.79 & 0.32$\pm$0.92  & 1.35 & 0.01 & 0.83  & 1.00  & N  \\  
$(J0410-J0660)_{0}$  & 0.84$\pm$0.02  & 0.51$\pm$0.02 & 0.77$\pm$0.27  & 0.85$\pm$0.03 & 0.75$\pm$0.20 & 0.42$\pm$0.13 & 771 & 0.79 & 0.13$\pm$0.72  & 0.99 & 0.01 & 0.92  & 1.00  & N  \\  
$(J0410-J0861)_{0}$  & 1.14$\pm$0.02  & 0.56$\pm$0.01 & 0.01$\pm$0.51  & 1.16$\pm$0.19 & 0.22$\pm$0.12 & 0.54$\pm$0.08 & 762 & 0.98 & 2.76$\pm$1.24  & 0.01 & 0.77 & 0.17  & 0.60  & N  \\  
$(J0430-J0515)_{0}$  & 0.42$\pm$0.02  & 0.46$\pm$0.02 & 0.23$\pm$0.71  & 0.42$\pm$0.08 & 1.07$\pm$0.39 & 0.40$\pm$0.16 & 559 & 0.95 & 0.25$\pm$2.02  & 4.15 & 0.01 & 0.88  & 1.00  & N  \\  
$(J0430-J0660)_{0}$  & 0.84$\pm$0.02  & 0.48$\pm$0.02 & 0.84$\pm$0.47  & 0.84$\pm$0.05 & 0.77$\pm$0.37 & 0.37$\pm$0.22 & 566 & 0.80 & 0.01$\pm$1.32  & 2.85 & 0.01 & 1.00  & 1.00  & N  \\  
$(J0430-J0861)_{0}$  & 1.14$\pm$0.02  & 0.55$\pm$0.02 & 0.91$\pm$1.00  & 1.14$\pm$0.38 & 1.36$\pm$0.49 & 0.52$\pm$0.16 & 559 & 0.98 & 0.22$\pm$2.67  & 1.87 & 0.01 & 0.88  & 1.00  & N  \\  
$(J0515-J0660)_{0}$  & 0.43$\pm$0.01  & 0.28$\pm$0.01 & 0.40$\pm$0.02  & 0.51$\pm$0.07 & 0.21$\pm$0.03 & 0.40$\pm$0.06 & 749 & 0.29 & 0.36$\pm$0.25  & 1.48 & 0.01 & 0.82  & 1.00  & N  \\  
$(J0515-J0861)_{0}$  & 0.73$\pm$0.01  & 0.37$\pm$0.01 & 0.67$\pm$0.07  & 1.06$\pm$0.47 & 0.33$\pm$0.08 & 0.43$\pm$0.11 & 740 & 0.14 & 1.02$\pm$1.60  & 0.46 & 0.01 & 0.60  & 0.99  & N  \\  
$(J0660-J0861)_{0}$  & 0.30$\pm$0.01  & 0.24$\pm$0.01 & 0.29$\pm$0.13  & 1.53$\pm$0.53 & 0.23$\pm$0.05 & 0.06$\pm$0.15 & 763 & 0.01 & 7.47$\pm$3.21  & 1.81 & 0.01 & 0.01  & 1.00  & N  \\   
\hline

\hline
	\end{tabular}
    \end{scriptsize}
    \end{center}    
                \begin{small}
    \tablecomments{(1) Color. 
    (2-3) Mean and standard deviation of the first peak in the double-Gaussian model. 
    (4-5) Mean and sigma of the second peak in the double-Gaussian model. 
    (6) Total number of GCs. 
    (7) Fration of \ngc\ associated with the second peak. 
    (8) Separation of the means relative to their widths. 
    (9) GMM $p$-values based on the likelihood-ratio test $p$($\chi^{2}$), 
    peak separation $p$(DD), and 
    Kurtosis $p$(kurt) (lower $p$-values are more significant).
    (10) Kurtosis of the colors distribution. 
    (11) Bimodality final evaluation: Y (confirmation), N(discard).}
                \end{small}
\end{table*}

\begin{table*}
    \setlength\tabcolsep{4pt}
	\centering
    \begin{center}
	\caption{GMM fitting values for colors distribution at differents \rv.}
	\label{tabla:colores2}
    \begin{scriptsize}
	\begin{tabular}{l c c c c c c c c c r c c c c c c} 
\hline
\multicolumn{4}{c}{Unimodal results} & \multicolumn{7}{c} {Bimodal results} \\
Color  &     Peak&   $\sigma$&   Peak1&   Peak2 &  $\sigma1$ & $\sigma2$&   \ngc & $f_{2}$ &    D &  Kurtosis & \multicolumn{3}{c}{$p$-values}  &   Bi  \\ 
(1)&        (2)  &   (3)  &   (4)   &      (5) &    (6)&     (7)&   (8) & (9) & (10) &   (11)  &   \multicolumn{3}{c}{(12)}  &   (13) \\
\hline
\multicolumn{13}{c}{0.5~\rv} \\
\hline
$(g-i)_{0}$   & 0.88$\pm$0.02  & 0.19$\pm$0.01 & 0.67$\pm$0.06  & 0.96$\pm$0.10 & 0.04$\pm$0.04 & 0.17$\pm$0.03 & 97 & 0.71 & 2.38$\pm$0.50  & -0.60 & 0.01 & 0.44  & 0.11  & Y  \\  
$(g-z)_{0}$   & 1.07$\pm$0.03  & 0.28$\pm$0.02 & 0.99$\pm$0.11  & 1.42$\pm$0.19 & 0.22$\pm$0.05 & 0.20$\pm$0.08 & 97 & 0.19 & 2.07$\pm$1.08  & -0.07 & 0.71 & 0.56  & 0.63  & Y  \\  
\hline
\multicolumn{13}{c}{1~\rv} \\
\hline
$(g-i)_{0}$   & 0.89$\pm$0.01  & 0.20$\pm$0.01 & 0.68$\pm$0.05  & 0.98$\pm$0.09 & 0.04$\pm$0.03 & 0.18$\pm$0.03 & 301 & 0.72 & 2.26$\pm$0.37  & -0.42 & 0.01 & 0.31  & 0.06  & Y  \\  
$(g-z)_{0}$   & 1.07$\pm$0.02  & 0.28$\pm$0.01 & 0.99$\pm$0.04  & 1.45$\pm$0.13 & 0.22$\pm$0.03 & 0.18$\pm$0.05 & 301 & 0.17 & 2.32$\pm$0.59  & -0.21 & 0.09 & 0.30  & 0.27  & Y  \\  
\hline
\multicolumn{13}{c}{2~\rv} \\
\hline
$(g-i)_{0}$   & 0.87$\pm$0.01  & 0.20$\pm$0.01 & 0.68$\pm$0.01  & 0.96$\pm$0.01 & 0.05$\pm$0.01 & 0.18$\pm$0.01 & 836 & 0.70 & 2.13$\pm$0.08  & -0.27 & 0.01 & 0.20  & 0.06  & Y  \\  
$(g-z)_{0}$   & 1.03$\pm$0.01  & 0.28$\pm$0.01 & 0.97$\pm$0.03  & 1.44$\pm$0.11 & 0.24$\pm$0.02 & 0.18$\pm$0.04 & 836 & 0.13 & 2.17$\pm$0.44  & -0.11 & 0.03 & 0.17  & 0.30  & Y  \\  
\hline
\multicolumn{13}{c}{3~\rv} \\
\hline
$(g-i)_{0}$   & 0.87$\pm$0.01  & 0.20$\pm$0.01 & 0.67$\pm$0.01  & 0.95$\pm$0.01 & 0.05$\pm$0.01 & 0.18$\pm$0.01 & 1550 & 0.70 & 2.15$\pm$0.05  & -0.34 & 0.01 & 0.17  & 0.01  & Y  \\  
$(g-z)_{0}$   & 1.03$\pm$0.01  & 0.29$\pm$0.01 & 0.89$\pm$0.05  & 1.28$\pm$0.10 & 0.22$\pm$0.03 & 0.22$\pm$0.03 & 1550 & 0.36 & 1.73$\pm$0.30  & -0.24 & 0.01 & 0.21  & 0.01  & Y  \\  
\hline
	\end{tabular}
    \end{scriptsize}
    \end{center}    
                \begin{small}
    \tablecomments{Same code as in Table~\ref{tabla:colores}.}
                \end{small}
\end{table*}

\section{Transformation to Johnson-Cousin System }
Majority of the photometric data for GCs in the MW and external
galaxies are reported in the standard Johnson–Cousins UBVRI system
\citet{Bessell:1990}. Hence, in order to compare the results obtained
from our study to that obtained in other galaxies, it is necessary
to transform our u, g, r, i, and z 
magnitudes and
colors into the standard system. 
The corresponding transformation equations are discussed in detail by \citet{Jordi:2006}. 
Transformation between ugriz and Johnson–Cousins systems are given by the equations:

\label{apendice_C}

\begin{equation}
 u-g  = (0.750 \pm 0.050)\times(U-B)+ (0.770 \pm 0.070)\times(B-V) \\ + (0.720 \pm 0.040)
 \label{equ:to_transformation1}
\end{equation}

\begin{equation}
 g-V  = (0.596 \pm 0.009)\times(B-V)  - (0.148 \pm 0.007) \\
 \label{equ:to_transformation2}
\end{equation}

\begin{equation}
 g-B  = (-0.401\pm 0.009)\times(B-V)  - (0.145 \pm 0.006)  \\
 \label{equ:to_transformation}
\end{equation}

\section{GCLFs}
\label{apendice_E}

\begin{figure} 
\centering
    \includegraphics[width=0.9\columnwidth]{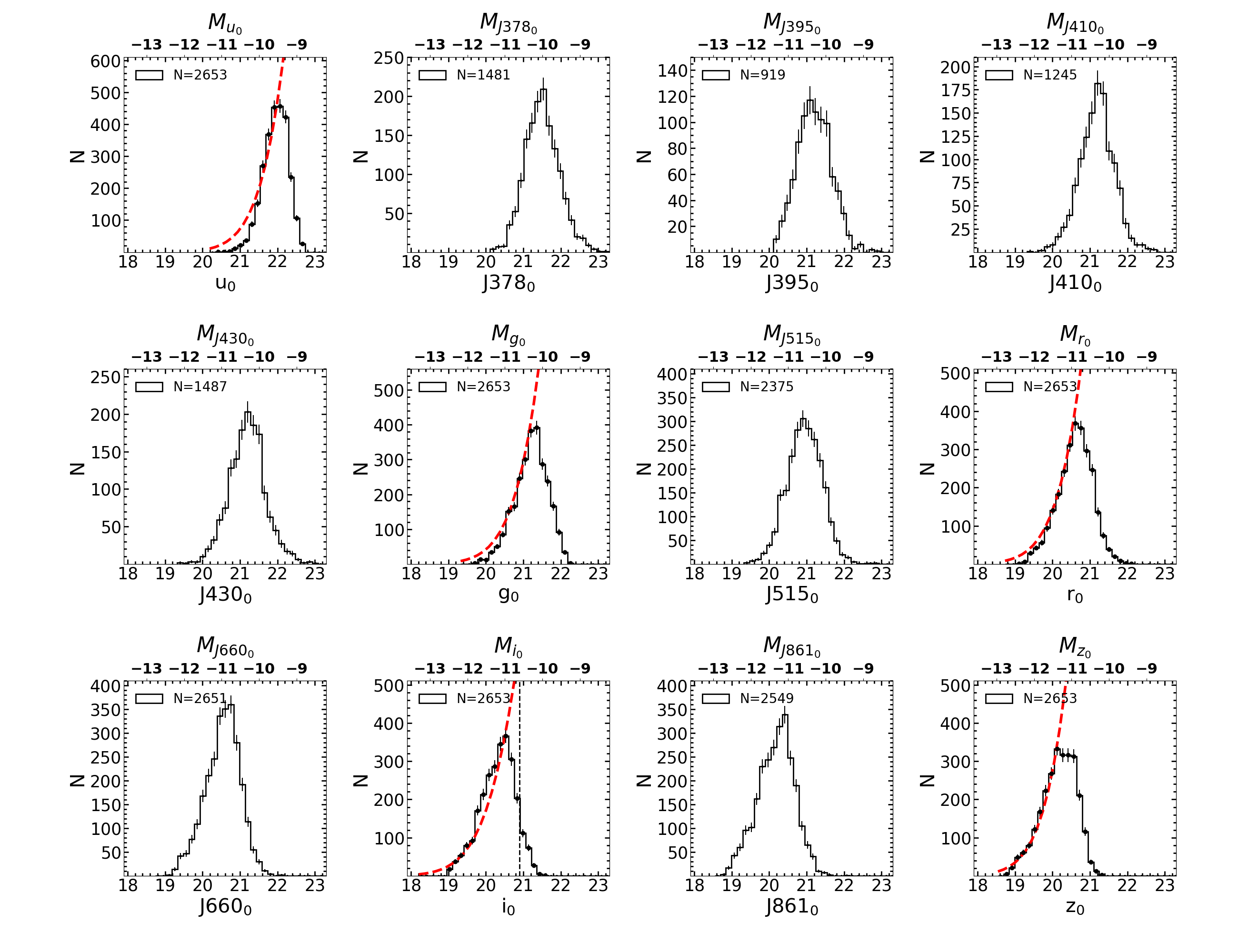}      
    \caption{Same color code as in Figure~\ref{figura:gclf}. 
    } 
    \label{figura:gclf_apendice} 
\end{figure}

For completeness we show
the 12-bands GCLFs in Figure~\ref{figura:gclf_apendice}. 

\bibliography{example}{}
\bibliographystyle{aasjournal}



\end{document}